\documentclass[preview,5p]{elsarticle}
\usepackage{lineno,hyperref}
\modulolinenumbers[5]

\usepackage[utf8]{inputenc}
\usepackage[T1]{fontenc}
\usepackage{graphicx}
\usepackage{textcomp}
\usepackage{balance}

\usepackage{paralist}
\usepackage{dblfloatfix}
\usepackage{tabularx}
\usepackage{multirow}
\usepackage{multicol}
\usepackage[para]{footmisc}
\usepackage{url}
\usepackage{listings}
\usepackage{color}
\usepackage{lipsum}             
\usepackage{xargs}   
\usepackage{subfigure}
\usepackage[pdftex,dvipsnames]{xcolor}

\usepackage{booktabs}
\usepackage{colortbl}

\usepackage{dsfont}
\usepackage{relsize}
\usepackage{natbib}

\usepackage[colorinlistoftodos,prependcaption,textsize=tiny]{todonotes}
\usepackage{xspace}
\usepackage{hyperref}

\newcommand{\al}{\textit{et al.}\xspace}
\newcommand{\ie}{\textit{i.e.,}\xspace}
\newcommand{\eg}{\textit{e.g.,}\xspace}

\definecolor{dkgreen}{rgb}{0,0.6,0}
\definecolor{gray}{rgb}{0.5,0.5,0.5}
\definecolor{mauve}{rgb}{0.58,0,0.82}

\newcommandx{\unsure}[2][1=]{\todo[linecolor=red,backgroundcolor=red!25,bordercolor=red,#1]{#2}}
\newcommandx{\change}[2][1=]{\todo[linecolor=blue,backgroundcolor=blue!25,bordercolor=blue,#1]{#2}}
\newcommandx{\info}[2][1=]{\todo[linecolor=OliveGreen,backgroundcolor=OliveGreen!25,bordercolor=OliveGreen,#1]{#2}}
\newcommandx{\improvement}[2][1=]{\todo[linecolor=red,backgroundcolor=red!25,bordercolor=red,#1]{#2}}
\newcommandx{\thiswillnotshow}[2][1=]{\todo[disable,#1]{#2}}

\newcommand{\rqOne}[1]{\emph{RQ1: What refactoring-related topics have been investigated in secondary studies?}}

\newcommandx{\rqTwo}[1]{\emph{RQ2: What smells-related topics have been investigated in secondary studies?}}

\newcommandx{\rqThree}[1]{\emph{RQ3: Which tools have been mentioned for code smell detection and refactoring support?}}

\newcommandx{\rqFour}[1]{\emph{RQ4: Which RQs have been studied on code smells and refactoring? What are the highest cited secondary studies?}}

\newcommandx{\rqFive}[1]{\emph{RQ5: What are the annual trends of types, quality, and the number of primary studies reviewed by the secondary studies?}}

\def\isdraft{1}
\if\isdraft1
	\newcommandx{\petrillo}[1]{\todo{#1}}
	\newcommandx{\guilherme}[1]{\todo{#1}}
	\newcommandx{\YANN}[1]{\textbf{$>>>$ Yann says: #1 $<<<$}}
\else
	\newcommandx{\petrillo}[1]{}
	\newcommandx{\guilherme}[1]{}
\fi

\newcommandx{\textblue}[1]{\textcolor{blue}{#1}}

\hypersetup{
    colorlinks=true,
    linkcolor=blue,
    citecolor=red
}

\journal{Journal of Systems and Software}

\begin{document}

\begin{frontmatter}

\title{Code Smells and Refactoring: A Tertiary Systematic Review of Challenges and Observations}

\author[unisinos,ufrgs]{Guilherme Lacerda\corref{mycorrespondingauthor}
\cortext[mycorrespondingauthor]{Corresponding author}}
\author[uqac]{Fabio Petrillo}
\author[ufrgs]{Marcelo Pimenta}
\author[concordia]{Yann Gaël Guéhéneuc}

\address[unisinos]{University of Vale do Rio dos Sinos\\
  Polytechnic School\\
  São Leopoldo, RS, Brazil\\ E-mail: guilhermeslacerda@gmail.com}

\address[uqac]{University of Quebec at Chicoutimi\\
Department of Computer Science \& Mathematics
\\Chicoutimi, Quebec, Canada \\ E-mail: fabio@petrillo.com}

\address[ufrgs]{Federal University of Rio Grande do Sul\\
  Institute of Informatics\\
  Porto Alegre, RS, Brazil\\ E-mail: mpimenta@inf.ufrgs.br}

\address[concordia]{Concordia University\\
Departement of Computer Science and Software Engineering
\\Montreal, Quebec, Canada \\ E-mail: yann-gael.gueheneuc@concordia.ca}
   
\begin{abstract}

Refactoring and smells have been well researched by the software-engineering research community these past decades. Several secondary studies have been published on code smells, discussing their implications on software quality,their impact on maintenance and evolution, and existing tools for their detection. Other secondary studies addressed refactoring, discussing refactoring techniques, opportunities for refactoring, impact on quality, and tools support.

In this paper, we present a tertiary systematic literature review of previous surveys, secondary systematic literature reviews, and systematic mappings. We identify the main observations (\emph{what we know}) and challenges (\emph{what we do not know}) on code smells and refactoring. 
We perform this tertiary review using eight scientific databases, based on a set of five research questions, identifying 40 secondary studies between 1992 and 2018.

We organize the main observations and challenges about code smell and their refactoring into: smells definitions, most common code-smell detection approaches, code-smell detection tools, most common refactoring, and refactoring tools. We show that code smells and refactoring have a strong relationship with quality attributes, \ie{} with understandability, maintainability, testability, complexity, functionality, and reusability. We argue that code smells and refactoring could be considered as the two faces of a same coin. Besides, we identify how refactoring affects quality attributes, more than code smells. We also discuss the implications of this work for practitioners, researchers, and instructors. We identify 13 open issues that could guide future research work.

Thus, we want to highlight the gap between code smells and refactoring in the current state of software-engineering research. We wish that this work could help the software-engineering research community in collaborating on future work on code smells and refactoring.
\end{abstract}


\end{frontmatter}

\section{Introduction}
\label{sec:introduction}

Software maintenance is an essential activity for any software system. According to Lowe \cite{Lowe:2005} and Telea \cite{Telea:2011}, 50\% to 80\% of software costs are related to maintenance activities: repairing design and implementation faults, adapting software to a different environment (hardware, OS), and adding or modifying functionalities. Software maintenance is difficult because of the lack of helpful documentation, large and complex source code becomes the only reliable source of information about a system \cite{Lowe:2005}. 

Several studies provided a broad overview of pitfalls \cite{Webster:1995}, anti-patterns \cite{Brown:1998}, and smells \cite{Fowler:1999}. Although there are many contexts where smells can be found, like models \cite{Misbhauddin:2015}, tests \cite{Garousi:2016}, requirements \cite{Alves:2016}, architecture \cite{Besker:2018, Li:2015}, and  services \cite{Sabir:2018}, our main focus is smells found in source code, a.k.a. code smells, because these impact maintainability negatively \cite{Hall:2014}.

Code smells are violations of coding design principles \cite{Fowler:2019}. They increase technical debt \cite{Kruchten:2012}, affecting software maintenance \cite{Sjoberg:2013, Yamashita:2013}, and evolution \cite{Abbes:2011, Hall:2014}. They contribute negatively to software understanding and potentially lead to the introduction of flaws \cite{Khomh:2009, Khomh:2012}. In general, developers introduce code smells in software systems when modifications and enhancements are performed to meet new requirements. The code becomes complex and the original design is broken, lowering software quality. 

Refactoring can remove code smells \cite{Fowler:1999}. Refactoring is a process of improving software systems by applying transformations that should preserve their observable behavior \cite{Wake:2003,Joshua:2004,Fowler:2019}. One of the major challenges of software-engineering research is to provide strategies for determining which refactoring to apply and when they should be applied \cite{Fowler:1999}. There are many opportunities to use refactoring \cite{AlDallal:2012, Bian:2014, Chatzigeorgiou:2017, Vedurada:2017, Terra:2018} to remove code smells. 

However, there are many problems with refactoring and the refactoring process; among other problems: (a) How to detect smells, either about code or design? (b) After detecting these smells, which refactoring should be applied? (c) What are the steps to apply these refactoring? (d) What are the gains when applying these refactoring to remove code smells? According to Mealy \cite{Mealy:2006,Mealy:2007}, these questions, without automated support, are difficult to answer \cite{Moha:2010}. Looking at the entire process and these problems, we claim that the relationship between code smells and refactoring should be further investigated.

Tertiary systematic literature reviews (SLRs) are reviews of reviews, using secondary studies, in a given area, to provide an overview of the state of the evidence in that area. The software-engineering community accepted secondary and tertiary studies as useful and helpful \cite{Kitchenham:2010, Petersen:2008}. We chose to perform a tertiary study to investigate the relationship between code smells and refactoring because it is one of the ways used by researchers to provide evidence in a specific area, which can be integrated with practical experience regarding software development and maintenance. There is a relatively high number of secondary and tertiary studies in software engineering \cite{Nurdiani:2016, Garousi:2016, Hoda:2017, Rios:2018}. Other studies \cite{Kitchenhamb:2010} reported the usefulness and value of these studies.

Thus, we perform a tertiary SLR \cite{Kitchenham:2010}, from surveys, systematic mappings, and secondary SLRs, to understand and report on the relationship (or lack thereof) between code smells and refactoring. There are large numbers of studies on code smells and refactoring, respectively, evaluate their implications and contexts, usually in the form of secondary studies, to explore these topics together. In addition to the relationship between code smells and refactoring, we also study and discuss \emph{what we know} and \emph{what we do not know} about code smells and refactoring, such as the detection of smells (types, techniques, tools) and applications of refactoring (opportunities, tools).

Our study answers five research questions (RQs):

\begin{itemize}
\item \rqOne

\item \rqTwo 

\item \rqThree

\item \rqFour

\item \rqFive 

\end{itemize}

\begin{figure*}[ht]
    \centering
    \includegraphics[width=13.5cm]{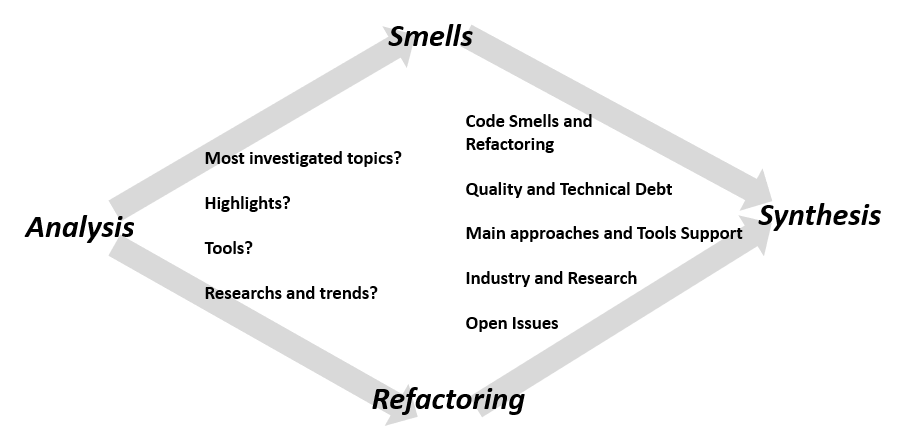}
    \caption{The strategy used in this research, from the analysis process to the consolidation of results}
    \label{fig:strategy_used}
\end{figure*}

We identify 40 secondary studies on code smells and refactoring. We analyze the secondary studies to identify the most discussed topics. We then explore these topics in detail, summarizing the observations and challenges related to these topics. We name an observation (\emph{what we know}) a consensus in the literature, whereas a challenge (\emph{what we do not know}) is a topic still open and to be better explored. Figure \ref{fig:strategy_used} summarizes our research method.

Thus, we answer our research questions and provide the following contributions:

\begin{itemize}
\item We cross-reference the most frequent code smells with their detection approaches, detection tools, suggested refactoring, and refactoring tools (Table \ref{tab:main_solutions}).

\item We report the relationships of the top 10 code smells with their refactoring and their impact on quality (Figure \ref{fig:quality_smells_refactoring_sankey}). Also, we relate internal attributes with external quality attributes using the QMOOD model \cite{Bansiya:2002}. Thus, we show that refactoring affect quality more than code smells (Figure \ref{fig:quality_smells_refactoring_qmood_sankey}).

\item We present the implications of this study from the perspective of practitioners, researchers, and instructors (Section \ref{sec:implications}).

\item We report on 13 open issues about code smells and refactoring (Section \ref{sec:open_issues}).
\end{itemize}

This paper is organized as follows: Section \ref{sec:background} defines code smells and refactoring. Section \ref{sec:methodology} presents the structure of our tertiary systematic literature review, with goal and RQs, identification of relevant literature, selection criteria, quality assessment, data extraction, and execution. Section \ref{sec:findings} shows the main findings and discussion about the observations and challenges. Section \ref{sec:qsr} discusses the relationship between quality, code smells, and refactoring. Section \ref{sec:implications} presents the implications of our research from the perspective of practitioners, researchers, and instructors. Section \ref{sec:open_issues} presents open issues useful for the future works. Section \ref{sec:threats} discusses the main threats to validity of our study. Finally, Section \ref{sec:conclusion} concludes with future work.
\section{Background}
\label{sec:background}

In this section, we present some key concepts needed to deepen the discussion about smells and refactoring.

\subsection{Smells}
\label{subsec:smells}

Although ``smell'' is a well-known practical concept, there is not a rigorous definition nor an agreement on how to categorize it and organize it. Next, we summarize the main definitions, taxonomies, and categories related to smells.

\subsubsection{Definitions}
\label{subsubsec:definitions}

The term ``smell'' refers to some internal problem in the software either at a lower level, known as code level \cite{Fowler:1999} or higher, \emph{design} \cite{ Brown:1998} describing symptoms observed in components that impair software evolution. According to such level, a smell is respectively named \emph{code smell} or \emph{design smell}.

Differently from a bug, a smell does not necessarily cause a fault in the application but may lead to other negative consequences, impacting on software maintenance and evolution. 

It is undeniable that the concept of smells was adopted, first, by the agile software development community as a way of pointing out something wrong or an improvement point \cite{HighsmithFowler:2001, Beck:2004, Amr:2007}. Currently, the industry has also adopted this term to represent anomalies in software elements.

The use of the term ``smell'' became popular mainly due to the original work of Fowler \al \cite{Fowler:1999}, who used it to identify code patterns that contain structural problems and, therefore, should be improved. Fowler \al \cite{Fowler:1999} were pioneers in identifying and discussing code smells and providing a practical guide to techniques to resolve them.

Brown \al \cite{Brown:1998} present 40 anti-patterns, which describes common occurrences for a problem that generates negative consequences. Anti-patterns are categorized in development, architecture, and project management. In Table \ref{tab:smells_brown}, the main smells related to code are presented.

\renewcommand{\arraystretch}{2}
\newcommand{\myrowcolour}{\rowcolor[gray]{0.925}}

\begin{table*}[ht]
	\caption{List of design smells presented by Brown \al~ \cite{Brown:1998}}
	\label{tab:smells_brown}
    \tiny
	\scriptsize
	\begin{center}
	   \tiny{
		\begin{tabular}{p{3cm} p{13cm} }
			\toprule
			\textbf{Smell} & \textbf{Description} \\
			\hline
		    \emph{Blob} & as know as \emph{God Class}, is a style of procedural design procedural which brings an object to have too many responsibilities (\emph{Controller}) and attributes with low cohesion, while others only save data or execute simple processes \\
		    \myrowcolour
    		\emph{Lava Flow} & dead code and forgot information frozen with design \\ 	

    		\emph{Functional Decomposition} & a procedural code in a technology that implements the OO paradigm (usually the main function that calls many others), caused by the previous expertise of the developers in a procedural language and little experience in OO \\
		    
		    \myrowcolour
		    \emph{Poltergeist} & classes that have a role and life cycle very limited, frequently starting  a process for other objects \\ 	
    
    		\emph{Spaghetti Code} & use of classes without structures, long methods without parameters, use of global variables, in addition to not exploiting and preventing the application of OO principles such as inheritance and polymorphism \\ 	
            
            \myrowcolour
    		\emph{Cut and Paste Programming} & reused code by a copy of code fragments, generating maintenance problems \\ 	

    		\emph{Swiss Army Knife} & exposes the high complexity to meet the predictable needs of a part of the system (usually utility classes with many responsibilities) \\ 	
            \bottomrule
		\end{tabular}}
	\end{center}
\end{table*}

Fowler \al \cite{Fowler:1999} described 22 smells and associated sequences of refactoring that could be applied to mitigate each smell. Such work made an important contribution by organizing and cataloging a list of smells, presented in Table \ref{tab:smells_fowler}.
Recently, Fowler updated information on smells, introducing the other 6 smells to the original list \cite{Fowler:2019}.

Wake \cite{Wake:2003} extended that list (Table \ref{tab:smells_wake}), taking into account some problematic aspects, often identified by software developers in practice. Kerievsky \cite{Joshua:2004} brought this discussion from the perspective of the application of design patterns. Gamma \al proposed design patterns \cite{GoF:1994} that provide targets for refactoring. Design Patterns represent solutions commonly used by developers to solve recurrent problems in a specific context. 
Kerievsky broadened that list, suggesting specific refactoring for the implementation of design patterns, and including a few more smells (Table \ref{tab:smells_kerievsky}).

Martin \cite{Bob:2008} seeks to identify possible problems in code from the standpoint of cleaning heuristics, that is, what is needed in terms of style rules, good practices, and discipline to keep code clean. The smells described are not only related to the code structure, but also comments, building environment, error handling, formatting, element naming, and even testing.

Although there is an agreement concerning many smells, we can also find distinct points of view, depending on the practical experience of each author. For example, Mihancea \cite{Mihancea:2006}, inheritance is considered both a good practice of OO design and, at the same time,  a problem for software maintenance and evolution. In other's works \cite{Offutt:2001, Mihancea:2009a}, particular modes of use of inheritance and polymorphism it is related to comprehension pitfalls and repetitive patterns, which can easily deceive developers during the activities of software understanding and evolution.

The way usually adopted to describe smells is the original description proposed by Fowler \al \cite{Fowler:1999}. However, Zhang \al \cite{Zhang:2009} has attempted to define a distinct approach to represent some specific smells (\eg~\emph{Data Clumps, Middle Man, Message Chains, Speculative Generality, Switch Statements}).

Among the smells previously described, it is necessary a more detailed explanation regarding \emph{Code Clones}. Although clone studies have grown in recent years, with specific communities dedicated to the subject, we have chosen to keep clone studies within our research. 
A clone \cite{Koschke:2007} is something that appears to be a copy of an original form. It is a synonym of duplicate. Although cloning leads to redundant code, not every redundant code is a clone. There may be cases in which two code segments that are no copy of each other happens to be similar or even identical by accident. Also, there may be redundant code that is semantically equivalent but has an entirely different implementation.
There are four types of clones, describe as follows \cite{Sheneamer:2016}: a) \emph{Type-1 (Exact Clone)} are the clones which look like an original code; b) \emph{Type-2 (Renamed/parameterized Clone)} is the clones where variations come in the name of literals, keywords, variables, among others; c) \emph{Type-3 (Near-miss Clone)} are changed to persist in the code in the form of addition, deletion, and modification of statements; and finally d) \emph{Type-4 (Semantic Clone)} are function or behavior of the clone remains same, but the syntax or coding of the program is different.

\begin{table*}[ht]
	\caption{List of code smells presented by Fowler \al~ \cite{Fowler:1999, Fowler:2019}}
	\label{tab:smells_fowler}
    \tiny
	\scriptsize
	\begin{center}
	    \tiny{
		\begin{tabular}{p{3cm} p{13cm} }
			\toprule
			\textbf{Smell} & \textbf{Description} \\
			\hline
			\emph{Duplicated Code} & consists of equal or very similar passages in different fragments of the same code base \\

            \myrowcolour
			\emph{Long Method/Long Function} & very large method/function and, therefore, difficult to understand, extend and modify. It is very likely that this method has too many responsibilities, hurting one of the principles of a good OO design (\emph{SRP: Single Responsibility Principle} \cite{Bob:2007})		\\

			\emph{Large Class} & class that has many responsibilities and therefore contains many variables and methods. The same \emph{SRP} also applies in this case   \\

            \myrowcolour
			\emph{Long Parameter List} & extensive parameter list, which makes it difficult to understand and is usually an indication that the method has too many responsibilities. This smell has a strong relationship with \emph{Long Method} \\ 	

			\emph{Divergent Change} & a single class needs to be changed for many reasons. This is a clear indication that it is not sufficiently cohesive and must be divided \\ 	
			
			\myrowcolour
			\emph{Shotgun Surgery} & opposite to \emph{Divergent Change}, because when it happens a modification, several different classes have to be changed \\ 	
			
			\emph{Feature Envy} & when a method is more interested in members of other classes than its own, is a clear sign that it is in the wrong class \\ 	
            
            \myrowcolour
			\emph{Data Clumps} & data structures that always appear together, and when one of the items is not present, the whole set loses its meaning \\ 	

			\emph{Primitive Obsession} & it represents the situation where primitive types are used in place of light classes \\ 	
			
			\myrowcolour
			\emph{Switch Statements/Repeated Switches} & it is not necessarily smells by definition, but when they are widely used, they are usually a sign of problems, especially when used to identify the behavior of an object based on its type \\ 	

			\emph{Parallel Inheritance Hierarchies} & existence of two hierarchies of classes that are fully connected, that is, when adding a subclass in one of the hierarchies, it is required that a similar subclass be created in the other \\
			
			\myrowcolour
			\emph{Lazy Class} & classes that do not have sufficient responsibilities and therefore should not exist \\ 	
			
			\emph{Speculative Generality} & code snippets are designed to support future software behavior that is not yet required \\ 	
			\myrowcolour
			\emph{Temporary Field} & member-only used in specific situations, and that outside of it has no meaning \\ 	
			
			\emph{Message Chains} & one object accesses another, to then access another object belonging to this second, and so on, causing a high coupling between classes \\ 	
			
			\myrowcolour
			\emph{Middle Man} & identified how much a class has almost no logic, as it delegates almost everything to another class \\ 	
			
			\emph{Inappropriate Intimacy} & a case where two classes are known too, characterizing a high level of coupling \\ 	
			
			\myrowcolour
			\emph{Alternative Classes with Different Interfaces} & one class supports different classes, but their interface is different \\ 	
			
			\emph{Incomplete Class Library} & the software uses a library that is not complete, and therefore extensions to that library are required \\ 	
			
			\myrowcolour
			\emph{Data Class} & the class that serves only as a container of data, without any behavior. Generally, other classes are responsible for manipulating their data, which is a case of \emph{Feature Envy} \\ 	

			\emph{Refused Bequest} & it indicates that a subclass does not use inherited data or behaviors \\ 	
			
            \myrowcolour
			\emph{Comments} & it cannot be considered a smell by definition but should be used with care as they are generally not required. Whenever it is necessary to insert a comment, it is worth checking if the code cannot be more expressive \\ 	
        
        	\emph{Mysterious Name} & non-significant names that do not represent the software elements \\ 	
			
            \myrowcolour
			\emph{Global Data} & it can be modified from anywhere in the code base, and there's no mechanism to discover which bit of code touched it \\ 	
        
        	\emph{Mutable Data} & it changes to data can often lead to unexpected consequences and tricky bugs \\ 	
			
            \myrowcolour
			\emph{Lazy Element} & software elements designed to grow, but do not conform with software evolution \\ 	

        	\emph{Insider Trading} & coupling problems caused by trade data between modules \\ 	
        \bottomrule
		\end{tabular}}
	\end{center}
\end{table*}

\begin{table*}[ht]
	\caption{List of code smells presented by Wake \cite{Wake:2003}}
	\label{tab:smells_wake}
    \tiny
	\scriptsize
	\begin{center}
	    \tiny{
		\begin{tabular}{p{3cm} p{13cm} }
			\toprule
			\textbf{Smell} & \textbf{Description} \\
			\hline
			\emph{Type Embedded in Name} & names used, usually defined with duplication, such as \emph{schedule.addCourse(course)} instead of \emph{schedule.add(course)}. This category also included the use of Hungarian notation and variables that reflect their type in counterpoint to their purpose or function \\ 	
            
            \myrowcolour
			\emph{Uncommunicative Names} & names used in software elements (usually attributes and local variables) that do not communicate their name/intent enough, such as \emph{x} or \emph{value1}. It is even more critical when used in methods and classes		\\

			\emph{Inconsistent Names} & same name used in different places, for different purposes  \\ 	
			
            \myrowcolour
			\emph{Dead Code} & characterized by a variable, attribute, or code fragment that is not used anywhere. It is usually a result of a code change with improper cleaning \\ 	

			\emph{Null Check} & occurrences that repeatedly appear, verifying the null values of objects \\ 	

            \myrowcolour
			\emph{Complicated Boolean Expression} & code snippets involving boolean operators such as \emph{and, or} and \emph{not} \\ 	

			\emph{Special Case} & complex conditional statements \\
            
            \myrowcolour
			\emph{Magic Numbers} & numeric values that appear deliberately in the code and that invariably do not change \\ 	
			 	
        \bottomrule
		\end{tabular}}
	\end{center}
\end{table*}

\begin{table*}[ht]
	\caption{List of code smells presented by Kerievsky \cite{Joshua:2004}}
	\label{tab:smells_kerievsky}
    \tiny
	\scriptsize
	\begin{center}
	    \tiny{
		\begin{tabular}{p{3cm} p{13cm} }
			\toprule
			\textbf{Smell} & \textbf{Description} \\
			\hline
			\emph{Conditional Complexity} & it describes that although conditional structures are not problems in themselves, the exaggerated use of them is a smell that must be tackled \\ 	
            
            \myrowcolour
			\emph{Indecent Exposure} & it happens when clients have too much access to the classes they use. It unnecessarily increases the complexity of the system	\\

			\emph{Solution Sprawl} & similar to the \emph{Shotgun Surgery}, where an update causes changes in several parts of the system  \\ 

            \myrowcolour
			\emph{Combinatorial Explosion} & it is a more subtle form, but very similar to \emph{Duplicated Code}, where several code snippets execute the same function but in objects of different types \\ 	
            
			\emph{Oddball Solution} & it occurs when there are two ways to solve the same problem on the same system, which is usually a subtle sign of \emph{Duplicated Code} \\ 	
			 	
        \bottomrule
		\end{tabular}}
	\end{center}
\end{table*}

\subsubsection{Categorization}
\label{subsubsec:categorization}

An interesting way to understand smells is through categorization, based on possible relationships between them, aiming to achieve better comprehension \cite{Mantyla:2003a}. For example, Wake \cite{Wake:2003} proposed a classification of the  smells cataloged by Fowler \al~ \cite{Fowler:1999}, with the following division:

\begin{itemize}
\item \emph{Smells within Classes}: smells identified with simple metrics (\emph{comments, long method, large class, long parameter list}), names that need to be improved (\emph{type embedded in name, uncommunicative name, inconsistent names}), unnecessary complexity (\emph{dead code, speculative generality}), code snippets that need to be removed (\emph{magic numbers, duplicated code, alternative classes with different interfaces}), and problems in conditional logic (\emph{null check, complicated boolean expression, special case, simulated inheritance}); and

\item \emph{Smells between Classes}: in this category, we find smells that represent data like lost objects, with the absence of appropriate behavior (\emph{primitive obsession, data class, data clump, temporary field}), relationship between class hierarchies (\emph{refused bequest, inappropriate intimacy, lazy class, combinatorial explosion}), balancing responsibilities (\emph{feature envy, message chains, middle man}), code changes (\emph{divergent change, shotgun surgery, parallel inheritance hierarchies}), and the lack of an incomplete library class.
\end{itemize}

Mäntylä \al~\cite{Mantyla:2003b} proposed another taxonomy of smells, presented as follows:

\begin{itemize}
    \item \emph{Bloaters:} a bloater represents any element in the code that has become very large and can not be effectively handled. In general, bloaters are difficult to understand and modify. Smells belonging to this category are \emph{Long Method, Large Class, Primitive Obsession, Long Parameter List} and \emph{Data Clumps};
    \item \emph{Object-Orientation Abusers:} workaround solutions used in the code, without exploring principles of a good OO design \cite{Bob:2007}. Smells in this category is \emph{Switch Statements, Temporary Field, Refused Bequest, Alternative Classes with Different Interfaces} and \emph{Parallel Inheritance Hierarchies};
    \item \emph{Change Preventers:} software structures very difficult to modify; in general, this difficulty may occur at one or several points. In this category we find \emph{Divergent Change} and \emph{Shotgun Surgery};
    \item \emph{Dispensables:} smells that are unnecessary and, therefore, should be deleted. Smells in this category are \emph{Duplicated Code, Lazy Class, Data Class}, and \emph{Speculative Generality};
    \item \emph {Couplers:} smells characterizing a high coupling, like \emph{Feature Envy} and \emph{Inappropriate Intimacy}.
\end{itemize}

In addition to the proposed taxonomy, Mäntylä presents in another work \cite{Mantyla:2003a} a set of metrics supporting the identification of smells, as well as a study of how effective is the use of these metrics, suggesting techniques to carry out measurements. Developers' opinions on these smells and their perceptions can vary significantly due to some factors, like experience, theoretical knowledge, and familiarity with the code in question, among others.

Perez \cite{Perez:2011} proposes another smell classification, according to problem levels. The smells are categorized as low-level and high-level smells. The low-level smells are related to particular problems in the code, such as \emph{Large Class} and \emph{Long Method}. The high-level smells relate to more complex problems that may be detected in the code structure, such as \emph{Blob} (see Table \ref{tab:smells_brown}), for instance.

Some of the low-level smells could be considered equivalent to code smells, whereas some of the high-level smells could be regarded as equivalent to the architectural/design smells.
Sometimes, high-level smells manifest by the composition of low-level smells.

In addition to several definitions of the term ``smell'' definitions, there are several ways to detect smells, ranging from human perception to metrics, rule-based strategies, search-based methods, and software visualization. In practice, tools are also built to support such detection mechanisms, an issue explored by some studies of our research.

\subsection{Refactoring}
\label{subsec:refactoring}

Refactoring is the primary approach to remove smells. Next, we summarize the main concepts, the refactoring process, and some automation aspects regarding refactoring.

\subsubsection{Definition}
\label{subsubsec:definition}

The term ``refactoring'' came from the work of Opdyke \cite{Opdyke:1992}, which defines it as reorganization strategies that support a change in a software element.
Refactoring helps to make the code more readable and eliminating possible problems, as well as improving the internal quality attributes of the software \cite{Mens:2004}. Refactoring is also used for reengineering, allowing to turn more modular and structured a specific system (legacy or decayed code) \cite{Fanta:1998}.

There are different levels of abstraction and types of software artifacts that one can apply the refactoring. For instance, it is possible to apply refactoring in UML models, database schemas, requirements, software architecture, and structures of a language \cite{Mens:2004}. So refactoring focuses not only on the source code but also on other artifacts, and for this reason, there is a need to keep all the artifacts synchronized.
Because refactoring does not change the behavior of a program, the order of application of the techniques may vary according to distinct criteria. Often, some techniques can be used in sequence, as long as they satisfy the preconditions for the techniques used \emph{a posteriori}. Other times, the sequence is arbitrary. 

The refactoring typically occurs at two levels: high and low level.  High-level refactoring (or composite refactoring) can be defined such as those referring to significant (usually macro or architectural) design changes, and low-level (or primitive refactoring) are those for small (ordinary) changes. The work of Opdyke \cite{Opdyke:1992} also introduces a fundamental element for refactoring: the precondition,  a to guarantee that the transformation preserves the program behavior. Preconditions are checked before applying the transformation to make sure that it will not introduce compilation problems or change the behavior of the program. For example, the \textit{Extract Method refactoring} checks the selected code fragment contains broken elements before to perform the refactoring. 
Opdyke states, to perform high-level refactoring, it is necessary to perform low-level refactoring.

The point is that low-level refactoring are rarely executed in isolation. Generally, they are used together, when the developers already have in mind a defined objective to apply the techniques to achieve the desired design. The pioneer thesis of Opdyke \cite{Opdyke:1992} defined 23 primitive refactoring techniques and presented three examples of composite refactoring, formed by primitive techniques. Since then, a lot of work \cite{Lopez:2014, Lahtinen:2016, Haendler:2018} has been made to improve refactoring adoption, mainly concerning the refactoring process and automation.

\subsubsection{Process and Automation}
\label{subsubsec:process}

Every refactoring can be composed of a set of simple basic steps. If a developer does not know where to start or if she feels overwhelmed, these basic steps are a good way to start. 
Sometimes such a set of basic steps is considered as a process called \textit{refactoring process}.
Mens and Tourwé \cite{Mens:2004} identified a common process in which refactoring operations: 

\begin{enumerate}
    \item Detect pieces of code with refactoring opportunities;
    \item Determine which refactoring can apply in the selected code snippet;
    \item Ensure that the selected refactoring preserve the behavior;
    \item Apply the chosen refactoring in the respective locations;
    \item Assess the effect of the refactoring on quality characteristics of the software;
    \item Maintain the consistency between the refactored artifact and other software artifacts.
\end{enumerate}

At the same time, refactoring can become a continuous improvement tool for software, especially if the team is made up of developers concerned about the quality of the implemented code.
According to Parnin \al~\cite{Parnin:2008}, one problem-related to refactoring is the benefits of software quality gained through these practices are often diluted by the high costs and low priority when compared to the urgency of bug fixes and the implementation of new features. 

This problem is because 40\% of the time invested in software maintenance is the cost to understand the software and architecture will evolve \cite{Telea:2011}.

Murphy-Hill and Black \cite{MurphyHill:2008} presents two terms that can summarize the posture concerning refactoring: \emph{floss refactoring}, that is, to adopt day-to-day refactoring techniques in a healthy and disciplined way and \emph{root channel refactoring} when there is no habit of cleaning and this can be very costly over time, with the necessity to plan.

Although the refactoring scenario initially was proposed for object-oriented languages, other researchers have applied the idea of refactoring to various paradigms, such as functional \cite{Li:2008}, logic programming \cite{Saadeh:2009}, aspects \cite{Piveta:2006}, among others. The general idea is similar to that of object-oriented languages, but the conditions and mechanics are different.

Roberts \cite{Roberts:1999}, another pioneer in this field, extended the work of Opdyke, addressing the optics of creating refactoring support tools that are fast and reliable for developers. For this, Roberts added the post-conditional element in refactoring, which describes what should or should not be valid after the application of the technique.
The \emph{Refactoring Browser} tool, created by Roberts, implements some refactorings inside the Smalltalk programming language.

Some tools proposed for various programming languages, such as Java, Smalltalk, C++, C\#, Python, and others. The goal is double: (i) to reduce possible errors and code inconsistencies and (ii) to automate the refactoring's practice. Through the tool, the developer can select the code snippet, which technique should be applied, and which parameters required for execution. The tool automatically checks the preconditions and, if all is correct, uses refactoring.
Some tools, whether commercial or academic, also have been proposed to support refactoring in the form of IDE plugins.  

For the Java language, there is \emph{JDeodorant} \cite{JDeodorant:2018} and \emph{JRefactory} \cite{JRefactory:2018}. \emph{Refactory}  \cite{Refactory:2013} allows developers to apply refactorings from UML diagrams. \emph{CppRefactory} \cite{CppRefactory:2001} is an open-source refactoring tool that automates the refactoring process in C++ projects. For C\#, \emph{Visual Assist X} \cite{VisualAssist:2018}, \emph{Code Rush} \cite{CodeRush:2018}, and \emph{ReSharper} \cite{ReSharper:2018}. \emph{XRefactory} \cite{XRefactory:1998} is a refactoring \emph{browser} for Emacs, XEmacs, and jEdit. For Visual Basic, the tool is \emph{Project Analyzer} \cite{ProjectAnalyzer:2018}. And, for Python, the tools are \emph{Rope} \cite{Rope:2018} and the \emph{Bicycle Repair Man} \cite{Bicycle:2018}.

Tsantalis \cite{Tsantalis:2010} advocates the automatic detection approach. Thus, his research group analyzes \emph{JDeodorant} as a tool that automatically detects smells. 
Tsantalis and Chatzigeorgeou \cite{Tsantalis:2011} proposed to analyze the repository of code versions, to classify the refactoring according to the number, proximity, and extent of the changes crossing with the corresponding smells. 
Another approach based on the selection of techniques using technical debt as a metric in which the debt and interest rate to pay are calculated \cite{Zazworka:2011}.
Of course, it is necessary to have some mechanism to select and rank the refactoring techniques \cite{Piveta:2009, Bruch:2009, Lee:2013}.

\section{Study Design}
\label{sec:methodology}

\begin{figure*}[htp]
    \centering
    \includegraphics[width=\linewidth]{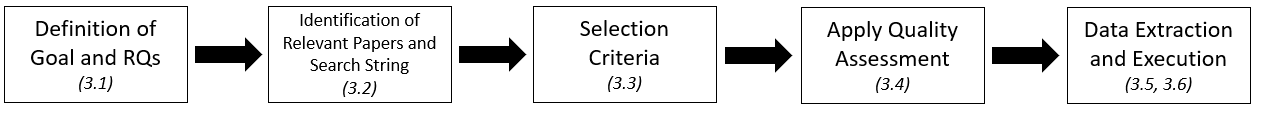}
    \caption{Steps defined of SLR}
    \label{fig:process_plan}
\end{figure*}

We realized a tertiary systematic literature review (SLR), adopting the Kitchenham \al \cite{Kitchenham:2007, Kitchenham:2009, Kitchenham:2010} SLR guidelines. The SLR followed five main steps (Figure \ref{fig:process_plan}): (1) definition of goal and research questions; (2) identification of relevant papers; (3) selection criteria, (4) quality assessment, and (5) data extraction. These steps are detailed as follows.

\subsection{Goal and Research Questions}
\label{subsec:goals}

The main goal of this work is to examine the current research works and the most important contributions of the smells and refactoring fields. At the same time, a comprehensive and systematic view can be a contribution to the research community as it can facilitate assessments and discussions of future directions related to these issues.

To structure our research, we studied and evaluated other tertiary studies \cite{Imtiaz:2013, Hoda:2017, Khan:2019}. We also wanted to understand how the research area evolved. Thus, we studied the research questions (RQs) asked in other tertiary studies in SE \cite{Kitchenham:2010, Garousi:2016, Rios:2018} to inspire our work. We built the RQs using this information.

The purpose of this SLR is to answer the following RQs:

\begin{itemize}
    \item \rqOne
    
    \item \rqTwo~Answering RQ1 and RQ2 will enable us to determine the refactoring and smells topics covered and not covered by secondary studies. Knowing the topics not covered will pinpoint the need for conducting secondary studies in those topics.
    \item \rqThree~Answering RQ3, we are investigating which tools have been mentioned to aim to smell detection. Furthermore, we are also searching for supporting tools for refactoring.
    \item \rqFour~Responding to RQ4, we analyze which aspects discussed in the studies and what the correlation between them is. Given the importance of citations to determine scientific merit, we decided to investigate what secondary studies are the most cited.
    \item \rqFive~Answering RQ5, will allow us to get a big picture of the landscape in this field and on the several studies about it.
\end{itemize}

\subsection{Identification of Relevant Literature}
\label{subsec:relevant_literature}

The search involved in eight digital libraries (see Table \ref{tab:databases}), aiming to identify relevant secondary studies, published in journals, and conferences about software engineering, software development, software maintenance and evolution, and software quality.

We used the PICOC process, the search string construction, the search engines, and the selection criteria for the returned studies. Each of them is described hereafter. PICOC is a process that aids in structuring SLRs \cite{Kitchenham:2007, Kitchenham:2009, Kitchenham:2010}. This process consists of identifying the population (P), the intervention (I), the comparison (C), the expected outcomes (O), and the (C) context of a SLR. They are:  
\begin{itemize}
    \item Population (P): Conferences and journals about Software engineering, software development, software maintenance and evolution, software quality;
    \item Intervention (I): systematic literature, systematic mapping, survey, systematic literature review; 
    \item Comparison (C): Not applicable, since the purpose of the study is to characterize the secondary studies available in the literature; 
    \item Outcomes (O): Methods, techniques, practices, application, problems, strategies, and tools described in Surveys, Systematic Mappings, and Systematic Literature Reviews;
    \item Context (C): Domain of smells and refactoring.
\end{itemize}

The strategies for the final search string were: a) derivation of terms used in the research question (example: \emph{smells}, \emph{refactoring}) and related to the RQs; b) list of keywords of papers consulted; c) use of the Boolean operator OR to incorporate synonyms; d) use of the AND operator to make the conjunction between the different keywords. The search string was built based on the following terms:

\noindent \textit{survey, systematic mapping, systematic literature review, smell, refactoring, tool, technique, method, practice, application, problem, software}

The digital libraries chosen are presented in Table \ref{tab:databases}. 
We build specific search strings to each digital library, taking into account its characteristics. Some criteria have been configured in the search engine itself. The terms used in the queries were also prioritized, depending on each search engine in the databases.

\begin{table*}[htp]
	\caption{Digital libraries, criteria considered, and search queries}
	\label{tab:databases}
    \tiny
	\scriptsize
	\begin{center}
		\begin{tabular}{p{2.5cm} p{4.5cm} p{3.5cm} p{4.5cm} }
			\toprule
			\textbf{Database} & \textbf{URL} & \textbf{Criteria} & \textbf{Query} \\
			\hline	
			 ACM Digital Library & \scriptsize http://dl.acm.org/ &  \tiny Published since 1993, Content Format: PDF &  \tiny \fontfamily{pcr} \selectfont acmdlTitle:(smell refactoring) AND recordAbstract:(survey "systematic literature" "systematic literature review" "systematic review" review "systematic mapping" "systematic study" "mapping study") AND (tool* technique* method* practice* problem*)
 \\ 	
			\myrowcolour
			IEEE Xplore  & \scriptsize http://ieeexplore.ieee.org/Xplore/ home.jsp  & \tiny Filters Applied: 1992-2018, Conferences Journals and Magazines &  \tiny \fontfamily{pcr} \selectfont (("Document Title":smell OR refactoring) AND ("Abstract":systematic literature OR systematic mapping OR mapping study OR systematic review OR survey) AND ("Publication Title":software engineering OR software quality OR software maintenance OR software evolution))
  \\ 	
			Science Direct  & \scriptsize http://www.sciencedirect.com/  &  \tiny Year 1992-2018, Find articles with these terms (1), Title, abstract and keywords (2), Publication title: Journal of Systems and Software; Information and Software Technology
            &  \tiny \fontfamily{pcr} \selectfont (1) smell OR refactoring (2) survey  OR  "systematic literature"  OR  "systematic literature review"  OR  "systematic review"  OR  review  OR  "systematic mapping"  OR  "systematic study"  OR  "mapping study"
 \\ 	
			\myrowcolour
			 Wiley InterScience & \scriptsize http://www.interscience.wiley.com  &  \tiny Date Range: 01/1992 and 12/2018, Computer Science
            &    \tiny \fontfamily{pcr} \selectfont survey OR 'systematic literature' OR 'systematic review' OR review OR 'systematic mapping' OR 'systematic study' OR study OR mapping" in Abstract and "refactoring OR smell" in Abstract AND software AND (technique* OR tool* OR method* OR practice* OR application OR problem) \\ 	

            Scopus & \scriptsize https://www.scopus.com/  &  \tiny English, Computer Science Software Engineering, Conference Paper and Chapter, before to 1992
 &  \tiny \fontfamily{pcr} \selectfont (TITLE(survey OR "systematic literature" OR "systematic literature review" OR "systematic review" OR review OR "systematic mapping" OR "systematic study" OR "mapping study") AND TITLE-ABS(refactoring OR smell)) AND ALL(tool* OR technique* OR method* OR practice* OR problem* AND software) AND ((PUBYEAR > 1992) AND (PUBYEAR < 2019)) AND (LIMIT-TO(SUBJAREA, "COMP")) AND (LIMIT-TO(DOCTYPE, "cp") OR LIMIT-TO(DOCTYPE, "ar")) AND (LIMIT-TO(LANGUAGE, "English")) \\ 	
			\myrowcolour
			AIS eLibrary & \scriptsize http://aisel.aisnet.org/  &  \tiny  Date Range: 1992-01-01 and 2019-01-01, Limited search to: All Repositories, Format: Links, Computer Sciences
            & \tiny \fontfamily{pcr} \selectfont (survey OR 'systematic literature' OR 'systematic review' OR review OR 'systematic mapping' AND software) AND abstract:(refactoring OR smell) AND (software AND (technique* OR tool* OR method* OR practice* OR application OR problem*))
 \\ 	
		Google Scholar & \scriptsize https://scholar.google.com  & \tiny 1992-2018 & \tiny \fontfamily{pcr} \selectfont  allintitle:(smell OR refactoring) AND ("systematic literature" OR "systematic mapping" OR "systematic study" OR "literature review" OR survey) \\ 	
			\myrowcolour
			Springer & \scriptsize http://link.springer.com/  & \tiny  English, Computer Science, Software Engineering, Conference Paper and Chapter, 1992 and 2018  &  \tiny \fontfamily{pcr} \selectfont (systematic literature OR mapping study OR systematic mapping OR literature review) AND (smell OR refactor*) AND (tool* OR technique* OR method* OR practice* OR problem*)
        \\ 	
        \bottomrule
		\end{tabular}
	\end{center}
\end{table*}

\subsection{Selection criteria}
\label{subsec:selection_criteria}

To select the studies returned with the search strings, we elaborate on a list of criteria for exclusion and inclusion.

The exclusion criteria used were as follows: 
\begin{itemize}
    \item articles not related to the software engineering area (development, quality, maintenance, and evolution of software);
    \item articles/studies not written in English;
    \item works presented in non-academic events in the area of computing (\eg~\emph{Agile Conference, Agiles Conf});
    \item studies such as tutorials, position papers, theses, and dissertations.
\end{itemize}

The criteria for inclusion include: 
\begin{itemize}
    \item secondary studies (Systematic Literature Reviews, Systematic Mappings, and Surveys) about smells and refactoring;
    \item articles describing the use/development/evaluation of tools, methods, practices for smells detection;
    \item articles describing the use/development/evaluation of refactoring tools, methods and techniques;
    \item works published between 1992 and 2018. We defined 1992 as the beginning of the research, because of the doctoral thesis published by Opdyke \cite{Opdyke:1992}, the first detailed written work on refactoring \cite{Fowler:2019}; 
    \item only full papers (more than 6 pages);
    \item be available online for download.
\end{itemize}

To avoid missing any potentially relevant studies then we applied the snowballing technique by checking the references of each selected study \cite{Wohlin:2014}.

\subsection{Quality Assessment}
\label{subsec:quality_assessment}

Each candidate Survey, Systematic Mapping, or Systematic Literature Review was evaluated using the same criteria adopted by previous research studies (\eg~by Kitchenham \al)~in tertiary studies \cite{Kitchenham:2009, Kitchenham:2010}. These criteria were defined by the Centre for Reviews and Dissemination (CDR) Database of Abstracts of Reviews of Effects (DARE), of the York University \cite{DARE:2018}. The criteria are four quality assessment questions, described as follows:

\begin{enumerate}
    \item Are the review’s inclusion and exclusion criteria described and appropriate? 
    \item Is the literature search likely to have covered all relevant studies?
    \item Did the reviewers assess the quality/validity of the included studies?
    \item Were the primary data/studies adequately described? 
\end{enumerate}

Kitchenham \al \cite{Kitchenham:2009, Kitchenham:2010} proposed a score for these questions. For each candidate secondary study in our pool, the quality score was calculated by assigning \{0, 0.5, 1\} to each of the four questions and then adding them up.

\subsection{Data Extraction}
\label{subsec:data_extraction}

We structured the Google Forms for data extraction. In this form, we find the main information that we consider relevant regarding the papers. In general, we consider:

\begin{itemize}
	\item Paper's information: title, authors, author\'s institution and country, year of publication, initial and final year of research, where the paper published, abstract and keywords, type of publication (journal or conference);
	\item Main contributions, evidence, and type of method research (Survey, Systematic Mapping, or Systematic Literature Review);
	\item Databases used and Research Questions defined;
    \item Amount of papers considered and analyzed;
	\item Categorization of smells research (definition, detection options, support tools, technical debt, and others) and categorization of refactoring research (techniques, opportunities, support tools, tests, and others);
    \item Research approaches (case studies, surveys, experimental and empirical studies);
    \item Type of projects (FLOSS, commercial, toy/academic) and repositories and programming languages used;
	\item Context: refactoring techniques presented, tools more cited, and smells (design/code) described. 
\end{itemize}

In addition to the spreadsheet, Mendeley\footnote{http://www.mendeley.com/} is also used to assist in the cataloging, structuring and searching for papers.
All artifacts produced from our research are available on the replication package \cite{Lacerda:2019}.

\subsection{Execution}
\label{subsec:execution}

With the research protocol defined, we started filtering these studies. As there were a large number of papers identified in the search phase, the filtering process consisted of four steps. Each step used the inclusion and exclusion criteria, and relevance of the study according to its content. We describe these steps as follows: 

\begin{enumerate}
 \item Search and delete studies based on criteria defined through reading the title, abstract and keywords;
    \item Remove duplicate papers and full-text analysis using inclusion/exclusion criteria;
    \item For selected studies, we apply the snowballing process;
    \item Define the context and categorization of the works, saving this information using the adopted tools (spreadsheet and Mendeley).
\end{enumerate}

\begin{figure*}[ht]
    \centering
    \includegraphics[width=14.5cm]{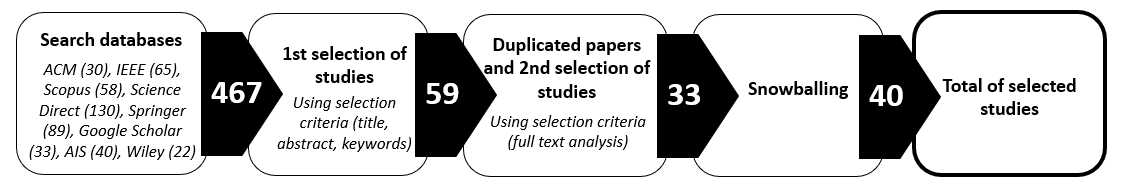}
    \caption{Steps of execution research and number of selected studies on each databases}
    \label{fig:execution_slr}
\end{figure*}

In total, there were 467 secondary studies, and we selected 59 studies. Of these 59, 26 (six duplicates and 20 based on the selection criteria) were discarded, leaving 33 selected secondary studies. After applying to snowball, another seven studies were added, totaling 40 selected studies. For each paper selected, the data were extracted and analyzed (Figure \ref{fig:execution_slr}). 

Considering all databases, we obtained an average accuracy of 12.63\% in the search, in which Google Scholar showed better accuracy (33.33\%), while Springer had the lowest accuracy (5.62\%).

\begin{figure}[ht]
 \centering
 \includegraphics[width=\linewidth]{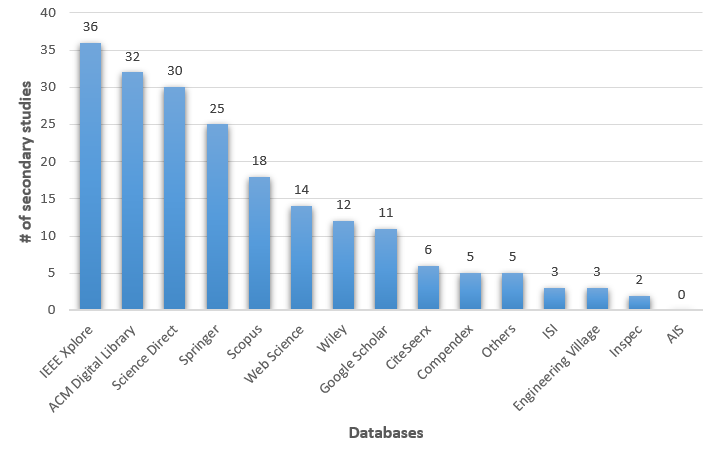}
 \caption{Databases used in the secondary studies}
 \label{fig:database_used_research}
\end{figure}

In the selected secondary studies, 19 different databases (Figure \ref{fig:database_used_research}) were used for research, highlighting IEEE Xplore, used in 90\% of the selected studies, followed by ACM Digital Library and Science Direct (80\% and 75\% respectively).

Selecting all these libraries together can lead to overlapping results, which requires the identification and removal of redundant results; however, this selection of libraries increases confidence in the completeness of the review. Our search considered all years between 1992 and 2018 to increase the comprehensiveness of the review, considering 1992 as a mark of the year of pioneer publication about refactoring \cite{Opdyke:1992}. The list of selected secondary studies found is presented in Table \ref{tab:list_studies}.

\begin{table*}[htp]
	\caption{List of selected secondary studies}
	\label{tab:list_studies}
	\scriptsize
	\begin{center}
	    \tiny{
		\begin{tabular}{p{0.3cm} p{14cm} p{0.8cm} }
			\toprule
			\textbf{\#} & \textbf{Title} & \textbf{Ref} \\
            \hline	
			S1 & A systematic literature mapping on the relationship between design patterns and bad smells & \cite{Sousa:2018} \\ 	
			\myrowcolour
			S2 & A review-based comparative study of bad smell detection tools & \cite{Fernandes:2016} \\

            S3 & UML model refactoring: a systematic literature review &
            \cite{Misbhauddin:2015} \\
			\myrowcolour
            S4 & Identifying Various Code-Smells and Refactoring Opportunities in Object-Oriented Software System : A systematic Literature Review & \cite{Singh:2018} \\

            S5 & Trends, Opportunities and Challenges of Software Refactoring: A Systematic Literature Review &  \cite{Abebe:2014} \\
			\myrowcolour
            S6 & Classification and Summarization of Software Refactoring Researches: A Literature Review Approach & \cite{Abebe:2014a} \\

            S7 & Empirical Evaluation of the Impact of Object-Oriented Code Refactoring on Quality Attributes: A Systematic Literature Review & \cite{AlDallal:2018} \\
			\myrowcolour
            S8 & Bad Smells in Software Product Lines: A Systematic Review &
            \cite{Vale:2014} \\

            S9 & The vision of software clone management: Past, present, and future & \cite{Roy:2014} \\
			\myrowcolour
            S10 & A survey of software refactoring & \cite{Mens:2004} \\

            S11 & Code Bad Smells: a review of current knowledge & \cite{Zhang:2011} \\
			\myrowcolour
            S12	& A Systematic Literature Review: Code Bad Smells in Java Source Code & \cite{Gupta:2017} \\

            S13 & A systematic literature review: Refactoring for disclosing code smells in object oriented software & \cite{Singh:2017} \\
			\myrowcolour
            S14	& A systematic mapping study on software product line evolution: From legacy system reengineering to product line refactoring & \cite{Laguna:2013} \\

            S15 & A survey on software smells & \cite{Sharma:2018} \\
			\myrowcolour
            S16 & Smells in software test code: A survey of knowledge in industry and academia & \cite{Garousi:2018} \\

            S17	& A survey of search-based refactoring for software maintenance & \cite{Mohan:2018} \\
			\myrowcolour
            S18	& A review of code smell mining techniques & \cite{Rasool:2015} \\

            S19 & Clone evolution: a systematic review & \cite{Pate:2013} \\
			\myrowcolour
            S20 & Software clone detection: A systematic review & \cite{Rattan:2013} \\

            S21 & Managing architectural technical debt: A unified model and systematic literature review & \cite{Besker:2018} \\
			\myrowcolour
            S22 & Identifying refactoring opportunities in object-oriented code: A systematic literature review & \cite{AlDallal:2015} \\

            S23 & Identification and management of technical debt: A systematic mapping study & \cite{Alves:2016} \\
			\myrowcolour
            S24 & A systematic review on the code smell effect & \cite{Santos:2018} \\

            S25 & A systematic review on search-based refactoring & \cite{Mariani:2017} \\
			\myrowcolour
            S26 & A systematic mapping study on technical debt and its management & \cite{Li:2015} \\

            S27 & Analyzing the concept of technical debt in the context of agile software development: A systematic literature review & \cite{Behutiye:2017} \\
			\myrowcolour
            S28 & Co-ocurrence of Design Patterns and Bad Smells in Software Systems : A Systematic Literature Review & \cite{Cardoso:2015} \\

            S29 & Non-Source Code Refactoring: A Systematic Literature Review & \cite{Rochimah:2015} \\
			\myrowcolour
            S30 & Survey of Research on Software Clones & \cite{Koschke:2007} \\

            S31 & A Survey of Software Clone Detection Techniques & \cite{Sheneamer:2016} \\
			\myrowcolour
            S32 & A Systematic Literature Review of Code Clone Prevention Approaches & \cite{Ali:2014} \\

            S33 & Systematic Mapping Study of Metrics based Clone Detection Techniques & \cite{Rattan:2016} \\
			\myrowcolour
            S34 & A systematic literature review on the detection of smells and their evolution in object-oriented and service-oriented systems & \cite{Sabir:2018} \\

            S35 & The Impact of Code Smells on Software Bugs: a Systematic Literature Review & \cite{Cairo:2018} \\
			\myrowcolour
            S36 & Refactoring UML Models of Object-Oriented Software: A Systematic Review & \cite{Sidhu:2018} \\

            S37 & Impact of Code Smells on the Rate of Defects in Software: A Literature Review & \cite{GRADISNIK:2018} \\
			\myrowcolour
            S38 & Empirical evidence of code decay: A systematic mapping study & \cite{Bandi:2013} \\

            S39 & Software Design Smell Detection: a systematic mapping study & \cite{Alkharabsheh:2018} \\
            \myrowcolour
            S40 & A systematic literature review on bad smells - 5 W's: which, when, what, who, where & \cite{Sobrinho:2018} \\
            
            \bottomrule
		\end{tabular}}
	\end{center}
\end{table*}

We queried smells and refactoring separately to find secondary studies from these two fields because, although they are closely related, researches have studied smell and refactoring separately. For better distribution, we organize these studies into themes like refactoring (13 studies), smells (25 studies), and both (2 studies covering both themes).

Within each theme, papers are categorized into topics.
As we show in Figure \ref{fig:strategy_used}, the topics emerged from the analyzed studies. 
To organize the topics, we used a card sorting technique \cite{Barrett:1995, Sheikh:2011}. Card sorting is a knowledge elicitation method commonly used for capturing information about different ways of representing domain knowledge. It has been used in various fields such as psychology, knowledge engineering, and software engineering.
We discuss the most cited topics in detail in Section \ref{sec:findings}\footnote{We make all categories and topics and their organization in other documents used during the research, available at \cite{Lacerda:2019}. However, We will not discuss the topics were having already distinct research communities, such as refactoring models and clones. Also, some topics that not appeared in secondary studies maybe nowadays considered as essential by the academical community: thus, a preliminary discussion of some implications of our work to the academic community is presented in Section \ref{sec:threats}.}

\begin{figure*}[ht]
\centering
\includegraphics[width=14.5cm]{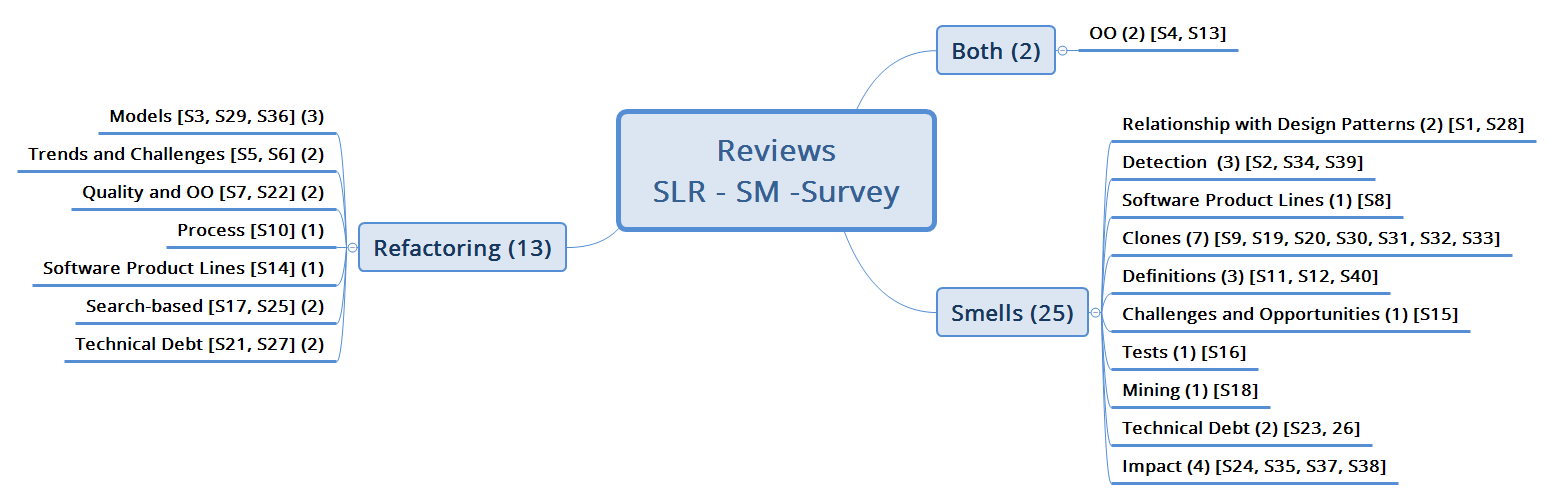}
\caption{Mind-map of secondary studies, categorized by smells, refactoring and both (in the parentheses are the numbers of studies and in the brackets their references)}
\label{fig:mindmap_studies}
\end{figure*}

The topics related to refactoring are trends and challenges, quality and object-orientation (OO), process, software product lines (SPL), search-based and technical debt (TD), and models.
The topics related to smells are relationship with design patterns, detection tools, SPLs, clones, definitions, challenges, tests, mining, technical debt (TD), and impact and effects. The topic covered by both refactoring and smells is related to object-orientation. In Figure \ref{fig:mindmap_studies}, we present a mind-map with this distribution and quantity by topics. Some abstract topics such as models and SPL are not discussed here, because our research is more focused on code-related topics like code smells and refactoring.

In Table \ref{tab:list_studies_by_journals_conf}, the list of studies by publication source and venue (Journal, Conference) is displayed. The emphasis is on Systems and Software, Information and Software Technology, and Transactions on Software Engineering, which receive secondary studies in their submissions.

\begin{table*}[htp]
	\caption{Distribution of studies by venues}
	\label{tab:list_studies_by_journals_conf}
    \tiny
	\scriptsize
	\begin{center}
		\begin{tabular}{ p{8.5cm} p{1.8cm} p{1cm} p{3.2cm} }
			\toprule
			\textbf{Source} & \textbf{Venue} & \textbf{Amount} & \textbf{Studies} \\
			\hline	
			Systems and Software  & Journal & 6 & S15, S16, S20, S21, S24, S26 \\ 	
			\myrowcolour
			Information and Software Technology  & Journal & 5 & S20, S22, S23, S25, S27 \\ 	

			IEEE Transactions on Software Engineering & Journal & 3 & S7, S10, S40 \\
			\myrowcolour
			Journal of Software: Evolution and Process & Journal & 2 & S18, S19 \\

			CSMR-WCRE (IEEE Conference on Software Maintenance, Reengineering, and Reverse Engineering) & Conference & 2 & S9, S38 \\
			\myrowcolour
			Advanced Science and Technology Letters & Journal & 1 & S6 \\

			Empirical Software Engineering & Journal & 1 & S3 \\
			\myrowcolour
			International Journal on Future Revolution in Computer Science \& Communication Engineering & Journal & 1 & S4 \\

			International Journal of Software Engineering and Its Applications & Journal & 1 & S5 \\
			\myrowcolour
			Journal of Software Maintenance and Evolution: Research and Practice & Journal & 1 & S11 \\

			Journal of Software Engineering Research and Development & Journal & 1 & S17 \\
			\myrowcolour
			Journal of Software Engineering Research and Development & Journal & 1 & S17 \\

			Science of Computer Programming & Journal & 1 & S14 \\
			\myrowcolour
			International Journal of Software Engineering and Its Applications & Journal & 1 & S29 \\

			International Journal of Computer Applications & Journal & 1 & S31 \\
			\myrowcolour
			International Journal of Software Engineering and Technology & Journal & 1 & S32 \\

            Journal of Software: Practice and Experience & Journal & 1 & S34 \\
			\myrowcolour
            Journal of Information & Journal & 1 & S35 \\
            
            Journal of Software Quality & Journal & 1 & S39 \\
            
            \myrowcolour
            IJSEKE (International Journal of Software Engineering and Knowledge Engineering) & Journal & 1 & S36 \\
			
			SAC (Symposium of Applied Computing) & Conference & 1 & S1 \\ 
            
            \myrowcolour
            EASE (International Conference on Evaluation and Assessment in Software Engineering) & Conference & 1 & S2 \\ 

            SBCARS (Brazilian Symposium on Software Components, Architectures and Reuse) & Conference & 1 & S8 \\ 
            \myrowcolour            
            SBSI (Brazilian Symposium on Information Systems & Conference & 1 & S28 \\

            IBIF (Internationales Begegnungs und Forschungszentrum fur Informatik) & Conference & 1 & S30 \\
            
            \myrowcolour
			Ain Shams Engineering Journal  & Journal & 1 & S13 \\

            ICCSA (International Conference on Computational Science and its Applications) & Conference & 1 & S12 \\
            
            \myrowcolour
            AICTC (International Conference on Advances in Information Communication Technology \& Computing) & Conference & 1 & S33 \\

            SQAMIA (Workshop of Software Quality) & Conference & 1 & S37 \\
            \bottomrule
		\end{tabular}
	\end{center}
\end{table*}

\section{Findings}
\label{sec:findings}

We compile the results from 40 selected studies to show which approaches used to detect code smells, which tools have supported this detection process, which refactoring techniques used and which are being supported by tools. 
Table \ref{tab:main_solutions} summarizes the ten most-cited smell, with the following information: the main approaches to smell detection,  the most cited smell detection tools, the most suggested refactoring for each smell, and most cited refactoring tools. We sorted each item by the number of references. Thus, we organized the main findings to help researchers and practitioners to address their research activities (we discuss some implications about practitioners, researchers, and instructors in the Section \ref{sec:implications}). \emph{Duplicated Code} was already the smell considered by the agile community as being the most cited smell \cite{Fowler:1999, Joshua:2004, Feathers:2004, Bob:2007, Bob:2008}. Also, it was the most smell studied and the most referenced smell in secondary studies [S2, S4, S11, S15, S18, S24]. Here, we grouped \emph{Code Clones} with \emph{Duplicated Code} following the classification proposed by some works [S9, S19, S20, S30, S31, S32, S33].

\begin{table*}[htp]
	\caption{List of Top-10 smells and strategies to identify/remove/mitigate}
	\label{tab:main_solutions}
	\begin{center}
	\tiny{
	\begin{tabular}{p{1.8cm}  p{3cm}  p{4.7cm}  p{4.3cm}  p{2cm} }
			\toprule
			\textbf{Smells} &  \textbf{Main approaches to detect} & \textbf{Most cited detection tools} & \textbf{Most suggested refactoring} & \textbf{Most cited refactoring tools} \\
			\hline 
			  Duplicated Code/Clones & \textbf{textual} [S2, S19, S20, S30, S31, S32, S33], \textbf{token} [S2, S19, S20, S30, S31, S32], \textbf{metrics-based} [S2, S18, S20, S30, S32, S33], tree [S2, S20, S31, S32, S33], strategies/rules [S2, S20, S31], probabilistic/search-based [S20, S31, S32, S33], visualization [S9, S30, S31] 
			  
			  & \textbf{CCFinder} [S2, S9, S19, S20, S30, S31], \textbf{CloneDr} [S2,S9, S20, S31], \textbf{Nicad} [S2, S20, S31, S40], \textbf{JCD} [S2, S30, S31], PMD [S2, S9, S40], Checkstyle [S2, S18, S40], CloneDigger [S2, S9], Duploc [S20, S31], inFusion [S2, S18, S39, S40], CP-Miner [S19, S31], Jplag [S30, S31], ConQAT/CloneDectetive [S9, S20, S40], DECKARD [S19, S20], Clever [S9, S19], DECOR [S13, S39],  inCode [S2, S39, S40], iPlasma [S2, S40], Gendarme [S2], Clone Miner [S9], SYMake [S2], CloneBoard [S9], CPC [S9], CloneScape[S9], SHINOBI [S9], JSync [S9], Cleman [S9], CeDAR [S9], CloneTracker [S19], Cyclone [S19], CloneInspector [S19], Columbus [S19],  Datrix [S19], SimScan [S19], Simian [S19, S20], SmallDude [S19], iClones [S19], CLAN/Covet [S20, S31], ARIES [S33], JCodeCanine [S13], JDev [S13], JCCD, SAME, Borland Together, SonarQube, Stench Blossom, InsRefactor [S40] 
			  
			  & \textbf{Extract Method} [S4, S7, S9], \textbf{Pull Up Method} [S4, S9], \textbf{Rename Method} [S4, S9], \textbf{Replace Constructor With Factory} [S4, S30],
			  Extract Class [S4],  Form Template Method [S4],  Push Down Method [S9], Push Up Method [S4], Substitute Algorithm [S13], Move Method [S9], Extract Superclass [S9], Extract utility-class [S9] 
			  
			  & \textbf{JDeodorant} [S39, S40], Wrangler [S2], CodeRush [S9] \\ 	
            \myrowcolour
            Large Class/God Class 
            
            &  \textbf{metrics-based} [S2, S15, S18], \textbf{strategies/rules} [S2, S15, S38], probabilistic/search-based [S15, S38], history-based [S15], optimization-based [S15], visualization [S13]  
            
            & \textbf{DECOR} [S2, S18, S38, S39, S40], PMD [S2, S18, S39, S40], Gendarme [S2], inCode [S2, S39, S40], inFusion [S2, S39, S40], iPlasma [S2, S39, S40], Checkstyle [S2, S18, S40],  SDMetrics [S2, S40], Weka [S13], HIST [S4, S40], Stench Blossom [S13, S40], BSDT [S13], CodeNose [S18], JDev [S13], JCosmo [S13], 2D-DSL, NosePrints, Prodetection, P-EA, BLOP, History Miner, BBT, TACO, SMURF, EvolutionAnalyzer, Van, Borland Together, Understand, Pascal Analyzer, SCOOP/Organic, CodeVizard, IYC, MuLATo, SpIRIT, InsRefactor, JCodeOdor, JSNOSE, HULK, Paprika, Metrics, SourceMiner [S40] 
            
            & \textbf{Extract Class} [S4, S13], \textbf{Extract Subclass} [S4, S13], Replace
            Data Value with Object [S4], Extract Interface [S4], Duplicate Observed Data [S13]  
            
            &  \textbf{JDeodorant} [S2, S18, S39, S40], TrueRefactor [S17]   \\ 	
            
            Feature Envy 
            
            & \textbf{strategies/rules} [S2, S13, S15], \textbf{metrics-based} [S13, S18, S38],  history-based [S15], optimization-based [S15], probabilistic/search-based [S13, S15], visualization [S13]  
            
            & \textbf{iPlasma} [S2, S13, S39, S40], IntelliJ IDEA [S2], Stench Blossom [S13, S40], JSpIRIT [S2, S40], NosePrints [S2, S40], Weka [S13], HIST [S4, S40], JCosmo [S13], SACSEA [S13], JCodeCanine [S13], inFusion [S18, S39, S40], inCode [S13, S39, S40], CodeNose [S18], DECOR, Prodetection, P-EA, BLOP, TACO, Fluid Tool, Methodbook, Borland Together, Understand, SCOOP/Organic, MuLATo, SourceMiner [S40] 
            
            & Move Method [S4, S13], \textbf{Extract Method} [S4, S7, S13], Move Field [S4]  
            
            & \textbf{JDeodorant} [S13, S18, S40], JMove [S40]   \\ 	
            
            \myrowcolour
            Long Method 
            
            & \textbf{metrics-based} [S2, S15, S18], \textbf{strategies/rules} [S2, S13, S15],  
            probabilistic/search-based [S15]
            
            & \textbf{Checkstyle} [S2, S13, S18, S40], \textbf{PMD} [S2, S18, S39, S40], TACO [S24], Stench Blossom, \textbf{DECOR} [S18, S39, S40], inFusion, JSpIRIT, CodeNose [S18], Gendarme [S2], iPlasma [S39, S40], inCode, 2D-DSL, NosePrints, Prodetection, TACO, Teamscale, Borland Together, Understand, SCOOP/Organic, IYC, MuLATo, InsRefactor, ConQAT [S40]  
            
            & \textbf{Extract Method} [S4, S7, S13], \textbf{Replace Temp with Query} [S4, S7, S13], Introduce Parameter Object and Preserve Whole Object [S13], Replace Method with Method Objects [S7, S13], Decompositional Objects [S7, S13]  
            
            & \textbf{JDeodorant} [S2, S13, S18, S39, 40], TrueRefactor [S2, S17] \\ 	
            
            Long Parameter List 
            
            & \textbf{strategies/rules} [S2, S13, S15], \textbf{metrics-based} [S15, S18],  optimization-based [S15] 
            
            &  \textbf{PMD} [S2, S13, S18, S39, S40], Checkstyle [S2, S18, S40],  DECOR [S18, S39, S40], CodeNose [S18], JDev [S13], SACSEA [S13], iPlasma [S39, S40], inCode, inFusion, 2D-DSL, NosePrints, P-EA, BLOP, Borland Together, Understand, Pascal Analyzer, SCOOP/Organic, MuLATo, SpIRIT, InsRefactor, JSNOSE, Metrics, SDMetrics [S40] 
            
            & \textbf{Replace Parameter with Method} [S4, S13], Preserve the Whole Object [S4], Introduce Parameter Object [S4]  
            
            & JDeodorant [S40]   \\ 	
            
            \myrowcolour
            Divergent Change 
            
            & \textbf{strategies/rules} [S13, S15], metrics-based [S18], history-based [S15]  
            
            & \textbf{HIST} [S4, S13, S40], DECOR [S39], Borland Together, Understand, inFusion, inCode, SCOOP/Organic, MuLATo, SourceMiner [S40] 
            
            & \textbf{Extract Class} [S4, S13] 
            
            &  JDeodorant [S40]  \\ 	
            
            Data Clumps 
            
            & \textbf{metrics-based} [S2, S18], strategies/rules [S2], tree [S2], visualization [S13] 
            
            & CBSDetector [S2, S40], \textbf{inCode} [S2, S39, S40], \textbf{inFusion} [S2, S39, S40], IntelliJ IDEA [S2], Stench Blossom [S2, S40], NosePrints, Borland Together [S40]
            
            & Introduce Parameter Object [S4, S13], Extract Class [S4, S13], Preserve Whole Object [S4, S13]  
            
            & *   \\	
            \myrowcolour
            Refused Bequest 
            
            & \textbf{metrics-based} [S13, S15, S18], strategies/rules [S15], visualization [S13] 
            
            & \textbf{iPlasma} [S13, S39, S40], \textbf{inCode} [S2, S39, S40], \textbf{inFusion} [S2, S39, S40], IntelliJ IDEA [S2], Stench Blossom [S2], DECOR [S39, S40], 2D-DSL, NosePrints, Prodetection, Borland Together, SpIRIT [S40]
            
            & \textbf{Replace Inheritance with Delegation} [S4, S13], Push Down Method [S13], Push Down Field [S13]  
            
            & *   \\ 	
            
            Shotgun Surgery 
            
            & \textbf{metrics-based} [S15, S18], strategies/rules [S13, S38], optimization-based [S15], history-based [S15] 
            
            & \textbf{HIST} [S4, S13, S40], inFusion, inCode, iPlasma [S39, S40], DECOR [S39, S40], Prodetection, P-EA, BBT , Borland Together, Understand, SCOOP/Organic, CodeVizard, MuLATo, SpIRIT, JCodeOdor [S40]  
            
            & \textbf{Move Method} [S4, S13],Move Field [S4], Inline Class [S4],  Move Class [S13] 
            
            & TrueRefactor [S17]   \\
            
            \myrowcolour
            Lazy Class 
            
            & metrics-based [S18]
            
            & BSDT [S13], DECOR [S39] 
            
            & \textbf{Collapse Hierarchy} [S4, S13], Inline Class [S4, S13]  
            
            & TrueRefactor [S2, S17]   \\ 	
            
            \bottomrule
		\end{tabular}}
	\end{center}
\end{table*}

\begin{table}[htp]
 \centering
 \caption{Tools and secondary studies related}
 \label{tab:tools_studies}
 \includegraphics[width=8.4cm]{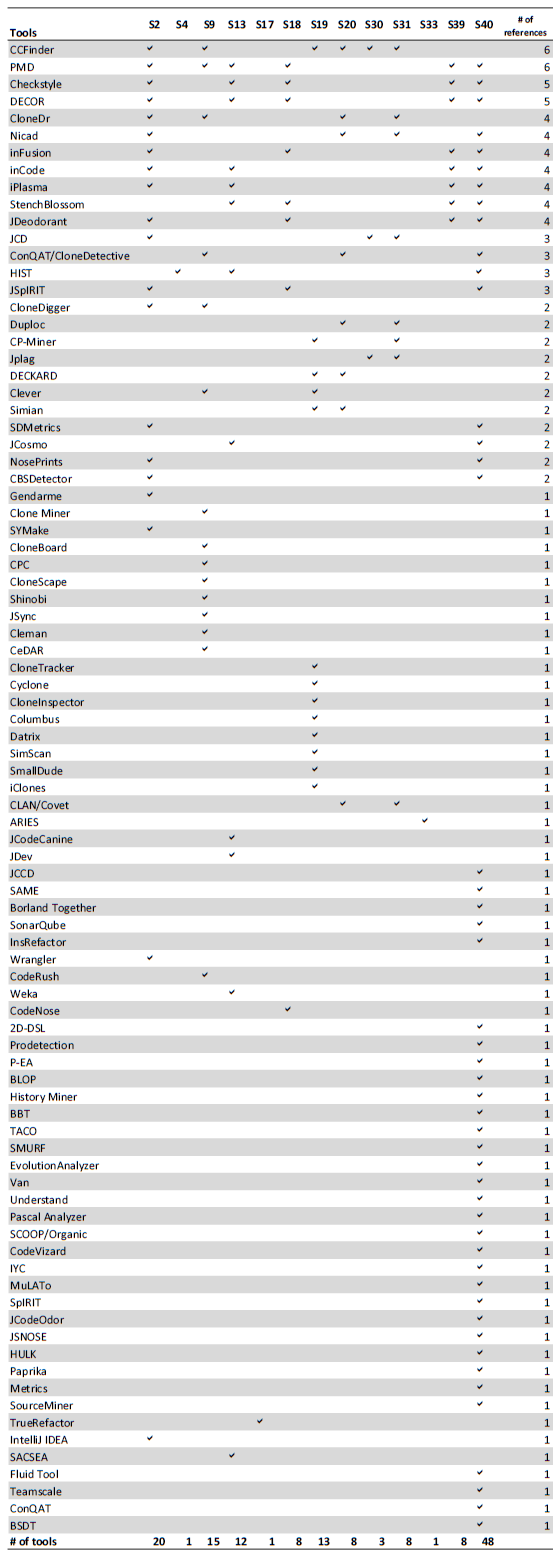}
\end{table}

Next, we present our findings in subsections grouped by RQs.

\subsection{RQ1: What refactoring-related topics have been investigated in secondary studies?}
\label{subsec:rq1}

In our research, as shown in Figure \ref{fig:mindmap_studies}, we found 13 secondary studies related to refactoring [S3, S5, S6, S7, S10, S14, S17, S21, S22, S25, S27, S29, S36].

\begin{figure}[ht]
 \centering
 \includegraphics[width=\linewidth]{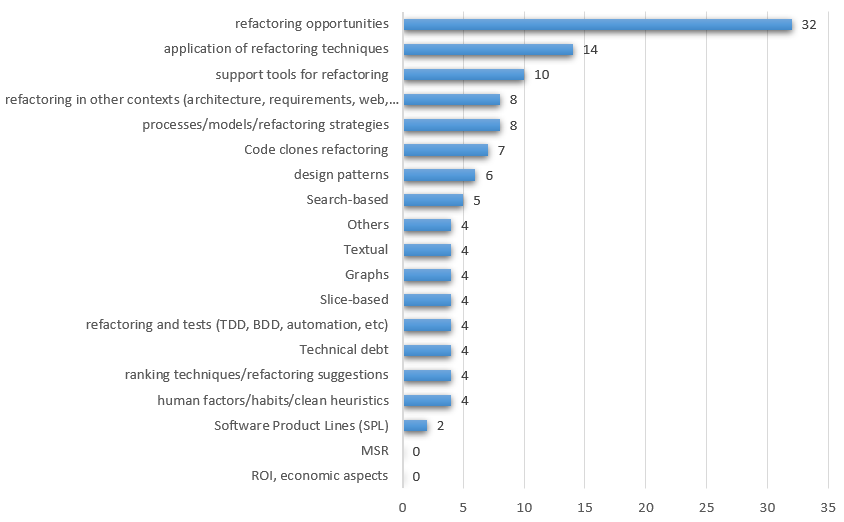}
 \caption{\textblue{Categorization of refactoring topics}}
 \label{fig:refactoring_context}
\end{figure}

We categorize the refactoring topics of such 13 studies as they were discussed in the literature, plotting the Figure \ref{fig:refactoring_context}. 
A summary of the main highlights found in these studies and a discussion of some points are presented here.

\subsubsection{Refactoring Techniques Highlights}
\label{subsec_rq1_techniques}

\begin{figure}[ht]
 \centering
 \includegraphics[width=\linewidth]{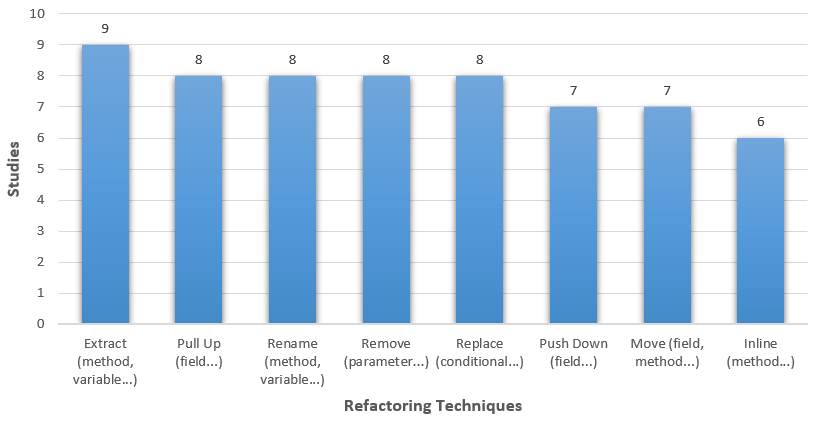}
 \caption{Top 10 refactoring techniques}
 \label{fig:top10_refactoring}
\end{figure}

Ten studies mention refactoring techniques [S4, S7, S8, S10, S13, S14, S22, S25, S30, S31]. We presented the top 10 refactoring more quoted on studies (Figure \ref{fig:top10_refactoring}).
The techniques that \textbf{most appear are extraction techniques (\eg~method, variable, class)} [S4, S7, S8, S10, S13, S14, S22, S25, S30]. 

Extraction techniques are quoted as a technique used in 5 smells more cited (Table \ref{tab:main_solutions}).
For instance, \emph{Extract Class} can be applied in \emph{Duplicated Code/Clones, Large Class/God Class, Divergent Change}, and \emph{Data Clumps}. We note that the same refactoring can be applied to more than one smell. Of course, in these cases, the developer should take context into account.

Also, the most cited refactoring are also the most studied, according to [S5, S7, S22, S25]. These studies show that the researchers are more interested in the \emph{Extract Class} and \emph{Extract Method} than in other refactoring techniques. \emph{Move Method} also deserves a highlight. The high interest in these techniques may indicate their significant importance in the software industry. However, although these techniques could be potentially more frequently applied during the refactoring process than other refactoring activities due to their influence [S7, S22], \emph{Extract Class} and \emph{Extract Superclass} is rarely used in practice, as also shown in [S22]. 
Other techniques such as \emph{Rename Field, Rename Method, Inline Temp}, and \emph{Add Parameter} are among the most used techniques in practice. Still, surprisingly we did not find studies that highlight opportunities for their appropriate application. Still, \emph{Rename Method} is the most commonly used automated refactoring [S22].

Most of the refactoring described in the studies are the same defined by Fowler \al~\cite{Fowler:1999}. However, the \textbf{number of techniques explored is still small}. 
Indeed, the studies [S7, S25] report a low number of considered techniques (20 and 27, respectively).

In practice, it is challenging for the developer to identify refactoring opportunities, that is, to determine which type of refactoring should be applied to correct a smell [S25]. 
The \textbf{relationship between smells and refactoring is not
a one to one relationship} [S7, S25]. We can apply more than one refactoring technique to a smell. It may even be necessary to combine more than one refactoring to remove it or reduce its impact. Also, some refactoring can be applied in more than one smell.  

These observations suggest that \textbf{there is a gap between refactoring practice and research for the topic of identifying refactoring opportunities}. These results point up opportunities to evaluate unexplored or underutilized refactoring, which could (or not) be applied together.

\subsubsection{Refactoring Opportunities}
\label{subsec_rq1_context}

The \textbf{refactoring opportunities, application of refactoring and tools support are the most studied} topics [S3, S5, S7, S10, S14, S17, S21, S22, S25, S27].

The most commonly used approaches to identifying refactoring opportunities are quality metrics oriented, pre-condition oriented, and clustering oriented [S7, S17, S22].

Quality metrics oriented approach is used to predict and identify refactoring opportunities (\emph{Extract Subclass, Extract Superclass, Extract Class, Move Method, Extract Method, Pull Up Method, Form Template Method, Parameterize Method}, and \emph{Pull Up Constructor}) [S7, S17, S22]. Most metrics are related to coupling, cohesion, and distance (similarity) between code elements, and the studies described several distinct ways to calculate such metrics. 

Pre-condition oriented approach is used to identify refactoring opportunities (\emph{Move Method}), mainly related to \emph{Feature Envy} and \emph{Code Clones} smells [S17, S22].
This approach firstly evaluates a condition just before applying a refactoring technique. Such a condition is also usually related to some metrics. 

Clustering oriented approaches use algorithms based on some similarity measure and combination of code elements (\eg~lines of code, attributes, methods, and classes), for refactoring opportunities (\emph{Extract Method, Move Method, Move Class, Move Field, Inline Class}, and \emph{Extract Class}) [S17, S22].

Graph-oriented approach, code slicing, and dynamic analysis are other approaches to the identification of refactoring opportunities (\eg~\emph{Move Method, Extract Method}, and \emph{Extract Class}) [S22]. Such approaches are mainly useful to discover different refactoring opportunities in SPL [S14], and models [S3, S29, S36]. 
Too, code slicing is a practical approach to small code snippets (\eg~methods), although having scalability issues [S22].

Moreover, the study [S22] relates that only a single approach for specific refactorings (Graph-oriented approach used to \emph{Extract Interface} and clustering-oriented approach used to \emph{Inline Class}).

Search-based approaches use to detect refactoring opportunities and to evaluate their applicability (application, behavior preservation, impact) [S17, S25].
The technique most used is the adoption of evolutionary algorithms (\emph{Genetic Algorithms}), with highlights for \emph{Hill-Climbing Search} [S17, S25]. 
The search-based approach has also explored this opportunity for pattern-oriented refactoring \cite{Joshua:2004}, with studies of patterns \emph{Template Method, Decorator, Abstract Factory}, and \emph{Factory Method} [S17].

Some studies (see [S7, S22]) indicated that researchers generally compare the results obtained from one refactoring with the results of other refactoring that use the same identification approach. 
One of the key open issues in this area is analyzing the results of applying different approaches for identifying refactoring opportunities for a specific activity to determine the best approach. 
Indeed, examining how refactoring techniques using different identification approaches can be further explored in future work. 

Another important aspect of refactoring is how to apply it. The application of refactoring can \emph{direct} or \emph{indirect} [S25]. In the direct approach, the refactoring is applied directly to the artifact (\eg~code and model), and thus it can be easily automated. In this case, the preservation of the behavior of refactoring is ensured.
In the indirect approach, a sequence of refactoring produced as an optimized intermediary solution, and later that sequence is applied to the artifact. Thus, the artifact is indirectly optimized.
Most of the work reported by [S25] uses \textbf{indirect approach and one possible reason  may be the difficulty in ensuring the preservation of behavior}.

Automation is one of the critical difficulties in performing comparative studies among approaches. According to [S25], the refactoring process addresses six tasks [S10]. However, there is not a fully automatic approach for the whole software refactoring activity by solving all these tasks.
Based on the results presented in [S25], the most difficult tasks are: (1) assure that the applied refactoring preserves behavior, (2) implement the refactoring; and (3) maintain the consistency between the refactored artifact. Still, according to [S25], \textbf{one of the problems of automating this task is the difficulty in preserving the behavior}.  

In addition to automation, another interesting topic is the application of refactoring with tools support, discussed later in Sub-section \ref{subsec:rq3}.

\subsubsection{Impact on Software Quality}
\label{subsec_rq1_impact}

The studies [S4, S5, S7, S22] indicated that different refactoring sometimes have an opposite impact on different quality attributes.
Performing unnecessary refactoring (changes in a code that does not need to refactored)  may unexpectedly cause the code quality to degrade instead of being improved.
Therefore, \textbf{refactoring does not always improve all software quality aspects}.

The studies [S7, S22] investigated the impacts of a few individual refactoring on some internal quality attributes such as cohesion, coupling, complexity, inheritance, and size. However, such studies were not able to identify impacts on external and other internal quality attributes.

Researchers were more interested in exploring the impacts of \emph{Move Method, Extract Class}, and \emph{Extract Method} on quality than the impact of any other refactoring [S7, S22]. Still, according to [S7, S13], researchers took two main approaches in studying the effect of refactoring on quality. The first approach is identifying refactoring opportunities, determining those required to remove code bad smells, performing refactoring when it is applicable, and comparing the code quality before and after refactoring. 
The second approach is analyzing the changes implemented on code during the maintenance phase, detecting the changes due to refactoring, and comparing before and after the code quality.  

Each refactoring scenario includes: (1) a summary of the situation where refactoring is necessary, (2) a motivation for the importance of performing the required refactoring, and (3) a mechanism describing how to implement the refactoring. The study [S7] relates some refactoring and quality impact.

\emph{Extract Class} was found to have a potentially positive impact on cohesion, inheritance, and size, and a potentially negative effect on complexity and coupling. 
\emph{Extract Subclass} has a potentially negative impact on complexity and an inconsistent impact on cohesion and coupling. 
\emph{Inline Class} has a potentially positive impact on cohesion, coupling, and complexity, but it has an opposite effect on inheritance. \emph{Extract Method} has a potentially positive effect on cohesion, complexity, and size, and it does not affect inheritance and coupling (in most cases). 
\emph{Move Method} has a potentially positive impact on cohesion and a potentially negative impact on coupling and complexity. 
\emph{Move Field} has a potentially positive effect on cohesion and a potentially negative effect on the coupling. 
\emph{Encapsulate Field} has a potentially positive impact on complexity, an inconsistent impact on coupling and cohesion, and does not affect inheritance. 
\emph{Replace Data Value with Object} has potentially positive implications for cohesion and a potentially negative impact on the coupling. 
Finally, \emph{Replace Method with Method Object} has a potentially positive impact on the coupling.

We observe such information about positive or negative impacts are very relevant to support developers in applying refactoring and assessing which refactoring used, not only to eliminate smells but also to improve quality aspects.
Additionally, it is beneficial for \textbf{expanding studies on the impacts on quality in other refactoring, not yet explored}.
In Section \ref{sec:qsr}, we discuss more deeply the relationship between refactoring and quality.

\subsubsection{Software Evolution and Technical Debt}
\label{subsec_rq1_evolution_td}

There is nowadays an agreement in SE community about technical debt: \textbf{refactoring are the primary approach to minimize the effects of technical debt} [S21, S26, S27]. 
Besides, if refactoring is overlooked, it can lead to a development crisis in the long run [S21, S27].

Decision-making about refactoring is a challenge because costs are concrete and immediate. In contrast, the benefits of refactoring are vague,  long-term, and historically very difficult for the developers to quantify or justify [S21, S23].  The identification and application of refactoring may introduce new problems, and therefore, complicating the analysis.

One strategy to identify refactoring candidates [S21] is to locate the architecturally relevant classes as they are the pillar classes of the software design. To this end, we need finding classes that have earlier been frequently refactored together with looking for classes that are harmful to the system's design. The classes are prioritized and sorted according to their impact on the overall system's quality. However, as reported by [S21, S23, S26, S27], there is a lack of studies that conclusively describe the data caused by such code changes.
Therefore, \textbf{architectural refactoring is risky, difficult to estimate, and very difficult to prioritize}.

\noindent
\fcolorbox{black}{lightgray}{
    \parbox{\linewidth}{%
        {\footnotesize \textbf{RQ1 Summary} 
        
        \emph{\textbf{Challenges:}} The relationship between smells and refactoring is not a one to one relationship. It brings us numerous challenges regarding refactoring, such as (i) which refactoring can combine, (ii) which can not combine, (iii) which have the most significant impact on quality, and (iv) which detracts from the quality of software. 
        
        Although we have a large amount of research associated with refactoring, we still need to bring it closer to practice, encouraging researchers to improve the results most commonly used in practice. Another challenge is how to analyze the results obtained in the application of the refactoring.
        
        Comparing how refactoring can use different identification approaches can be further explored in future work.
        
        \emph{\textbf{Observations:}} Extraction techniques are the most mentioned in the secondary studies. However, the number of techniques explored is still small (between 20 and 27 of 72). 
        Refactoring opportunities, application of refactoring, and refactoring tools are topics that most appear in the studies. Quality metrics-oriented approach, precondition-oriented approach, and clustering-oriented approach are the most cited approaches to identify refactoring opportunities. 
        
        Refactoring is the first approach to minimize technical debt effects. However, some refactoring, when applied, negatively affect the quality of the software.   
        }
        
    }%
}

\subsection{RQ2: What smells-related topics have been investigated in secondary studies?}
\label{subsec:rq2}

As noted in Figure \ref{fig:mindmap_studies}, most of the selected studies refer to the smells [S1, S2, S8, S9, S11, S12, S15, S16, S18, S19, S20, S23, S24, S26, S28, S29, S30, S31, S32, S33, S34, S35, S36, S37, S38, S39, S40].
In this section, we focused on design and code smells, because they are the most cited subjects in the selected studies and also due to their direct relationship with the code. The following are the main points.

\subsubsection{Design Smells Highlights}
\label{subsec_rq2_designsmells}

\begin{figure}[ht]
 \centering
 \includegraphics[width=\linewidth]{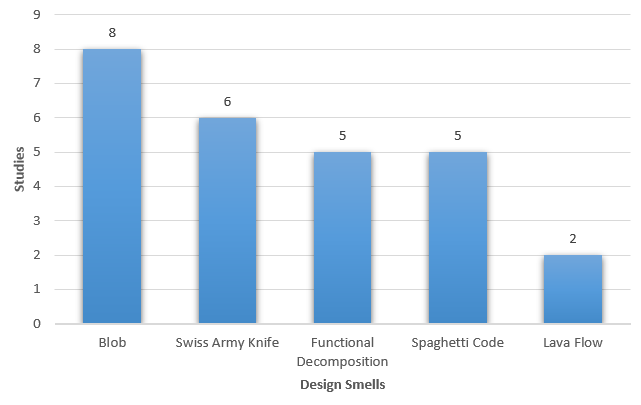}
 \caption{Top 5 design smells}
 \label{fig:top5_design_smells}
\end{figure}

Although studies primarily focus on code smells, papers about design smells are also found.
Fourteen studies quoted design smells [S1, S2, S3, S13, S15, S19, S24, S28, S34, S35, S37, S38, S39, S40]. The top 5 design smells (Figure \ref{fig:top5_design_smells}) were defined by Brown \al \cite{Brown:1998}. 

\emph{Blob} is the most mentioned design smell in the studies [S1, S3, S13, S24, S28, S34, S35, S37, S39, S40]. 
Some reasons can justify this mention. 
Blobs are easy to detect, and there are a variety of tools that identify this type of design smell. Table \ref{tab:main_solutions} presents a list of detection approaches and tools to detect \emph{Blob}.
Also, \emph{Blob} is used in some studies as synonymous with \emph{Large Class} \cite{Fowler:1999} or \emph{God Class} \cite{Riel:1996}. 
Previously, we presented the \emph{Blob} definition by Brown (see Table \ref{tab:smells_brown}) and the \emph{Large Class} definition by Fowler (see Table \ref{tab:smells_fowler}).
However, in other works, \eg [S34], this differentiation is made. Here, we have maintained this differentiation by considering the original definitions of distinct authors.

\textbf{We can observe, in these cases, problems related to smells nomenclature}. We also note that these naming and definition problems also occur with other smells.
For example, \emph{Copy and Paste Programming} has been used as a synonym for \emph{Duplicated Code/Clones}. We were identifying distinct studies [S11, S12, S15, S39, S40] using different smells names to describe the same problem in design and code. 

Some design smells usually appear together in many studies. For example, the following studies  [S1, S3, S13, S24, S34, S39, S40] that discussed \emph{Blob} also approached \emph{Spaghetti Code, Swiss Army Knife, Lava Flow, Functional Decomposition}, and \emph{Poltergeist}.
Many works often consist of evaluating a given design smell or even its relationship with other design smells.

The \textbf{design smells has also been studied together with code smells}. The studies [S28, S35, S37, S39] reported relations between the following pairs of design smells and code smells (\eg~\emph{Blob, Data Class}, and \emph{Blob, Large Class}).
Others pairs of design smell we found are: (\emph{God Class, God Method}), (\emph{God Class, Feature Envy}), (\emph{God Class, Data Class}), (\emph{God Class, Duplicated Code}), (\emph{Data Class, Data Clumps}), (\emph{Divergent Change, Shotgun Surgery}), and (\emph{Divergent Change, Shotgun Surgery, Feature Envy, Long Method}) are also related [S28, S39].

\subsubsection{Code Smells Highlights}
\label{subsec_rq2_codesmells}

\begin{figure}[ht]
 \centering
 \includegraphics[width=\linewidth]{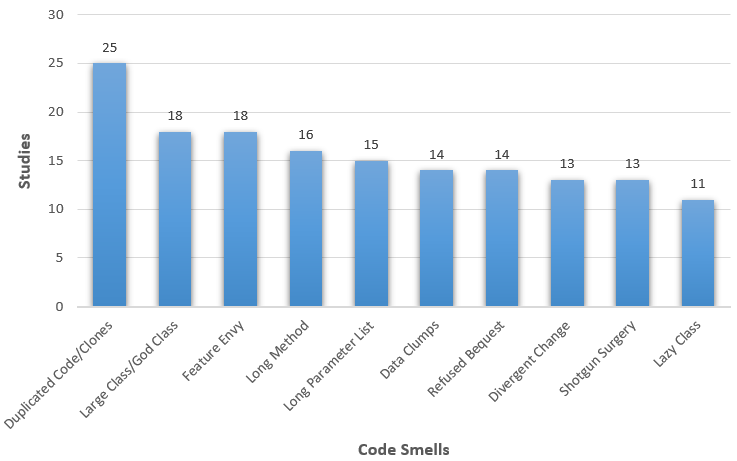}
 \caption{Top 10 code smells}
 \label{fig:top10_code_smells}
\end{figure}

The code smells initially defined by Fowler \al \cite{Fowler:1999} are the most mentioned. We presented top 10 most quoted code smells on studies (Figure \ref{fig:top10_code_smells}).

Twenty-eight studies quotes code smells [S1, S2, S3, S4, S8, S9, S11,  S12, S13, S15, S16, S18, S19, S20, S23, S24, S27, S28, S30, S31, S32, S33, S34, S35, S37, S38, S39, S40].
\emph{Duplicated Code/Clones} is the most studied code smell [S9, S19, S20, S30, S31, S32, S33] (7 studies from 28). We observe that the \emph{Duplicated Code/Clones} have been investigated separately and explored in different ways. 
\emph{Duplicated Code/Clones} cause code design problems, making maintenance difficult, and introducing subtle errors. Probably, It is the reason we find an active community  dedicated to \emph{Duplicated Code/Clones}.

Other code smells in the Top 10 most quoted list are \emph{God Class/Large Class, Feature Envy, Long Method, Long Parameter List, Divergent Change, Data Clumps, Refused Bequest, Shotgun Surgery,} and \emph{Lazy Class}. 

\textbf{When the technical debt was the subject, \emph{God Class/Large Class} has been the most investigated smell}. Such the smell is conceptually easy to understand, and, according to [S21], it is up to 13 times more likely to be affected by defects and up to 7 times more change-prone, which makes them a good candidate for TD mitigation.
Several reasons justify the higher prevalence of some smells than others: tools available for their detection, the frequency of smell occurrence, popularity among practitioners, representativity of design and code problems, and the incidence of one code smell in another.

However, rarely some code smells are investigated.
According to [S12, S18], \emph{Alternative Classes with Different Interfaces}, \emph{Incomplete Class Library} did not obtain the attention of researchers. The study [S12] still include \emph{Primitive Obsession, Inappropriate Intimacy}, and \emph{Comments}.
Perhaps, these smells are not so interesting, or they are complicated to identify, not justifying the carrying out of studies. The literature does not explain why researchers did not attempt to detect them.

\begin{figure*}[ht]
 \centering
 \includegraphics[width=13cm]{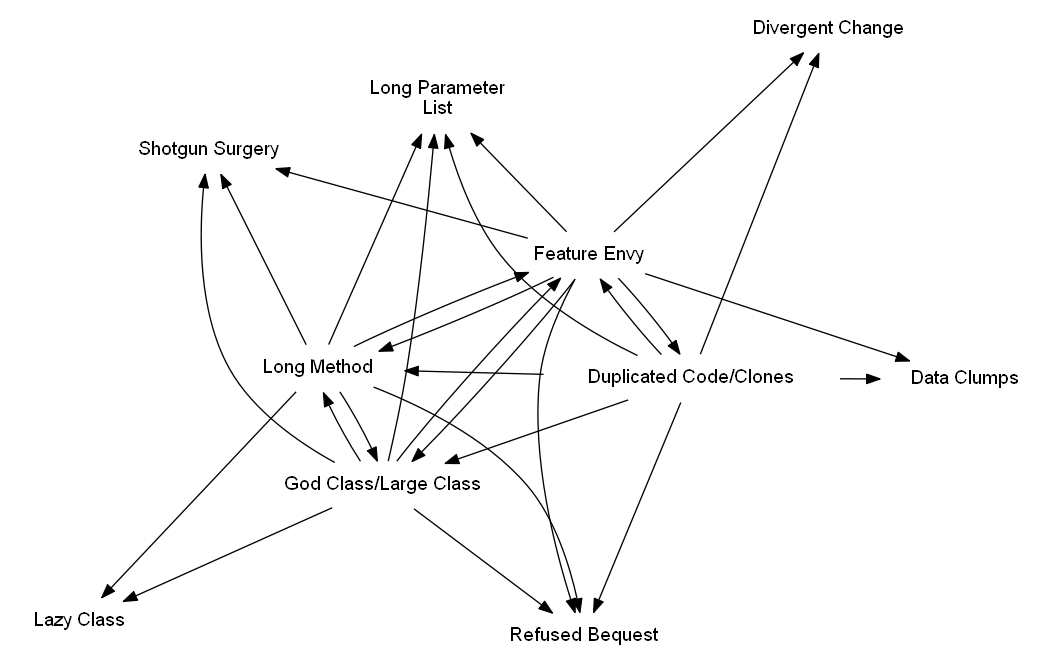}
 \caption{Co-occurrence of smells from the studies}
 \label{fig:co_occurrence_smells}
\end{figure*}

Some of the smells listed on the top 10 smells are related usually by co-occurrence. We observe in Figure \ref{fig:co_occurrence_smells} the relationship among smells reported by studies [S1, S28, S39, S40]. The nodes represent smells, and the edges are the relationships between them.

We observed that the code smells \emph{God Class/Large Class, Long Method, Feature Envy}, and \emph{Duplicated Code/Clones} co-occur in many selected studies.  
For example, the presence of a \emph{Long Parameter List} can result in a \emph{Long Method}. The presence of \emph {Long Method}, by its characteristic, can indicate a \emph{God Class/Large Class}. Also, the fact that we separate a \emph{Long Method} and it has many behaviors that are not related to the same class, can cause a \emph{Feature Envy}.

Other code smells, like \emph{Lazy Class, Refused Bequest, Shotgun Surgery, Long Parameter List, Divergent Change}, and \emph{Data Clumps} are mentioned in studies, but the relation between them is not mentioned, suggesting that this is still a topic deserving more attention. The current studies on the co-existence of smells in the code indicate an association with maintenance and design problems. \textbf{Co-occurrences can be more explored, such as the appearance of smell in consequence of another smell, smells that are always close (presence of one implies the presence of another), among others.}

In the same way, it occurs with design smells, naming problems are also found with code smells. 
Several studies [S11, S12, S15, S39, S40] claim that the \textbf{use of terms and classifications adopted  by different authors are not sound}. 
On the one hand, we found distinct definitions for the same smell name. On the other hand, the same smell definition is presented with different names. According to [S40], such fragmentation of definitions is due to the lack of a more systematic or formal taxonomy for code anomalies.

The standardization process is necessary to allow the unification of the terminology and its precise definition. A standard, cataloging all the smells (design/code) defined up to the present time should be possible, determining those that refer to the same smell with different names.
The study [S39] also suggests the creation of a unique catalog (in the same way as the \emph{Design Patterns Catalog}) with a unique entry in the catalog enriched with \emph{“other names”} or \emph{“also known as”}.
It is \textbf{important to increase efforts to the standardization of the concepts, which would also allow an increase in smells detection consistency}. 

\subsubsection{Smell Detection Approaches} 
\label{subsec_rq2_context}

\begin{figure}[ht]
 \centering
 \includegraphics[width=\linewidth]{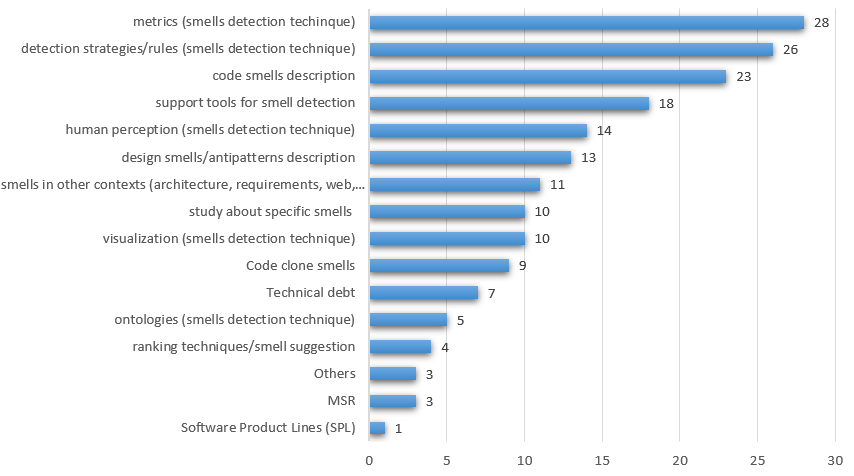}
 \caption{Smell topics, based on how are discussed in the literature}
 \label{fig:smells_context}
\end{figure}

The main topics for code smells are shown in  (Figure \ref{fig:smells_context}). 
The most discussed topics are smell detection approaches, appearing in six topics among the 16 most cited topics. There is a lack of standard agreed-upon definitions for code smell detection in the research community [S2, S11, S12, S39, S40].

We analyze the main approaches to smell detection, considering the top 10 most quoted smells (Table \ref{tab:main_solutions}). Among them, the use of human perception, metrics-based, detection rules, reverse engineering/static analysis, history-based, machine learning-based, and software visualization are the most mentioned approaches. We group them to facilitate our analysis. We defined the following classifications: human perception, metrics-based, strategy/rules-based, probabilistic/search-based, and visualization-based.

Human perception-based approach [S10, S12, S13, S15, S18, S19, S20, S24, S26, S34, S35, S38, S39, S40] is a fundamental manual approach to detect smells, usually based on different guidelines followed by developers to detect manually design defects  [S18, S39].
Manual techniques are human-centric, time-consuming, and error-prone. These techniques eliminate uncertainties in the detection process due to human involvement, but they are not useful for examining code smells within large systems. However, we do not eliminate human participation from the detection process.

Metrics-based approach [S2, S4, S5, S6, S7, S9, S12, S13, S14, S15, S16, S17, S18, S19, S20, S22, S24, S25, S26, S30, S31, S32, S33, S34, S35, S38, S39, S40] is usually the approach used to detect \emph{Large Class/God Class, Long Method, Data Clumps, Refused Bequest, Shotgun Surgery} and \emph{Lazy Class}. This approach was mentioned in all top 10 smells selected. It used to evaluate/measure source code elements (\eg~attributes, lines, parameters, methods, classes), allowing them to take some decisions.
The accuracy of metrics-based approaches is dependent on the proper selection of threshold values, which are usually empirical and not much reliable [S2, S18, S38, S39]. There is not yet a \textbf{consensus on the standard threshold values for the detection of smells, and consequently, there is a lot of disparity among results of different techniques}. One of the factors that can contribute to this finding is the lack of standardization/formalization of the smell definition.

Rule-based (or strategy-based) approach [S2, S4, S6, S9, S11, S12, S13, S14, S15, S17, S18, S19, S20, S22, S24, S25, S26, S30, S31, S32, S33, S34, S35, S38, S39, S40] is another approach which combines rules, logic expression, and metrics used typically to detect the following smells: \emph{Feature Envy, Long Parameter List}, and \emph{Divergent Change}.
Different smells are represented as detection rules. Each rule is specific to specific smells and can be defined manually or automatically using different techniques [S39]. 
The conversion of symptoms into detection rules requires analysis and interpretation effort to select the proper threshold values. There is not yet agreement on defining standard symptoms with the same interpretations, and thus the precision of the approach is low. Since rule-based approach makes intensive use of metrics, the same works that mention this type of approach also mention a metrics-based approach.  

Probabilistic/search-based approaches [S12, S15, S18, S20, S31, S32, S33, S34, S39] apply different algorithms and rules for the detection of smells directly from source code. Most techniques in this category apply machine learning algorithms and fuzzy logic. The study [S39] reinforces the use of the search-based approach to detect different types of smells. Several techniques and algorithms proposed for extracting specified rules to detect smells with techniques based on genetic and heuristic search algorithms.
These techniques learn from the standard design and coding practices and examine how the code deviates from these practices. The success of these techniques depends on the dataset's quality and training [S2]. These techniques are very limited for dealing with unknown and varying definitions of code smells, but it is one of the approaches to be more explored in future work, not only for smell detection but also to support refactoring recommendations.

Visualization-based approach [S9, S12, S13, S15, S18, S20, S26, S34, S39, S40] integrate the capability of human expertise with the automated detection process. In some cases, when the software is very complicated, the graphical representation of the software artifact arises as a solution to deal with complexity. Such an approach has scalability problems for large systems, and it is error-prone because of wrong human judgment depending on visualization type. However, this approach could help developers to identify points of code to be  improved, reducing the technical debt.

According to [S2, S39, S40], \textbf{smell detection approaches and their corresponding produced results are highly inconsistent}. 
\textbf{In general, generic approaches are used for all types of smells, while some specific approaches are used for more specific smells}. That is the case for approaches such as history-based, optimization-based, and probabilistic/search-based.
The study [S40] reports that different detectors for the same smell produce different answers, which is coherent with the need for new strategies to identify smells (or even approaches) in a more efficient/effective way than current approaches. 
\textbf{It is also necessary to explore whether the approaches could be combined or individually used for the detection of a set of smells}. We suggest, as future work, the assessment of which approach combination is better and in which context and conditions.

\subsubsection{Impacts and Effects}
\label{subsec_rq2_impacts}

The studies [S11, S15, S24] do not differentiate between a smell and a definite quality problem. The community believes existing smell detection methods suffer from high false-positive rates. Also, existing methods cannot define, specify, and capture the context of a smell adequately. Undoubtedly, the \textbf{impact of smells causes decay in the overall design, affecting the quality attributes} [S24, S35, S37, S38], including maintainability (the effort to change the code), understandability, and extendability (the effort to add new functionality). 
A deeper discussion of the relationship among quality attributes, smells and refactoring is presented in Section \ref{sec:qsr}.

However, some studies [S15, S23, S24] do not establish an explicit connection between smells and their impact on the productivity of a software development team. 
Several studies [S21, S23, S26, S27, S38, S39, S40] presents different factors in how smells affect the architecture decay, both developer-focused and development process-focused, concentrating on the relation between design/code smells.
Developer-focused issues involve difficulties related to inexperienced/novice developers focused on functionality build, lack of a system’s architecture knowledge, apprehension due to system complexity, and carelessness.
Development process-focused issues include difficulties related to missing functionalities, violation of object-oriented concepts (abstraction, information hiding, modularity, and hierarchy), project deadline pressures, changing and adding new requirements, updating new software and hardware components, and ad-hoc modifications without documentation.

The studies suggested that developers should promptly identify and address the code smells upfront. Otherwise, code anomalies increase modularity violations and cause architecture degradation. To achieve such skills, introducing new approaches to developers' education could be necessary. We do not find secondary studies discussing ways to teach such practices while developers are coding. Therefore, we suggest the development of mechanisms and tools which help developers, recommending practices in such context, as seen at Section \ref{sec:implications}.

\noindent
\fcolorbox{black}{lightgray}{
    \parbox{\linewidth}{%
        {\footnotesize \textbf{RQ2 Summary} 

        \emph{\textbf{Challenges:}} The literature does not explain why researchers did not attempt to detect \emph{Alternative Classes with Different Interfaces, Incomplete Class Library, Primitive Obsession, Inappropriate Intimacy}, and \emph{Comments}. It is necessary to evaluate the reason for the lack of interest in these smells. 
        
        A smells naming standardization is necessary, allowing the terminology and its precise meaning to be unified. With this standardization, cataloging the smells defined up to the present time should be possible, determining those that refer to the same smell with different names. This process will undoubtedly have repercussions on detection approaches. Another question that can investigate is the appearance of smell in consequence of another or the existence of smells that are always related, sometimes co-occurring. 
        
        Furthermore, it is necessary to explore which approaches can be complementary or explicitly used for a specific smell.
        There is a variety of approaches to revealing smells, with high false-positive rates. Thus, there are open possibilities to explore and to improve methods capable of defining, specify, and capture the smell context.
        
        \emph{\textbf{Observations:}} \emph{Blob} and \emph{Duplicated Code/Clones} are the most mentioned design and code smell, respectively. Related to technical debt, \emph{God Class/Large Class} has been the most investigated smell. Design smells have also been studied together with code smells. There are simple and composite smells (the combination of simple smell can lead to a composite smell). 
        
       It is a consensus that manual detection is difficult, time-consuming, and prone to errors. However, there is no consensus on the standard threshold values for the detection of smells, which are the cause of a disparity in the results of different techniques. 
        
        Over time, there has been a significant number of detection approaches, like metrics-based and strategies/rules. They have been the most cited. Other approaches, such as history-based, optimization-based, probabilistic-search-based, and visualization, have also been used to smells detection. Besides, smell detection approaches and the corresponding produced results are highly inconsistent. The impact of smells causes decay in the overall design, affecting quality attributes.
        
        }
  
    }%
}

\subsection{RQ3: Which tools have been mentioned for code-smell detection and refactoring support?}
\label{subsec:rq3}

Manual detection of code smells in the early days was very time consuming, error-prone, and costly. Smell detection tools automate specific smell detection techniques. To address these problems, researchers developed many semi-automated and automated code smell detection tools and refactoring support.
To answer RQ3, we investigated which platforms/programming languages are used and which tools have been aiming for smell detection. Furthermore, we  also investigated supporting tools for refactoring.

The tools investigated in studies [S2, S13, S18, S39, S40] have many characteristics, summarized as follows: if they are free or not; whether they are open source or proprietary; their supported languages;  the terms used to describe the smells; the internal representation of the software artifact; the degree of automation; the ability to also perform refactoring; the way to run the tool; their ability to generate metrics; the type of input source; the output format; the facility to work with \emph{Command Line Interface} (CLI); \emph{Graphical User Interface} (GUI)/plugged on IDEs, and the list of smells the tool can detect. 

\subsubsection{Platforms/Programming Languages}
\label{subsec_rq3_lps}

\begin{figure}[ht]
 \centering
 \includegraphics[width=\linewidth]{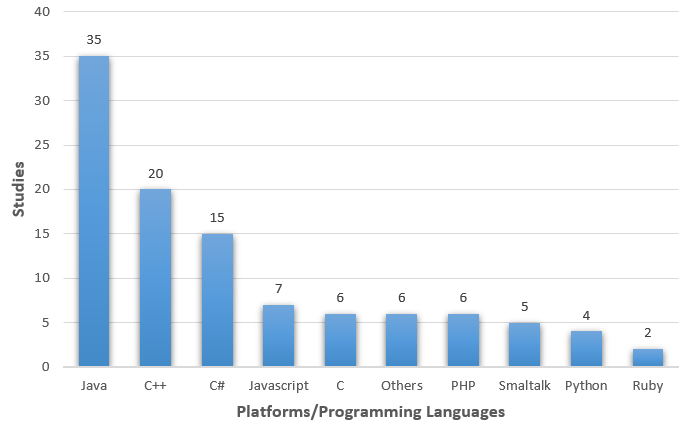}
 \caption{Platforms/programming languages more used}
 \label{fig:Platforms_PL_more_used}
\end{figure}

The majority of studies [S1, S2, S3, S4, S5, S7, S8, S9, S10, S11, S12, S13, S14, S15, S16, S17, S18, S19, S20, S22, S23, S24, S25, S26, S29, S30, S31, S32, S33, S34, S35, S36, S37, S39, S40] refer to \textbf{Java as platform/programming language(PL) most used to develop tools} (Figure \ref{fig:Platforms_PL_more_used}).

Thirty-five studies that mentioned platforms/PLs have suggested Java as the most used technology, following by C++ and C\# (57.14\% and 42.85\%, respectively). 
According to GitHub\footnote{more info: https://octoverse.github.com/, accessed December, 09 2019}, Java is the third most popular programming language used in project repositories, whereas according to the Tiobe Index\footnote{more info: https://www.tiobe.com/tiobe-index/, accessed December, 09 2019}, Java is the most popular programming language. 
Java has been the most widely used platform for tool development and also as a target language for experiments. Few tools (\eg~\emph{PMD, Borland Together, CCFinder, inFusion, inCode}, and \emph{iPlasma}) listed in the studies [S2, S18, S39, S40] work with more than one programming language. Therefore, \textbf{there is an opportunity to develop tools that support more than one programming language}. For instance, Javascript is more and more adopted by the industry and is already the leader in FLOSS projects on GitHub. However, it appears only as of the fourth technology mentioned in the studies so far. Nevertheless, the second edition of the recently released Fowler refactoring book \cite{Fowler:2019} presents all examples in Javascript.

\subsubsection{Smells Detection Tools}
\label{subsec_rq3_smells_tools}

\begin{figure}[ht]
 \centering
 \includegraphics[width=\linewidth]{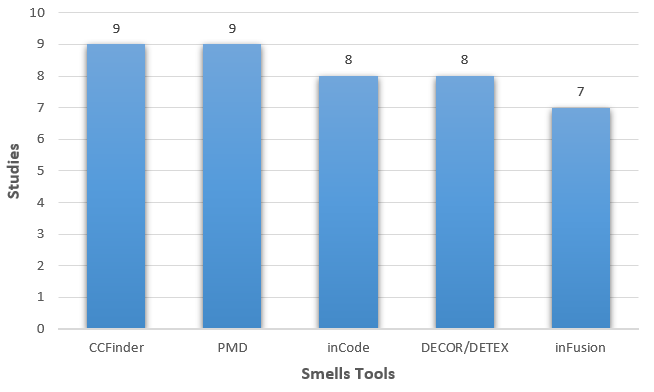}
 \caption{Top 5 smells tools}
 \label{fig:top5_smells_tools}
\end{figure}

Nineteen studies [S1, S2, S9, S12, S13, S17, S18, S19, S20, S26, S27, S30, S31, S33, S34, S35, S38, S39, S40] quoted smell detection tools. \textbf{In a set of more than 162 distinct tools, we presented the top 5 smell detection tools} (Figure \ref{fig:top5_smells_tools}). \emph{CCFinder} is the tool that most appears in studies [S2, S9, S19, S20, S30, S31, S34, S35, S39] for smells detection, together with \emph{PMD} [S1, S2, S9, S12, S13, S18, S26, S39, S40] (both with 47.36\% of studies). 
\emph{inCode} [S1, S2, S13, S18, S34, S35, S39, S40] and \emph{DECOR/DETEX} [S13, S18, S34, S35, S38, S39, S40] appears together with 42.10\%. \emph{inFusion} [S2, S13, S17, S18, S35, S39, S40] appears with 36.84\%.

\emph{CCFinder} is the most quoted tool to detect \emph{Code Clones} (\eg~\emph{Type-1} and \emph{Type-2}), using the token-based approach. The tool uses a suffix tree algorithm, and so it cannot handle statement insertions and deletions in \emph{Code Clones}. It applies several metrics to detect relevant clones. 
It also optimizes the sizes of programs to reduce the complexity of the token matching algorithm. It produces high recall, whereas its precision is lower when compared with some other techniques [S31]. 
\emph{CCFinder} probably is the most cited because detects \emph{Type-1} and \emph{Type-2} clones, which are the more commons clones. There is not a ``perfect'' clone detection technique, e.g., having high scores to all properties like precision, recall, ok, portability, and robustness [S9, S31]. Perhaps new clone detection approaches could do better by overcoming some of the limitations of existing techniques. Still according to [S31], \textbf{\emph{Type-4} clones require to solve an undecidable problem}. The clone detection technique can be improved by combining several different types of methods or re-implementing systems using a different programming language. Such the technique presents new challenges for software maintenance, refactoring, and clone management. \textbf{Clones introduce maintenance and evolution problems, but, in most of the cases, they do not affect quality} [S20, S35, S37].

\emph{PMD} is a static analysis tool used to detect code violations or bad development practices. Because of its broad action spectrum, it ends up detecting some smells like  \emph{Duplicated Code, Large Class, Long Method}, and \emph{Long Parameter List}. For \emph{Large Class} detection, the tool presents a low precision rate (about 14\%). On the other hand, it performed well with \emph{Long Method}, achieving 50\% to 67\% of recall and 80\% to 100\% of precision [S2]. 
In the study [S18] is discussed the use of \emph{PMD}, comparing with another coding standard/smell detection tool, called \emph{Checkstyle}\footnote{\emph{Checkstyle} was the sixth tool most mentioned in the studies}. Although \emph{Checkstyle} detects the same smells of \emph{PMD}, there is a difference in detection results, due to different threshold values used by these tools, as well as the metrics took in the account. Beyond the list of smells tools, the study [S18] presents a set of experiments realized.
For \emph{Large Class}, \emph{PMD} uses a threshold of 1000 NLOC\footnote{non-commented lines of code}, while \emph{Checkstyle} uses 2000. Still, for \emph{Long Method}, \emph{PMD} uses a threshold of 100 and \emph{Checkstyle}, 150. For \emph{Long Parameter List}, again we have differences of thresholds, with \emph{PMD} using ten and \emph{Checkstyle}, 7. PMD appears among the most cited due to its more general performance, although, as we have seen, it is not very accurate.

\emph{inCode} is an Eclipse plug-in that helps smell detection, detecting \emph{Duplicated Code, Large Class, Feature Envy, Data Clumps}, and \emph{Refused Bequest}, with visualization support. Although it is a tool that detects several smells [S2, S35, S39, S40], we do not found secondary studies discussing more details about functionalities, strategies used to detect, precision, and recall of operation. Since \emph{inCode} is an Eclipse plugin detects multiple smells and works with multiple programming languages, it is among the most commonly cited detection tools.

\textbf{\emph{DECOR/DETEX}}, proposed by Moha \al~\cite{Moha:2010}, \textbf{is the tool capable of detecting more than 10 smells} (\eg~\emph{Large Class/God Class, Lazy Class, Long Method, Long Parameter List, Refused Bequest, Speculative Generality, Message Chains, Shotgun Surgery, Duplicated Code, Comments, Data Class}) identified by Fowler \al~\cite{Fowler:1999}. Also, \emph{DECOR/DETEX} detects design smells proposed by Brown \al~\cite{Brown:1998}, like \emph{Swiss Army Knife, Blob}, and \emph{Functional Decomposition}. \emph{DECOR} is a method, and \emph{DETEX} is a tool that allows us to specify and to detect code and design smells, using a DSL\footnote{\emph{Domain Specific Language}} [S13]. The tool allows developers to make metrics threshold settings to detect smells. Although the study [S18] reports 50\% of precision and 100\% of recall, the study [S13] reports experiments indicating not such a high accuracy.  \emph{DECOR} is among the most cited detection tools probably due to the lightweight nature that allows researchers to employ it to detect several types of smells without the need of compiling anything each time.

\emph{inFusion} is another tool that detects \emph{Duplicated Code, Large Class, Feature Envy, Long Method, Data Clumps}, and \emph{Refused Bequest}. \emph{inFusion} has an open-source version called \emph{iPlasma}, which is quoted too, but with more limited functionalities. The tool presents the same numbers of \emph{PMD} [S2], with a 14\% recall rate for \emph{Large Class} detection. The authors report an experiment with just one project. Probably, it is necessary to achieve more experiments. Also, the tool performed well with \emph{Long Method}, achieving 50\% to 67\% of recall and 80\% to 100\% of precision.  

Smell detection tools use thresholds on metrics or ad-hoc rules to identify structures in code, at the price of some inaccuracy [S40].
The accuracy of a code smell detection tool is a key aspect of its validity. Some secondary \textbf{studies  [S2, S15, S18, S40] reveal approximately 30\% of the tools spotlight the accuracy, that is, precision and recall, of their technique or tool}. Also, the studies [S18, S40] describe that authors of code smell detection tools perform experiments on different systems, and the comparison of published results becomes difficult when tools are not available. 
\textbf{Standard benchmark systems for code smell detection tools are not available, which require the attention of the research community.}

Murphy-Hill \al~\cite{Murphy-Hill:2012} presents a list of guidelines as success factors related to usability for code smell detectors. The studies [S2, S40] contain a discussion about usability and detection tools. They define six features for tools analysis: a) easy exportation: results about the detected bad smells were easily exportable, for instance, to text, CSV or other file formats; b) highlighting of smell occurrences; c) configurability: allowing detection settings; d) graph visualization; e) detected smell filtering; f) Analysis of multiple versions. According to [S2], \emph{inFusion} is the only tool that supports five features (\emph{a} to \emph{e}), although two of these are available only in the full commercial version of the tool. In addition to the features mentioned above, the studies [S2, S40] describe that some usability issues could hinder the tool user experience. Some usability problems, such as difficulty in navigating between bad smell occurrences (in general, results showed in long lists without summarization), difficulty in identifying the source code related to a smell detection, and lack of advanced filters for specific bad smell detection.  
In general, tools do not provide data visualization through statistical analysis, counters of detection results, or result's presentation by charts.

\textbf{Most of the tools focus on the recovery of code smells from a single language}, that is, in most cases,  Java language [S2, S18, S39, S40].
None of the tools detect all 22 code smells identified by Fowler \al~\cite{Fowler:1999}. On average, tools cover three to four smells for detection [S2, S18]. The tool \emph{inFusion} claims to detect all code smells of Fowler, but it is a commercial tool, and it was not free and available for experiments realized in [S18, S39]. We identify an opportunity for the development of tools for the detection of relevant smells not yet explored. Additionally, tools that detect smells in more than one language should also receive attention from the SE community.

\subsubsection{Refactoring Tools}
\label{subsec_rq3_refactoring_tools}

\begin{figure}[ht]
 \centering
 \includegraphics[width=\linewidth]{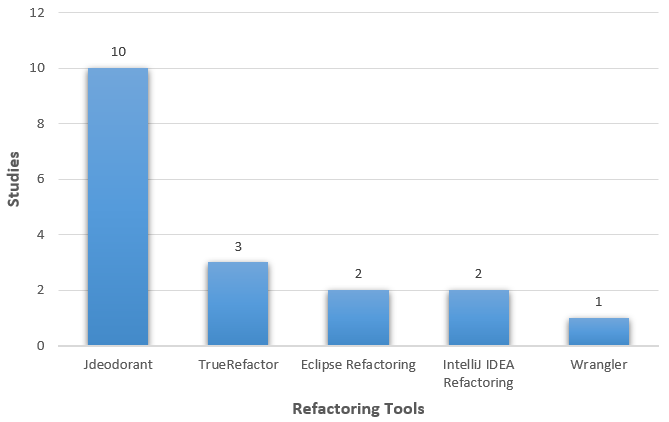}
 \caption{Top 5 refactoring tools}
 \label{fig:top5_refactoring_tools}
\end{figure}

Since refactoring tools are also essential to our work, and we did a study on tools that support refactoring. Thirteen secondary studies [S1, S2, S4, S9, S12, S13, S14, S17, S18, S29, S34, S39, S40] \textbf{presented 24 distinct tools that help developers applying refactoring. We select the top 5 refactoring tools} (Figure \ref{fig:top5_refactoring_tools}) for a detailed discussion.

\textbf{\emph{JDeodorant} is the tool that most appears in studies} [S1, S2, S4, S12, S13, S18, S29, S34, S39, S40] (71.42\% of studies), followed by \emph{TrueRefactor} [S2, S17, S25] (21.42\%), \emph{Eclipse Refactoring} [S18, S29] and \emph{IntelliJ IDEA Refactoring} [S2, S29] (14.28\% both), and finally \emph{Wrangler} [S13] (7.14\%). There are other 19 tools with a similar percentage, but we choice \emph{Wrangler}, because it is the first tool supporting refactoring for clones.

\emph{JDeodorant} [S13, S18, S40] is an Eclipse plug-in that automatically recognizes \emph{Large/God class, Feature Envy, Switch Statement/Type Check}, and \emph{Long Method} code smells from Java source code. It has support for refactoring and assists the user in refactoring transformations [S18, S40].

In the same study [S2] that also analyze \emph{PMD} and \emph{inFusion}, there was high agreement among these three tools concerning the detection results, although \emph{JDeodorant} points out more bad smell instances compared to the other tools, in its default configuration. \emph{JDeodorant} has too achieved low precision rate (about 14\%) in detecting \emph{Large Class}.
The tool uses both metrics and AST\footnote{\emph{Abstract Syntax Tree}} to detect bad smells.  
Considering that some tools apply unknown techniques, detection results may be different. The authors [S2] observe that \emph{JDeodorant} indicates the highest number of \emph{Large Class} and \emph{Long Method} instances, and scored the lowest results for both recall and precision. 
We note that \emph{JDeodorant} is one of the few tools that combine smells detection with the automatic application of refactoring. It may be the reason for being the most mentioned: even with some limitations presented, this shows that more tools with these characteristics are needed.

The \emph{TrueRefactor} [S17, S25] is an automated refactoring tool that significantly improves the comprehensibility of legacy systems \cite{Griffith:2011}. To detect code smells, each source file is parsed and then used to create a control flow graph to represent the structure of the software. For each code smell type, a set of metrics is calculated to identify whether a section of the code is an instance of a code smell type. The tool uses a genetic algorithm (GA) to search for the best sequence of refactoring that removes the highest number of code smells from the source code.
As an automated refactoring tool, \emph{TrueRefactor} does perform actual refactoring, but currently supports mainly the modification of UML rather than code. 
The study [S17] describes an example program with code smells artificially inserted to analyze the effectiveness of the tool. The number of code smells of each type over the set of iterations was measured jointly with the measure of a set of quality metrics. In both cases, the values increased initially before staying relatively stable throughout the rest of the process. A comparison of initial and final code smells shows that the tool removes a significative proportion of smells, and also metric values indicate that the surrogate metrics are improved.
Despite the limitations presented by \emph{TrueRefactor}, a positive aspect is a way adopted for sorting the most appropriate refactoring, based on a given smell and its impact. This shows that researchers can advance in more studies of classifying refactoring, taking into account their impact.

Beyond the \emph{JDeodorant} and \emph{TrueRefactor}, the rest of the tools do not present more details or analysis about use and accuracy. Two IDEs appear, too, as most mentioned. The \emph{IntelliJ IDEA} implements more than 40 refactoring, using a lexical and syntactic parser to convert the code into the form of AST, called \emph{Program Source Interface} (PSI) \cite{Jemerov:2008}. The PSI is used to validate any generated code. After code transformation,  the \emph{Formatter} is responsible for verifying the scope of the changes, adjusting the code with indentation, inserting blank lines, changing of qualified names, and imports of libraries. \emph{IntelliJ IDEA} still makes use of a built-in DSL to find fragments in the PSI using a defined lean notation. The use of DSL is also one of the paths suggested for future studies for code refactoring \cite{Zhang:2011, Jemerov:2008, Li:2012}.

\emph{Wrangler} \cite{Li:2006, Li:2008} is a tool that supports interactive refactoring of Erlang programs. 
It is integrated with Emacs as well as with Eclipse, through the ErlIDE plugin. Wrangler itself is implemented in Erlang. The tool supports a variety of refactoring, as well as a set of code smell inspection functionalities, and mainly a lot of  facilities to detect and eliminate \emph{code clones}.

\emph{Eclipse Refactoring} is also well-known for the constant improvements to the use of refactoring. The process consists of the phases of verification of preconditions, detailed analysis, and rewriting of the code. Guidelines help to seek a more straightforward code rewriting mechanism, also based on the AST form.  
Although \emph{Eclipse} supports more than 20 refactoring techniques, a lot of work has improved the use and application of the refactoring, resulting in more speed for developers \cite{Murphy-Hill:2007c}.

We note that \emph{Eclipse} and \emph{IntelliJ IDEA} appear here by bringing automation to refactoring, helping developers in the process. One of the advantages of using these tools is to ensure the application of refactoring (passed by refactoring preconditions and also by postconditions, ensuring that AST has not been broken). However, it is up to the developers to find the smells and also know the refactoring to apply. Murphy-Hill has argued in previous work \cite{MurphyHill:2008, MurphyHill:2010} on usability and habits of developers, showing that this is not a trivial process. We think tools taking advantage of such potentialities, guiding the developers in the process (for example, recommending a refactoring), can be explored in future studies.

\subsubsection{Tool Considerations}
\label{subsec_rq3_observation}

According to [S13, S18, S21], the \textbf{detection of code smell reduces the cost of maintenance if the failures found in the early stages of software development}.
The applicability of smell detection tools varies according to the goal of detection: the objective could be software quality management, or maintenance after smells detection, code quality improvement, and fault detection via refactoring.
One observation is the detection goal is highly related to the impact goal, since the investigation of the impact of smells occurs after the detection. Indeed, the study [S40] observes that the inconclusive knowledge about the negative impact of smells is partially attributed to tools/techniques used to detect them. 

Since there is a great variety of tools and discrepancies on tools findings, we cannot discard discrepant results in different studies because they were using different tools.
Still, according to [S39],  only a few tools can analyze very large-sized projects (millions of lines of code). Most of the tools did not take into account the expert feedback or other characteristics like the influence of the context,  project domain, and project status.

A \textbf{large number of tools are available for the detection and removal of code smells. However, evaluation frameworks that could help the user for appropriate selection of any tool for a given context are missing} [S18, S39].
The \textbf{currently available tools} [S2, S15, S39] \textbf{can detect only a very small number of smells}.  It is still a big problem to determine which code smell is effective in indicating the need for refactoring and what type of refactoring, and programmer involvement is still necessary [S2, S11]. 
Although refactoring has been proposed to remove smells, several subtleties make this activity inherently complicated [S40].

In the refactoring field, we suggest that developers of new detection tools should be aware of the possible usages of their tools, considering observations referenced by [S2]. According to [S5], \textbf{a refactoring tool developer can not provide custom refactoring fitting for all specific user needs because the possible number of refactoring is unlimited}. Therefore, customizable refactoring tools based on the demand of the developer are missing. Still, as reported by [S7], the existing refactoring tools are error-prone, and therefore, using these tools may result in producing incorrectly refactored pieces of code. As a result, tools usage sometimes negatively affect code quality.
Automating the refactoring process consists of automating the two following main steps [S22]: (1) identify refactoring opportunities, and (2) perform refactoring. It is necessary to automate effective techniques to identify opportunities for such refactoring and then perform it.

Tools should make it easier for programmers to refactor quickly and correctly. \textbf{Tools have to help analyze the impact of the smell: nowadays, many tools do a weak job of communicating errors triggered by the refactoring}.  

Furthermore, tools like \emph{Eclipse IDE} and \emph{IntelliJ IDEA} automate some pointed refactoring. The point is it depends on the developer knowing how to conduct it. In particular, the study of the human perception of what is a code smell and how to deal with it has been mostly neglected in the past [S15]. In the same way, human opinion remains important to decide where refactoring is worth applying [S5].
According to [S18, S39], there are few studies discussing such support for refactoring. A significant motivation for identifying code smells is source code refactoring, but most code smell detection tools focus on the detection/visualization of code smells. This finding is evidenced in our research (Table \ref{tab:main_solutions}, marked as *):  there is not a tool that automates refactoring, according to one indicated smell. It shows the need to deepen work on the constructions of refactoring support tools coupled to smell detection tools.

Moreover, from the perspective of tools evaluation, the unavailability of implementations hinder reproducibility and impose barriers to the underlying empirical studies, in particular for those aiming at comparing new approaches with the state-of-the-art.
\textbf{Experts from industry and academy need to assess the results of the tools regarding detecting false positives and false negatives}. It is also essential to have expert opinions concerning  smell prioritization, smell impact on product quality and technical debt, as well as evaluation of refactoring in the same terms.
In general, according to [S18, S39], tools \textbf{lack maturity and a lot of limitations restrict their use and adoption by the industry}.

\noindent
\fcolorbox{black}{lightgray}{
    \parbox{\linewidth}{%
        {\footnotesize \textbf{RQ3 Summary} 
        
        \emph{\textbf{Challenges:}} 
        We notice a preference for a programming language/platform, for both the tool construction and realization of experiments.
        It opens the opportunity to develop tools that support more than one programming language.
        
        It is necessary to expand the studies and experiments, evidencing the accuracy of the smell detection tools as well as refactoring tools. 
        
        We identify a lack of maturity of tools with limitations that restrict their use and adoption by the industry. Standard benchmark systems for results of code smell detection tools and refactoring tools are not available, which require the attention of the research community.
        Also, it is necessary to involve experts to assess the results of these tools. 
        
        \emph{\textbf{Observations:}} 
        Several aspects characterize tools: free or not, open-source or proprietary, supported languages, the terms used to describe the smells, the degree of automation, the ability also to perform refactoring, the way to run the tool, among others. 
        
        Java is the platform/programming language most used to develop tools. Also, most tools focus on the identification of code smells from a single language, where Java is predominant too.
        
        We have a large number of smell detection tools, using the most different detection approaches. Currently, available tools can detect only a tiny amount of smells (between 3 and 4). The number of tools that perform refactoring is small.
        
        The most quoted smell detection tool is \emph{CCFinder} because we have more studies related to the \emph{Duplicated Code/ Clones}. This smell has an impact on software maintenance and evolution, but we have not identified significant effects on quality. 
        The most cited refactoring tool is the \emph{JDeodorant}, used to apply specific refactoring for specific code smells. 
        }
    }%
}

\subsection{RQ4: Which RQs have been studied on smells and refactoring? What are the highest cited secondary studies?}
\label{subsec:rq4}

Responding to RQ4, we analyze which RQs discussed in the studies and the correlation among them. Also, we present a discussion of the most mentioned works. 

\subsubsection{Analysis of RQs}
\label{subsec_rq4_analysis}

RQs are essential points in any scientific work. They define the direction and conduct the focus of the studies. An explicitly stated RQ is one of the requisites for a review to be considered systematic.
We explore RQs from two points of view: general analysis and specific analysis.

In a more general analysis, \textbf{we had 181 RQs in the 40 selected secondary studies}. It gives an average of 4.5 RQs per study. \textbf{The study [S40] was the study with the highest number (13 RQs)}. Thus, it is a broad study, addressing several topics related mainly to smell definition, smells detection, tools, impact, trends, and focus on the research group and researchers.

\textbf{We had two studies [S28, S30] with only one RQ}. The study [S28] addresses the co-occurrence between smells and design patterns. The study [S30] is focused on \emph{Code Clones}.

Other studies, such as [S4, S9, S10, S18, S31], had no explicitly defined RQs.
The studies [S4, S18] are SLRs. They have defined goals, but not in the format of RQs, not following a SLR protocol and making difficult an analysis.
The studies [S9, S10, S31] are Surveys. Surveys do not explicitly have RQs and do not follow a defined protocol, unlike SLRs and SMs.

In a specific analysis, we used the RQ classification proposed by Easterbrook \al \cite{Easterbrook:2008} (Table \ref{tab:easterbrook_studies}). This classification has been used in other studies as well, \eg \cite{daSilva:2010, daSilva:2011, Petersen:2015, Garousi:2016}.
We ranked each RQ within the study, but we did not rank the studies based on their RQs, using the most specific RQ as the basis, as performed in Silva \al~\cite{daSilva:2010}.
We observed that the studies [S1, S14, S16, S21, S27, S40] have one major RQ is divided into secondary RQs. In these cases, we classified the secondaries RQs follow the Easterbrook RQ classification.
For brevity and space constraints, we are not reporting the entire list of RQs extracted from all studies, but the reader can find it in our online replication package \cite{Lacerda:2019}.

Revising the list of RQs can help researchers to use RQs for new secondary studies, creating better and more systematic RQs.
We note that Description-and-Classification (DC) RQs are the most popular, with a wide margin (75\% of studies). RQs of this type is mainly used in secondary studies. The second most frequent is the Frequency Distribution (FD), with Existence (E)  and Descriptive Process (DP) in the third.

We identified 152 RQs (83.97\%) about Exploratory and Base-rates Questions. In more detail, 34 studies (85\%) have at least one RQ in these categories\footnote{see the data in the replication package \cite{Lacerda:2019}}. Likewise found in SLRs, these categories of RQs are more common in SMs and Surveys \cite{Kitchenham:2009, Kitchenham:2010}. Although, according to Kitchenham \cite{Kitchenham:2010}, this distinction between SMs and SLRs is a bit confusing.

Relationship (R) and Causality (C) were a minority of the questions. It is the type of question one would ask to assess the effectiveness of treatments as in the traditional form of SLRs \cite{Kitchenham:2010}.

Our research evaluates the presence of empirical and evidence-based SE in secondary studies. We searched for terms like "empirical", "evidence", and "experimental" in all RQs, finding only 9 RQs (4.97\%) related to this purpose. As we know from the growth of primary studies with such objectives, we suggest more secondary studies crossing and analyzing this information.

As shown in Table \ref{tab:easterbrook_studies}, to our surprise, \textbf{there were not RQs in any of the secondary studies of type Causality-Comparative Interaction (CCI) nor Design (D)}. One could identify such RQs as more sophisticated ones compared to the others: \eg a CCI RQ may look like this: \textit{“Does smell A or B cause more maintainability problems under one condition but not others?”}. We hope to see secondary studies with such RQs in the future.

\begin{table*}[htp]
    \caption{Classification of RQs, as presented by \cite{Easterbrook:2008}, number of studies, number of RQs in the pool and examples}  
    \label{tab:easterbrook_studies}
    \tiny
    \begin{center}
        \begin{tabular}{p{2cm} p{4.0cm} p{0.5cm} p{1.5cm} p{1.5cm} p{6.5cm} }
            \toprule
            \textbf{RQ Category} & \textbf{Sub-Category} & \textbf{Code} & \textbf{\# of studies} & \textbf{\# RQs in the pool} & \textbf{Examples}
            \\
            \hline
            \myrowcolour
            Exploratory &  Existence &  E &  13 \textit{(32.5\%)} &  16 \textit{(8.83\%)} & \textit{Does X exist?} \\
            & & & & &  RQ1: What is the definition of a software smell? [S15] \\
            & & & & &  RQ3.3: Are there any analysis methods for detecting and/or evaluating ATD? [S21] \\
            & & & & &  RQ1.1: Are there bad smells significantly more studied than others? If so, is there any specific reason? Are bad smells studied alone or together with other bad smells (co-occurrences)? [S40] \\
                
            \myrowcolour
            &  Description and Classification &  DC &  30 \textit{(75\%)} &  90 \textit{(49.72\%)} & \textit{What is X like?} \\
            & & & & & RQ2.2: What co-occurrences have been identified by the studies? [S1] \\
            & & & & & RQ2: Which are the main features of these tools? [S2] \\
            & & & & & RQ3: What methods have been used to study Code Bad Smells? S[11] \\
            
            \myrowcolour
            & Description-Comparative & DCO & 7 \textit{(17.5\%)} & 11 \textit{(6.07\%)} & \textit{How does X differ from Y?} \\
            & & & & & RQ2: how to compare refactoring tools and techniques? [S6] \\
            & & & & & RQ2.1: What are different studies in semantic clone detection and their comparative analysis? [S20] \\
            & & & & & RQ1: What are the types of technical debt and what is not considered as technical debt? [S26] \\

            \myrowcolour
            Base-rate & Frequency Distribution & FD & 11 \textit{(27.5\%)} &  19 \textit{(10.49\%)} & \textit{How often does X occur?} \\
            & & & & & RQ3: Which are the most frequent types of bad smells these tools aim to detect? [S2] \\
            & & & & & RQ1: What refactoring scenarios were accounted for in the PSs? [S7] \\
            & & & & & RQ1: How many papers were published per year? [S17] \\
            
            \myrowcolour
             & Descriptive-Process & DP & 10 \textit{(25\%)} & 16 \textit{(8.83\%)} & \textit{How does X normally work?} \\
            & & & & & RQ2: What model smell detection strategies have been used to identify refactoring opportunities for model refactoring? [S3] \\
            & & & & & RQ2: What are the different approaches used for the detection of code smells and how the smells are removed using these approaches? [S13] \\
            & & & & & RQ4: How do smells get detected? [S15] \\

            \myrowcolour
            Relationship & Relationship & R & 12 \textit{(30\%)} & 13 \textit{(7.18\%)} & \textit{Are X and Y related?} \\
            & & & & & RQ1.3: RQ3: What is the correlation between the detection techniques based on bad smells? [S12] \\
            & & & & & RQ8: What tools are used in TDM and what TDM activities are supported by these tools? [S26] \\
            & & & & & RQ1.3: What research areas are emphasized in the literature that reports studies of TD (technical debt) in the context of ASD (Agile Sofware Development)? [S27] \\

             \myrowcolour
             Causality & Causality & C & 9 \textit{(22.5\%)} & 14 \textit{(7.73\%)} & \textit{Does X cause Y?} \\
             & & & & & RQ4: What evidence is there that Code Bad Smells indicate problems in code? [S11] \\
             & & & & & RQ4: Which quality attributes are compromised when technical debt is incurred? [S26] \\
             & & & & & RQ1: do all of the code smells equally impact software quality in terms of detected software defects? [S37] \\
             
             \myrowcolour
             & Causality-Comparative & CC & 5 \textit{(12.5\%)} & 6 \textit{(3.31\%)} & \textit{Does X cause more Y than does Z?} \\
            & & & & & RQ3.1: Attention level in the formal versus grey literature: How much attention has this topic received in the formal versus grey literature? [S16] \\
            & & & & & RQ2: How similar/different are the experimental settings of studies investigating smell effect? [S24] \\
            & & & & & RQ3.3: Considering the co-occurrence of bad smells in the papers of our dataset, how many of them actually study some relations between bad smells and what are the main findings of these co-studies? [S40] \\ 
            
             \myrowcolour
             & Causality-Comparative Interaction & CCI & 0 \textit{(0.0\%)} & 0 \textit{(0.0\%)} & \textit{Does X or Z cause more Y under one condition but not others?} \\
            
            \myrowcolour
             Design & Design & D & 0 \textit{(0.0\%)} & 0 \textit{(0.0\%)} & \textit{What's an effective way to achieve X?} \\
            
            \hline
             &  & & \textbf{Total} & 181 \textit{(100.0\%)} &  \\
            \bottomrule
        \end{tabular}
    \end{center}
\end{table*}

Also, We analyze the studies based on their RQs (and focus, when the study does not present RQs explicitly), and focus given on the defined RQs. We present the studies grouped by focus (Table \ref{tab:focus_studies}).

\begin{table}[ht]
    \caption{Distribution of studies based on RQs focus}
    \label{tab:focus_studies}
    \scriptsize
    \begin{center}
        \begin{tabular}{p{4cm} p{2.5cm} }
            \toprule
            \textbf{Focus} & \textbf{Studies}
            \\
            \hline    
            Co-occurrence between smells and relationship with design patterns & S1, S28, S39, S40 \\     
    
            \myrowcolour
           Smell detection tools & S2, S4, S11, S13, S15, S18, S19, S20, S31, S32, S35, S39, S40 \\ 
            Model refactoring & S3, S29, S36 \\     
            \myrowcolour
            Refactoring techniques applied on smells & S4, S6, S8, S13, S17, S21, S22, S25 \\    
    
            Trends, opportunities, challenges, gaps (refactoring and smells) & S5, S15, S17, S20, S21, S23, S24, S26, S27, S34, S37, S40 \\     

            \myrowcolour
            Refactoring tools & S6, S17, S25, S40 \\     

            Software quality and refactoring (impact, attributes, measures, scenarios) & S7, S11, S15, S37 \\     
           \myrowcolour            
            Product lines (refactoring, smells) & S8, S14 \\     
            
            Clones & S9, S19, S20, S30, S31, S32, S33 \\
            \myrowcolour            
            Refactoring process (human knowledge, mental model) & S10, S22 \\
            
            Smells definition & S2, S4, S8, S11, S12, S15, S40 \\
            \myrowcolour            
            Smell detection approaches & S8, S11, S12, S13, S15, S18, S19, S20, S23, S26, S30, S32, S33, S34, S38, S39, S40 \\     
            
            Introduction smells on systems, impact and affect/effect & S15, S24, S35, S37, S38, S39, S40 \\
            \myrowcolour
            Tests & S16 \\
            
            Search-based & S17, S25 \\
            \myrowcolour
            Technical debt & S21, S23, S26, S27 \\
            \bottomrule
        \end{tabular}
    \end{center}
\end{table}

Smell detection techniques are explored by several studies [S8, S12, S12, S13, S15, S18, S19, S20, S23, S26, S30, S32, S33, S34, S38, S39, S40]. There is a particular interest in smell detection approaches, using the most different approaches (we discussed the most previously mentioned). In some ways, some studies address techniques in specific contexts (\eg~studies [S20, S30, S32, S33]  in the context of clones).

Some studies have discussed techniques but not always ways to automate them [S8, S12, S23, S26, S30, S33, S34, S38].
Other studies have explored ways to automate detection techniques [S2, S4, S11, S13, S15, S18, S19, S20, S31, S32, S35, S39, S40].

Other studies have discussed trends and challenges in smells and refactoring topics [S5, S15, S17, S20, S21, S23, S24, S26, S27, S34, S37, S40]. Here, we do not differentiate them because we try to analyze both topics together, evaluating the relationship between them.  

There are also studies related to specific topics not commonly mentioned (such as tests [S16] and model refactoring [S3, S29, S36]).

There is no relation between study comprehensiveness and the number of RQs: the scope of the study is related to the scope of the RQ itself and not necessarily with the number of RQs. Such relation does not appear in studies that have a large number of RQs ([S20] has 12 RQs, [S25] has 12 RQs, and [S17] has 10 RQs), but study [S15] reinforces this finding. It has 5 RQs but it is related to 6 focuses.

\subsubsection{Ranking of Cited Secondary Studies}
\label{subsec_rq4_ranking}

To identify the highest-cited papers is becoming a popular subject not only in software engineering but in all computer science \cite{Wohlin:2007, Wohlin:2008, Garousi:2016a, Garousi:2016b}.
The reputation of the authors of a given paper could be a factor in our analysis. However, quantifying reputation is not easy, and discussing factors impacting the number of citations of a paper is outside the scope of our current work.
To investigate the number of citations we used two classifications.

First, we adopted a citation metric: Absolute (total) number of citations since its publication. To obtain such citation data, we use Google Scholar mechanism: for each paper selected, we use the Google Scholar Search and save the number of citations. We show the top-five list of secondary studies based on the metric defined in Table \ref{tab:ranking_citations_google}.

\begin{table*}[htp]
    \caption{Top 5 selected studies sorted by the number of citations of Google Scholar}
    \label{tab:ranking_citations_google}
    \scriptsize
    \begin{center}
    \tiny{
        \begin{tabular}{p{0.3cm} p{13cm} p{0.8cm} p{1.5cm} }
            \toprule
            \textbf{\#} & \textbf{Title} & \textbf{Year} & \textbf{Citations} \\
            \hline    
    S10    & A survey of software refactoring    & 2004 &    1301 \\
    \myrowcolour
    S30    & Survey of Research on Software Clones    & 2007 &    277 \\
    S20    & Software clone detection: A systematic review    & 2013 &    216 \\
    \myrowcolour
    S26    & A systematic mapping study on technical debt and its management    & 2015 &    208 \\
    S11    & Code Bad Smells: a review of current knowledge &    2011 &    113 \\
        \bottomrule
        \end{tabular}}
    \end{center}
\end{table*}

\begin{table*}[ht]
    \caption{Top 5 of selected studies sorted by the number of citations by year}
    \label{tab:ranking_citations_per_year}
    \scriptsize
    \begin{center}
    \tiny{
        \begin{tabular}{p{0.3cm} p{13cm} p{0.8cm} p{1.5cm} }
            \toprule
            \textbf{\#} & \textbf{Title} & \textbf{Year} & \textbf{Citations by Year} \\
            \hline    
    S10    & A survey of software refactoring    & 2004 &    86.72 \\
    \myrowcolour
    S26    & A systematic mapping study on technical debt and its management    & 2015 &    52.00 \\
    S14    & A systematic mapping study on software product line evolution: From legacy system reengineering to product line refactoring & 2017 & 42.50 \\
    \myrowcolour
    S20    & Software clone detection: A systematic review    & 2013 &    36.00 \\
    S23    & Identification and management of technical debt: A systematic mapping study  &    2016 &    27.00 \\
        \bottomrule
        \end{tabular}}
    \end{center}
\end{table*}

Second, we adopt another citation metric:  the number of citations per year, taking into account the year the work was published until now (Table \ref{tab:ranking_citations_per_year}). Comparing results considering the two metrics, we observed that the studies [S10, S20, S26] appear in both results. Also, among the topics most covered in the cited works are Code Clones [S20, S30] and technical debt [S23, S26]. Clones\footnote{the term code clone generated about 545,000 results (about 17,200 in the last five years) in Google Scholar} have deserved more and more prominence from the community, with studies directed at the topic, and often being recognized as a specific research area. The TD\footnote{the term technical debt generated about 2,490,000  (about 107,000 in the last five years) results in Google Scholar} has also grown in recent years with studies focusing on management policies, tools, and techniques to mitigate its impact on software maintenance and evolution. 

\textbf{Forty studies had a total of 2568 citations}, with an average of 64.2 citations per study and median=9. 
The difference between average and median gave by the fact that [S10] work has 1301 citations while [S30] has 272 ([S10] is almost 4.79 times more cited than [S30]).
Since [S10] published in 2004, the number of citations was significant. Recent studies (published between 2017 and 2018) still has a small number of citations.
Thus, it seems that primary studies more cited than compared to Surveys, SLR, and SM studies (Note that only 2\% of cited studies  refer to SLR/SM, as previously shown in Figure \ref{fig:type_researchpapers_used}, where they classified as "others").

Among the selected studies, [S10] remains the most cited with 14 citations. Next, the studies [S22, S3] appear with 13 and 12 quotes, respectively. Finishing the top 5 of the citations, we have the studies [S20, S11] with 11 and 10 citations. However, it is interesting to note  [S3] is the most cited study among selected secondary studies (38.70\% of the [S3] citations are from secondary studies).

\noindent
\fcolorbox{black}{lightgray}{
    \parbox{\linewidth}{%
        {\footnotesize \textbf{RQ4 Summary} 
        
        \emph{\textbf{Challenges:}} 
        Although the secondary studies aim at giving a panoramic view of the area, we identified the need for studies that contain more sophisticated RQs. With the increase of empirical studies on smells and refactoring, we consider new possibilities of secondary studies covering RQs about Causality (mainly CCI) and Design (D), comparing and evaluating phenomena, describing situations of efficacy and efficiency about methods, practices, and tools are open.

        \emph{\textbf{Observations:}} 
        We had 181 RQs in the 40 selected secondary studies. The study [S40] with the highest number (13 RQs) is indeed the most comprehensive.
        Besides, we find studies with only one RQ. And, finally, studies that did not explicitly have RQs.} It is the case of Surveys, which do not follow a defined protocol. Still, we had SLRs that did not follow a  protocol too. 

        The scope of the study is related to the scope of the RQ itself and not necessarily with the number of RQs. 
        
        Code clones and technical debt are the recurring themes in the most cited secondary studies.
       
        The study [S10] is the most cited among the secondary studies.
    }%
}

\subsection{RQ5: What are the annual trends of types, quality, and the number of primary studies reviewed by the secondary studies?} 
\label{subsec:rq5}

The growing number of secondary studies on smells and refactoring is a strong indication of the high interest in this field. Next, we present some main characteristics of these secondary studies.

\subsubsection{Paper types and references }
\label{subsec_rq5_typespapers}

\begin{figure}[htp]
 \centering
 \includegraphics[width=\linewidth]{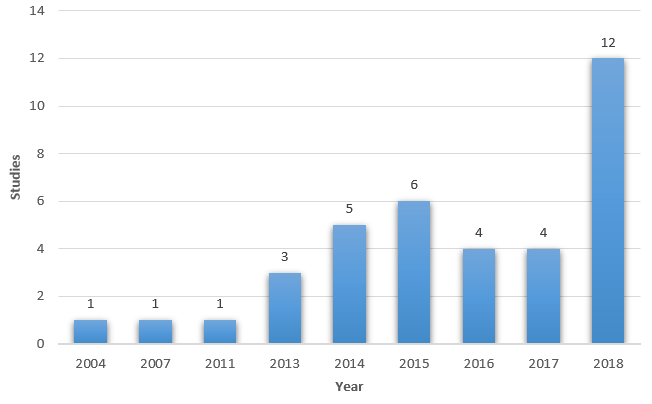}
 \caption{Distribution of publications per year}
 \label{fig:distribution_papers_year}
\end{figure}

Although our research defined as start reference the year 1992, the first secondary study [S10] published in 2004 (Figure \ref{fig:distribution_papers_year}). Then we had just two secondary studies publications until  2012 (2007 and 2011). After 2013, the number of publications of secondary studies increased. We note it is due to the popularization of SLR and SM in the SE field. Of these selected studies, 75\% published in journals and 25\% published in conferences.

\begin{figure}[ht]
 \centering
 \includegraphics[width=\linewidth]{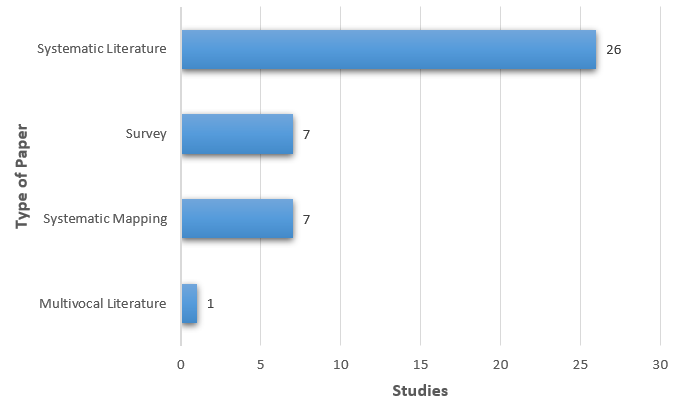}
 \caption{Type of secondary studies}
 \label{fig:type_of_paper}
\end{figure}

\textbf{Most of the studies refer to SLR (65\%)} [S2, S3, S4, S5, S6, S7, S8, S11, S12, S13, S18, S19, S20, S21, S22, S24, S25, S27, S28, S29, S32, S34, S35, S36, S37, S40], with 17.5\% refer to SM [S1, S14, S23, S26, S33, S38, S39] and 17.5\% refer to Survey [S9, S10, S15, S16, S17, S30, S31]. Only one study is multivocal literature [S16] (Figure \ref{fig:type_of_paper}). A multivocal literature mapping (MLM) \cite{Garousi:2016} is an SLR that include data from multiple types of sources, \eg~scientific literature and practitioners’ grey literature (\eg~blog posts, white papers, and presentation videos).
Multivocal mapping studies have just recently started to appear in SE literature. 

In the snowballing process, we revised 7141 references, finding 184 studies. Among them, we selected seven studies to complement our research, using our selection criteria. Most of the studies (65\%) did not use snowballing [S3, S4, S5, S6, S7, S8, S9, S10, S11, S12, S13, S14, S15, S19, S20, S22, S24, S28, S29, S30, S31, S32, S33, S36, S37, S38] and 35\% used snowballing [S1, S2, S16, S17, S18, S21, S23, S25, S26, S27, S34, S35, S39, S40] as a research mechanism \cite{Wohlin:2014}. We noted the \textbf{snowballing process appeared in studies published after 2015}. If we consider 2015 as a starting point, the use of snowballing is present in 50\% of the selected studies, showing the growth of the snowballing process in secondary studies.

\begin{figure}[ht]
 \centering
 \includegraphics[width=\linewidth]{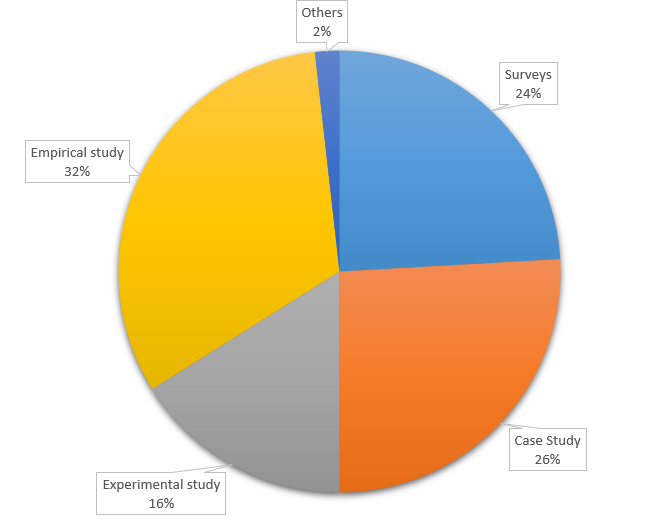}
 \caption{Distribution of studies types cited by secondary studies}
 \label{fig:type_researchpapers_used}
\end{figure}

Most of secondary studies used (around 58\%) empirical and case studies  (Figure \ref{fig:type_researchpapers_used}).
These types of studies are the preferred approaches for validating tools/prototypes. However, \textbf{there is a lack of validation using experts in this field}, qualifying the analysis. 
To increase confidence in empirical evaluation results in primary studies, we compile some information about secondary studies [S7, S22, S40]. According to such studies, it is necessary to pay attention to the following points:

\begin{enumerate}
    \item use a relatively large dataset implemented with different programming languages and considering a mixture of open-source and industrial systems; 
    \item clearly define the study goal (smell detections, refactoring techniques) considered;   
    \item fully identify the evaluation measures; 
    \item adequately describe the study participants; 
    \item clearly describe the scoring systems adopted; 
    \item compare the results with previous findings; and
    \item discusses validity threats.  
\end{enumerate}

According to the studies [S5, S6, S13, S22, S39, S40], although most researchers in this area are from academia, some participants in reported empirical studies are practitioners from industry, indicating they have some contact with each other.  We found three studies [S17, S39, S40], where the authors from the academic and industrial sectors worked together. In Section \ref{sec:implications}, we discuss in more detail the implications of researchers and practitioners on this type of work.

\subsubsection{Projects}
\label{subsec_rq5_projects}

\begin{figure}[ht]
 \centering
 \includegraphics[width=\linewidth]{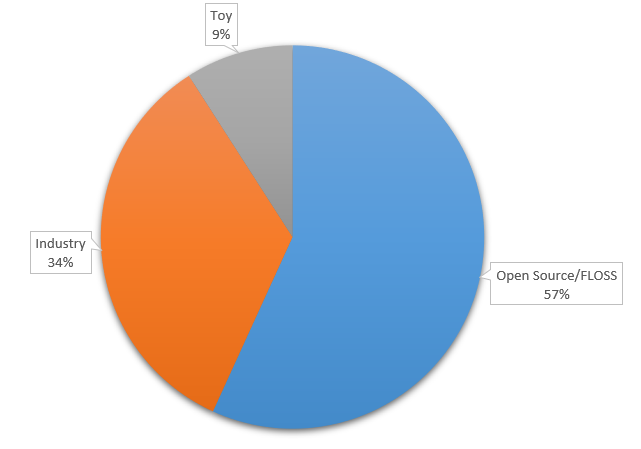}
 \caption{Type of referenced projects}
 \label{fig:type_of_project_referenced}
\end{figure}

Most of the academic researchers use nonindustrial datasets for the  studies [S22, S39, S40], usually open-source/FLOSS and Toy applications as illustrated by (Figure \ref{fig:type_of_project_referenced}). The \textbf{majority of projects realized experiments with FLOSS}, representing 57\% of the projects mentioned in the studies.

\begin{figure}[ht]
 \centering
 \includegraphics[width=\linewidth]{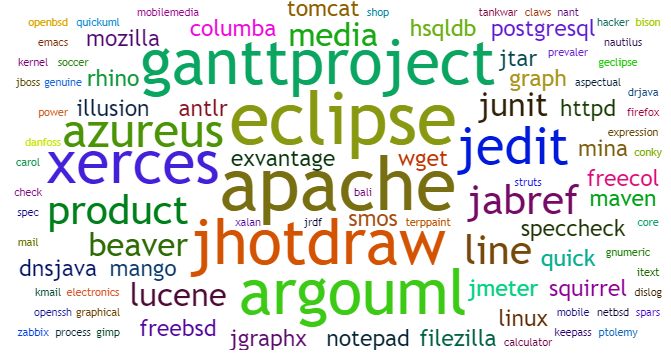}
 \caption{Word cloud of projects cited in secondary studies}
 \label{fig:cloud_tags_projects}
\end{figure}

A tags cloud containing keywords of selected studies provides a high-level picture of the cited software projects (Figure \ref{fig:cloud_tags_projects}). 
The projects more cited in studies are \emph{Apache, Eclipse, jHotDraw, ArgoUML}, and \emph{GanttProject}. Most of the more quoted  projects have some common features like a) they are long-lived projects (10+ years); b) they are structured in sub-projects; c) they are large projects, and d) they considered as marks in the open-source/FLOSS world.

According to [S18, S22], many authors perform experiments on open source projects for the evaluation of their techniques. Experiments on commercial/industrial projects are performed only by few authors.
On the one hand, it is easier to conduct experiments using FLOSS projects due to the availability of versions and constant evolution. More, many projects have often used as a basis for experiments. On the other hand, it is essential to put an emphasis on experiments carried out with industrial projects. It is a challenge to be faced in the coming years.
It may also be essential to analyze the ratio of code smells existing in open source versus industrial projects. \textbf{We do not have until now a benchmark of projects} [S22, S39, S40]. Some studies [S13, S18, S39, S40] point a lack of benchmark definitions for smells validated by experts. It happens with refactoring, too.
The large set of tools and systems used in the experimental settings suggest the lack of well-designed benchmarks should be better addressed.
The benchmarks could be constructed, having the same characteristics as the most used systems.

PROMISES\footnote{Research dataset repository is specializing in software engineering. 
More info: http://promise.site.uottawa.ca/SERepository/} is an example of a benchmark and an excellent initiative.
However, such a benchmark does not seem to be updated for some time and also does not have specific datasets for refactoring and smells. 
Similar initiatives should be encouraged to contribute to advances on this topic.

\subsubsection{Analysis of Selected Studies}
\label{subsec_rq5_analysis}

The selected studies have as the initial year of research 1990 (through 2010) and as the final year varying from 2004 to 2018. The average period used to search the selected secondary studies is 15.2 years, with an average of 1444 papers considered and 92.1 papers selected by the studies. The study [S8] considered the minor time was seven years (2007-2014) of a total of 165 studies, picking 18. The study [S40] considered the most extensive-time period of research (27 years, from 1990 until 2017), considering 9633 papers, selecting 351 for analysis and discussion.
The study [S34] has regarded as the most significant number of primary studies (13769), selecting 78. The study [S15] selected the most significant number of primary studies (445),  of a total of 1028. The study [S16] had the highest accuracy in the survey (52.86), selecting 166 studies out of a total of 314. The study [S34] had the lowest efficiency (0.56), choosing 78 studies out of a total of 13769. Probably, in all cases, this is a consequence of the selection criteria used by the authors.
The RQs used in the research has driven the study focus. \emph{A priori}, studies more comprehensive have more RQs, but in fact, it depends on the RQ comprehensiveness. The comprehensiveness of a study is directly related to the statement  of the RQs.

We also analyze if studies have search strings and inclusion/exclusion criteria explicit. Studies adopted different ways to explicit their search strings. Almost all studies without RQs also do not have explicit search strings and inclusion/exclusion criteria. Still, some studies have defined inclusion/exclusion criteria but did not have defined an explicit search string.
It shows that some authors did not explain the protocol of the study, making it difficult for its reproduction. A better decision would be to adopt a research protocol, such as those recommended by \cite{Kitchenham:2007, Kitchenham:2009}.

\begin{figure}[ht]
  \centering
  \includegraphics[width=\linewidth]{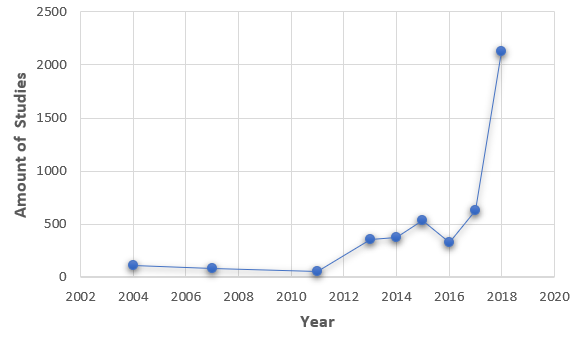}
  \caption{Number of primary studies referenced by secondary researches per year}
  \label{fig:number_of_prim_studies_year}
\end{figure}

The number of primary studies referenced can vary. A large number of regular surveys did not explicitly report the number of primary studies. In these cases, we counted the number of references of secondary papers, and use it as an estimate of the size of the study. The secondary studies have a sum of 4573 references, with an average of 114.32 references cited per study (median=99). Such data illustrated by  Figure \ref{fig:number_of_prim_studies_year}, which presents the number of primary studies analyzed in each secondary research, grouped by publication year.

As previously discussed (Sub-section \ref{subsec:quality_assessment}), each Survey, SM, and SLR was evaluated using a set of quality-related criteria used in earlier studies. 
For each study in our pool, the quality score calculated by assigning \{0, 0.5, 1\} to each of the four questions. The result value in the range of (0, 4) where 4 is the maximum score.

\begin{figure}[ht]
 \centering
 \includegraphics[width=\linewidth]{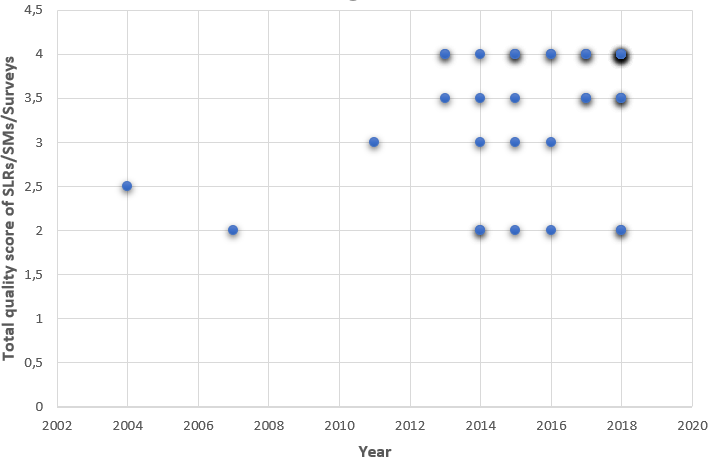}
 \caption{Total quality score per year}
 \label{fig:total_quality_score_year}
\end{figure}

As shown in Figure \ref{fig:total_quality_score_year}, all selected studies scored between 2 and 4. Surveys scored lower because they do not follow a formal protocol like SLR and SM do (even though some studies do not have RQs or selection criteria defined). Still, they are very cited in the literature. \textbf{More than 80\% of the studies we selected scored between 3 and 4}.

\noindent
\fcolorbox{black}{lightgray}{
    \parbox{\linewidth}{%
        {\footnotesize \textbf{RQ5 Summary} 
        
        \emph{\textbf{Challenges:}} 
        As we previously noticed on tools analysis, there is a lack of validation by experts in the field: experts should participate more effectively in experiments.
        
        Another challenge is the lack of benchmarks. It could enrich research and experiments, as well as support improved recall and accuracy of techniques and tools for both smells and refactoring.

        \emph{\textbf{Observations:}} 
        We noticed a growth of secondary studies in recent years. Considering the last three years, we have more than half of secondary studies than previously published.
        
        The vast majority of studies are SLR (65\%) — 75\% of the studies published in journals.
        The snowballing is not yet as present as the inclusion mechanism of new studies in the pool (it considering 2015 to 2018, 50\% of studies did not use the process).
        
        The preferred approaches for validating the proposed tools/prototypes are empirical studies and case studies. 
        
        The selected studies point to FLOSS projects (\ie~\emph{Apache, Eclipse, jHotDraw, ArgoUML}, and \emph{GanttProject}) as the most used for experiments. 
        
        We do not have a recognized benchmark of projects until now.
        
        }  
    }%
}
\section{The relationship between Code Smells and Refactoring}
\label{sec:qsr}

As we have seen in the studies reviewed, code smells have certain interesting features. Code smells are symptoms or design problems that may affect the evolution and maintenance of the software. Some of these code smells are small, and we call a simple smell. Often, the occurrence of one code smell may be related to or correlated (as shown in Figure \ref{fig:co_occurrence_smells}) with another code smell, deriving a composite smell, or design smell.
We have discussed in previous sections (see Sub-sections \ref{subsec:rq2} and \ref{subsec_rq3_smells_tools}) the main approaches for the detection of code smells, as well as the most cited tools to support this activity.

We also summarized some interesting characteristics of refactoring (see Sub-sections \ref{subsec:rq1} and \ref{subsec_rq3_refactoring_tools}). We note that there are simple refactoring, known as primitive refactorings. Often, for a given situation, we need to perform a sequence of refactoring, known as composite refactoring. Also, we present the more known tools supporting refactoring.

As we showed earlier, looking at code smells and refactoring in isolation, we noticed that there are some similar characteristics.

Code smells and refactoring affect software quality (see Sub-sections \ref{subsec_rq1_impact} and \ref{subsec_rq2_impacts}).
Quality is one of the most critical issues in software engineering, drawing attention from both practitioners and researchers. 
Developing software with quality is essential, but \textbf{preserving or increasing software quality during maintenance is even more critical}. 
Code smells produce software quality problems. Firstly, external quality attributes suffer from it over a long time, affecting the evolution of software, leading to increased technical debt.
Second, internal quality attributes are also affected by code smells. Some code smells produce problems like low cohesion, high coupling, encapsulation-related problems that influence design decisions and maintenance. 
Refactoring and code smells are linked, because refactoring are the main strategy to remove/mitigate code smells, improving the software quality (clarity, simplicity, comprehension). 
We know, as discussed in the Sub-sections \ref{subsec_rq1_evolution_td} and \ref{subsec_rq2_impacts}, that refactoring is the primary approach to mitigating technical debt. We also know that refactoring, if improperly applied, can generate new code smells and, consequently, affect the quality negatively.

\begin{figure*}[ht]
\centering
\includegraphics[width=14cm]{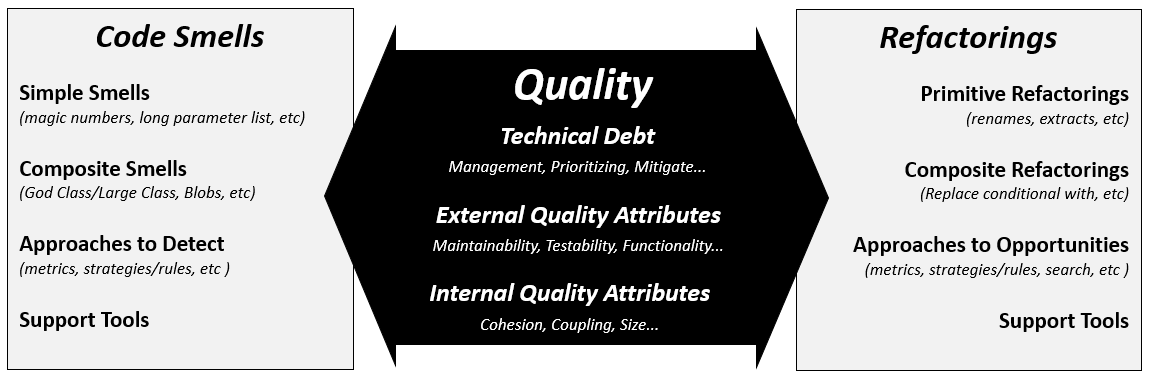}
\caption{Code smells and refactorings with some of their similar features. In the center, the aspects that connect them, by highlighting the quality}
\label{fig:relationship_smells_refactoring}
\end{figure*}

In Figure \ref{fig:relationship_smells_refactoring}, we present a scheme of this relationship, with the specific characteristics of code smells and refactoring, as well as the elements that connect them.
So, code smells and refactoring are closely related to software quality.

\subsection{Quality models, Code Smells and Refactoring}
\label{subsec_quality_models}

Different software quality models are found in the literature and referenced in the studies [S21, S23, S26, S27, S39]. Each model defines a set of main software quality attributes. Some attributes are common to different models. 
\textbf{The models mentioned in the studies were ISO/IEC 9126} \cite{iso:9126}, \textbf{FURPS} \footnote{https://en.wikipedia.org/wiki/FURPS}, and \textbf{McCalls Factor Model} \cite{MCcall:1977}.
However, the primary model of software quality factors mentioned in the selected studies is the ISO/IEC 9126.
The ISO/IEC 9126 model is the most comprehensive, providing six main features classified as external attributes (\eg~Functionality, Reliability, Usability, Efficiency, Maintainability, and Portability). Meanwhile, the current industry standard, called ISO/IEC 25010, mentioned in only one study [S26].

\textbf{Another difficult task investigated in the software refactoring field is the preference of code smells to be corrected, based on given importance} [S25]. 
A few studies [S15, S24, S35, S37, S38, S39, S40] identify the relationship between the detected types of smells and quality attributes. 
The nature of the relationship identified by the authors varies from one to another. \textbf{Most of the code smells}, in particular, defined by Fowler\al \cite{Fowler:1999} \textbf{affect more than one quality attribute}.
Therefore, \textbf{some quality attributes influence more than others}. The quality attributes most affected are maintainability, complexity, and understandability. They have a significant role in software maintenance costs. 
In this case, the \textbf{set of code smells related to these quality attributes will have the highest degree of priority for removal from the software}.

On the other hand, only one study [S7] has explored the impact of refactoring on quality attributes. The study [S7] presented external and internal quality attributes. 
The external quality attributes more often are maintainability, reusability, and understandability. Reliability and maintainability are attributes more studied.
The internal quality attributes more investigated are cohesion, coupling, complexity, inheritance, and size attributes, where coupling and size are the most and least considered attributes, respectively.  Coupling measures have also been one of the main approaches to evaluate decay [S38] and technical debt [S21].
It is important to note that estimated external quality attributes are quantified using combinations of internal quality measures such as cohesion, coupling, and inheritance. 
Therefore, studying \textbf{the impact of refactoring on a single internal quality attribute is potentially more straightforward and more accessible than addressing combinations of internal quality attributes}.
The study [S7] also recommends that \textbf{researchers conduct more empirical studies to explore the impact of refactoring on external quality attributes because these attributes are of direct interest to practitioners}.

According to studies [S7, S22], the researchers observed that \textbf{different refactorings potentially have different, and sometimes conflicting, impacts on quality}. 
It is difficult to distinguish between the effects of individual refactoring scenarios or to draw any conclusions regarding their impacts on quality. 
\textbf{One recommendation is must apply a set of refactoring that follow the same scenario and assess quality before and after refactoring}.

\subsection{Analysis}
\label{subsec_tool_analysis}

\begin{figure*}[ht]
 \centering
 \includegraphics[width=\linewidth]{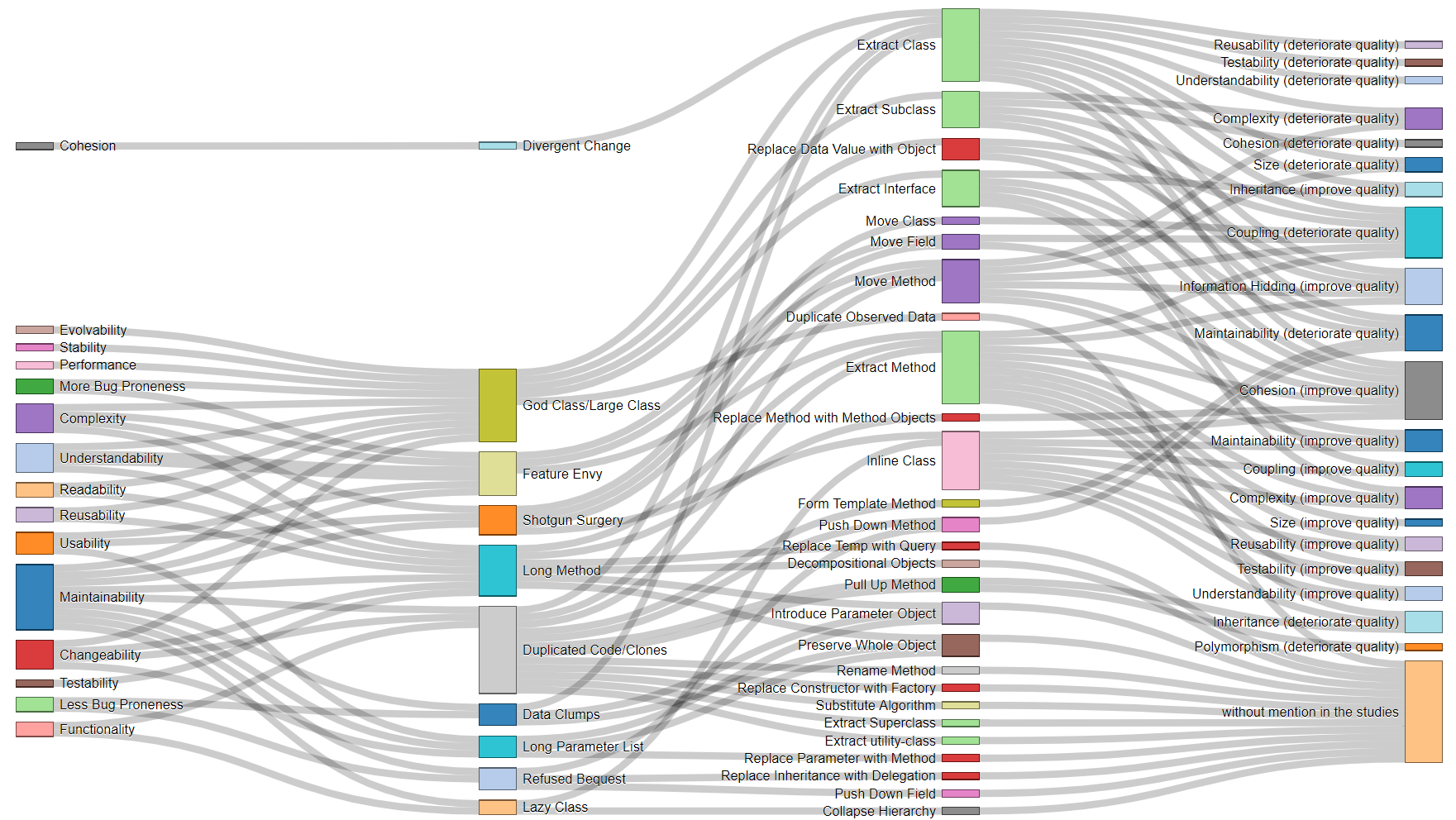}
\caption{The relationship among (external and internal) quality attributes and their impact on code smells and refactoring}
 \label{fig:quality_smells_refactoring_sankey}
\end{figure*}

We found studies [S15, S24, S35, S37, S38, S39, S40] focused on verifying how the occurrence of code smells impacts several quality attributes. 
In the same way, we found one study [S7] discussing refactoring and their impact on quality. 
With a base on results, we develop a visualization (Figure \ref{fig:quality_smells_refactoring_sankey}), that allows analyzing the relationship among quality attributes, which code smells affected them, which refactoring can be applying to these code smells, and what is the impact on quality when using such refactoring.

The \textbf{attributes of quality most affected by code smells are maintainability, understandability, and complexity}.
However, evolvability, stability, performance, and testability mentioned in only one code smell, each one.

\textbf{The code smells that most affect different quality attributes are \emph{God Class/ Large Class, Long Method}, and \emph{Feature Envy}. \emph{God Class/Large Class} is the most affect quality attributes.} It affects maintainability, complexity, evolvability, stability, performance, readability, reusability, and changeability. \emph{God Class/Large Class}  and \emph{Feature Envy} are more prone to bugs, as well as affecting complexity, understandability, usability, and maintainability. \emph{Long Method} affects complexity, understandability, readability, reusability, maintainability, changeability, and testability.

Some code smells have a larger set of refactoring than others (see Table \ref{tab:main_solutions}). It is the case of \textbf{\emph{Duplicated Code/Clones}, which have more refactoring that can be applied}. In this smell, we can use \emph{Pull Up Method, Rename Method, Replace Constructor with Factory, Form Template Method, Pull Up Method, Push Down Method, Substitute Algorithm, Extract Superclass, Extract Class, Extract utility-class}, and \emph{Move Method}.
\emph{Long Method} presents an alternative the refactoring \emph{Extract Method, Replace Temp with Query, Introduce Parameter Object, Preserve the Whole Object, Replace Method with Method Objects}, and \emph{Decompositional Objects}. \emph{God Class/Large Class} also presents some refactoring, such as 
\emph{Extract Class, Extract Subclass, Replace Data Value with Object, Extract Interface}, and \emph{Duplicate Observed Data}.

According to previously presented, the relationship between refactoring and code smell is not one-to-one [S7]. 
\textbf{Refactoring are flexible, can be applied in more than one code smell}. It is the case of \emph{Extract Class, Move Method}, and \emph{Extract Method}, which also have been more studied by researchers.

We also commented that refactoring does not always improve all quality attributes. \textbf{When quality improvement is the goal of refactoring, developers should be careful and check whether the application's proposed refactoring achieves the desired goal}.
Developers need to know they apply such refactoring on smell, which can lead to the introduction of other ones. 
It is important to note that by positively affected by refactoring on a quality attribute, we mean that the refactoring causes the value of measure that quantifies the quality attribute to increase, and vice versa.  
However, increase the value does not always mean that the quality is improved because, for some quality attributes (\eg~coupling, complexity, size), the improvement indicated by the decrease in the corresponding value. Of course, it depends on how being calculates such a metric.

Refactoring also affect different quality attributes. For instance, \textbf{\emph{Extract Method} and \emph{Extract Class} are refactoring that most affect different quality attributes} (10), with \emph{Inline Class} affected 8 attributes. \emph{Extract Method} affects inheritance and coupling negatively. The same refactoring affects complexity, cohesion, size, information hiding, maintainability, reusability, testability, and understandability positively.
\emph{Extract Class} affects inheritance, cohesion positively, and information hiding. Also, it affects coupling, complexity, size, maintainability, reusability, testability, and understandability negatively.

It is also important to mention that \textbf{14 refactoring options for the presented code smells do not have studies associated with their impact on quality}. It is the case of \emph{Replace Constructor with Factory, Substitute Algorithm, Extract Superclass, Extract utility-class, Introduce Parameter Object, Duplicate Observed Data, Replace Temp with Query, Preserve Whole Object, Decompositional Objects, Replace Parameter with Method, Replace Inheritance with Delegation, Push Down Field, Collapse Hierarchy}, and \emph{Rename Method}. Therefore, we do not know their impacts when applying these refactoring.

Besides, when evaluating the quality, \textbf{developers are advised to consider several quality attributes and not focus on one particular attribute}, ignoring others. Otherwise, the proposed refactoring may detract from quality rather than improve it.
We show that the impact of refactoring on most of the measured external quality attributes has not studied. Consequently, the impact of refactoring on measured external quality attributes requires more research and study.
We recommend that researchers conduct more studies to explore the impact of refactoring on external quality attributes because these attributes are of direct interest to practitioners.

The studies [S7, S24, S35, S37] are recent and \textbf{drive the necessity to answer if code smells harms the project, as well as the impact on refactoring}.  
Moreover, the \textbf{number of quality attributes and their possible combinations with distinct code smells are high, thus requiring different studies. The same happens with refactoring}. 
The study of impact/effect has been receiving so much attention until now. It suggests that there is still no comprehensive and sufficient evidence on the extent of adverse effects associated with code smells and positive impact on refactoring on software maintenance and evolution. 

\begin{figure*}[ht]
 \centering
 \includegraphics[width=\linewidth]{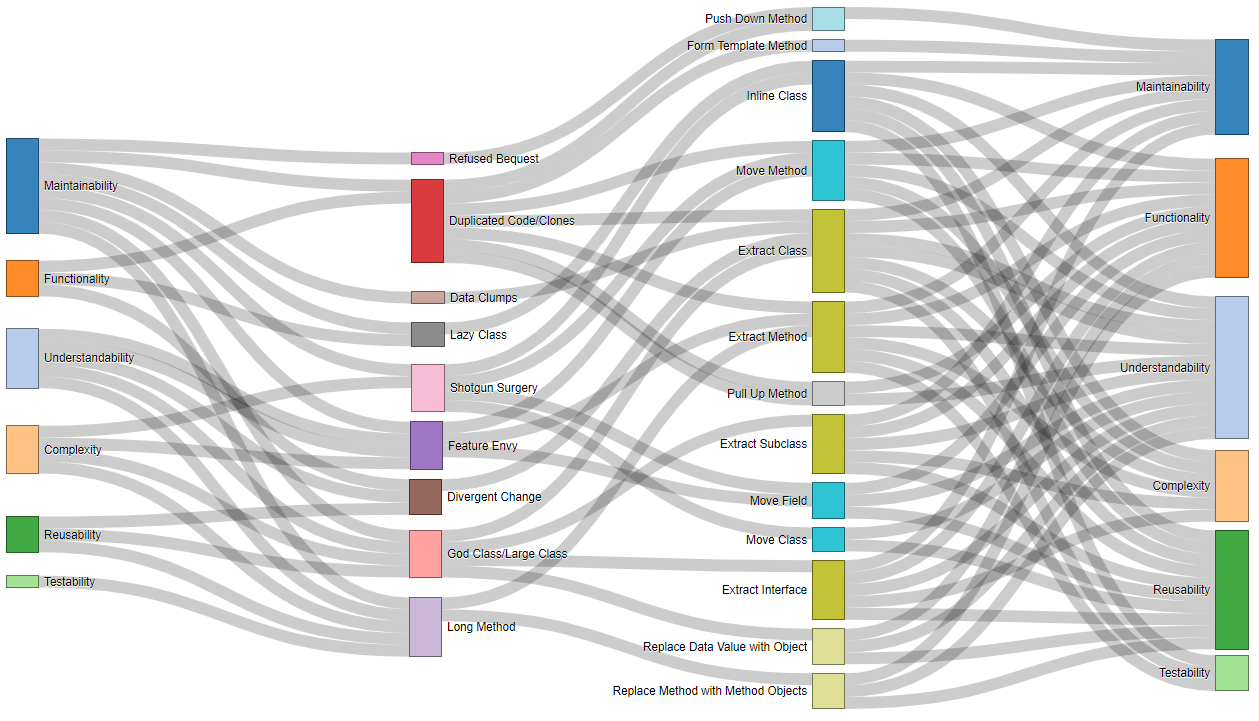}
\caption{Relationship among quality attributes using the QMOOD model, impact on code smells and refactoring}
 \label{fig:quality_smells_refactoring_qmood_sankey}
\end{figure*}

In studies on code smells, more external quality attributes referenced. In refactoring studies, there were internal and external attributes. Of course, about refactoring, we may experience code deterioration or improvement, as we discussed earlier.
To better represent this relationship between internal and external quality attributes, we have grouped internal attributes with their proper relation to external attributes. For this, we use the QMOOD model \cite{Bansiya:2002}.
We consider only the internal attributes found in the secondary studies and, based on QMOOD model, observe where they applied. We also ignored the quality attributes referenced by QMOOD model but not mentioned in the studies used in our research. Thus, we define that coupling used in reusability and understandability. We observe that both cohesion and size used in reusability, understandability, and functionality. Understandability uses information hiding. Functionality and understandability use Polymorphism.
As shown in Figure \ref{fig:quality_smells_refactoring_qmood_sankey}, we present a new view with a different arrangement of quality attributes.

\textbf{We note that the quality attributes most affected by code smells are also the ones most affected by refactoring.}
We identify that refactoring have more impacts on under\-stan\-da\-bi\-li\-ty, func\-tio\-na\-li\-ty, re\-usa\-bi\-li\-ty, and main\-tain\-a\-bi\-li\-ty.
With this new redistribution, we note that by grouping internal attributes with external attributes using the QMOOD model, \textbf{refactoring affect quality more than code smells}.
It demonstrates that \textbf{this relationship is not only due to the code smell refactoring link, but that the origin of the relationship is quality}. And that in the medium to long term, depending on the context of refactoring, can lead to new code smells, generating a cycle that can further deteriorate the software.
Therefore, we recommend that other studies explore this relationship of code smells and refactoring quality as their main aspect.

\section{Implications}
\label{sec:implications}

We now discuss the implications of this systematic literature review on code smells and refactoring for practitioners, researchers, and instructors because, as explained by Goues \al \cite{Goues:2018}, it is essential that SLRs provide some advice beyond their RQs.

\paragraph{Practitioners} Real software systems must be continually changed to meet the demands of the market and the expectations of their users. Thus, software development teams must be always concerned by the continuous improvement and quality of their systems \cite{Canfora:2011}. However, there are still many challenges related to software maintenance and evolution, including the need to understand the systems and the complexity involved in the development process. The lack of understanding of architectural deviations during software evolution compromises both the development process and the systems themselves.

Leppänen \al \cite{Leppanen:2015} present a decision-making framework, based on practitioners' perceptions. They claim that the need for refactoring is rather subjective and not necessarily rational. They show the empirical nature of the refactoring process. Exposure to the real world brings invaluable insights \cite{Griswold:2015, LeppanenMakinen:2015, Lahtinen:2016}. Developers are the real specialists who, with their perceptions and experiences, can say which structure is better than the others. Their knowledge should drive refactoring. We argue that refactoring should be a daily habit. The more refactoring are neglected, the greater the likelihood and need of doing larger refactoring, which are more problematic: they must be planned, they interrupt daily work, and often they must be justified to management. We reported in this paper that the most applied refactoring are primitive refactoring. Indeed, they are simpler than composite refactoring and are automated in tools.

The use of version control, testing, and reviews are encouraged as good practices \cite{LeppanenMakinen:2015}. Code reviews, for example, can help find targets for refactoring. We observed\footnote{One of the authors of this work worked for 20+ years in the industry, helping software development teams to improve code quality. This perception is based on his experiences.} that these practices help reduce code complexity, maintain quality, and improve the source code in the long run. New research discussing these practices in relation to refactoring could bring new perspectives on how developers should address such good practices.

Developers know the values of refactoring but are often prevented from applying them \cite{Tempero:2017}. One of the reason preventing the use of refactoring is the lack of measures showing their impacts. The lack of monitoring of refactoring was also pointed in several works \cite{MensDemeyerBois:2003, Leppanen:2015, LeppanenMakinen:2015}. We observed that, when prioritizing code smells, seeing the relationship and the impact of quality attributes to code smells helped their prioritization. When performing refactoring, it is also necessary to assess their impacts.

\paragraph{Instructors:} The implications described for practitioners are also valid for instructors.

The concept of ``quality'' is present in all software engineering knowledge areas (KAs) \cite{Swebok:2014}. We highlight that design, construction, testing, maintenance, models and methods, quality, and computing foundation KAs associating topics discussed in this work. We present a pragmatic way to support instructors in their classes.

The discussion about code smells and refactoring began with the introduction of eXtreme Programming in curricula \cite{Goldman:2004}. Some studies \cite{Smith:2006, Stoecklin:2007} described experiences in applying some lessons learnt based on self-documenting and functional tests, encapsulation, and unit testing, refactoring of constants and variables and to extract methods. These studies did not include all the lessons needed to learn refactoring, but reported some benefits, like the importance of self-documenting code, code smell recognition, testing comprehension, and improvement of code style. Then, several approaches were proposed to support instructors in learning/teaching refactoring, including tutoring systems \cite{Haendler:2019a}, agents \cite{Sandalski:2011}, gamification \cite{Haendler:2019b}, and on-line teaching \cite{Lopez:2014}. Yet, it is necessary to use real-world examples for learning and teaching. The use of a existing source code with real (or injected) problems promotes a collaborative environment to exchange knowledge.

We identified that these topics should be included in various courses in curricula, such as introduction to programming, software engineering, software quality, software design. We observed that the exposition to change brings students the perception of concepts that are progressive and intuitive. The reasoning behind refactoring is also essential (\textit{when do I refactor? What do I refactor? Why is this refactoring better?}). When refactoring are taught and used in class, they help students to acquire good programming practices and design principles.

We reported several aspects that instructors could consider in their classes because we are instructors ourselves. We have had the opportunity to apply and to discuss code smells and refactoring in classes. We taught this topic in undergraduate\footnote{Object-Oriented Programming, Software Engineering, and Software Architecture at UniRitter; Techniques of Program Construction at UFRGS} and graduate courses\footnote{Smells, Patterns and Refactoring at UniRitter; Agile Development Introduction and Agile Development with eXtreme Programming at Unisinos}. One of the practices that we use and recommend is to perform \textit{Coding Dojos}\footnote{http://codingdojo.org/}. A \textit{Coding Dojo} allows students to learn practices, such as test-driven development, refactoring, code review, pair programming, and the use of tools such as static analysis, code coverage, and build tasks. We encourage instructors to incorporate both code smells and refactoring into their classes to develop students' technical skills. New developers should be aware of code smells and how to remove them.

\paragraph{Researchers} We sought to show the close relationship between code smells and refactoring. We compiled the main results of the secondary studies and highlighted several challenges.

We understand that refactoring exist because code smells indicate that something is not right in the source code. Thus, a standardized form of code smell definitions, detection methods, and tools is needed. Dig \cite{Dig:2017} showed that there is growing interests on automation, prioritization, inference, and recommendation of refactoring. Researchers are encouraged to explore such interests together with the practice of refactoring. They should focus their research on refactoring that are often applied in practice rather than study refactoring opportunities already considered in the literature and--or rarely applied in practice. Researchers should also explore improving the results of applying refactoring that are commonly used in practice as well as propose refactoring tools.

Also, researchers should analyze the impacts of code smells and refactoring on software quality and technical debt. They must strive to bring their research closer to the software industry. Recent studies \cite{Garousi:2019a, Garousi:2019b, Garousi:2019c} on industry and education in software engineering showed a large gap in several areas, including quality and design. These areas are closely related to the topics discussed in this study. Researchers do work with industry: almost 50\% of collaborations started in industry and 90\% generated at least one paper \cite{Garousi:2019c}, which show the importance of industry-academic collaboration. 
Researchers should strive to create partnerships, showing how they can be useful for both areas.

\section{Open Issues}
\label{sec:open_issues}

We now present some open issues, \ie{} questions for future works, concerning code smells detection, refactoring techniques, support tools, impact on quality, and academic research and its relationship with industry.

Mens \al \cite{MensDemeyerBois:2003} presented some future trends in research in 2003. After more than 15 years, some questions remain open, interesting topics for future works. We discuss in detail each identified issue, summarized in Table \ref{tab:open_issues}.

As we noted in relation to studies [S11, S12, S15, S39, S40], there is a problem of naming and organizing code smell (see Subsection \ref{subsec_rq2_codesmells}). Code smells have definitions that are sometimes complex and not informal \cite{MurphyHill:2008}. Therefore, researchers could research the standardization of code smells.

As shown in Table \ref{tab:main_solutions}, several studies [S2, S13, S15, S18, S19, S20, S30, S31, S32, S33] presented many approaches to code smell detection (see Subsection \ref{subsec_rq2_context}). However, we still must understand which approaches are the most effective. Some code smells have more than one approach while others have none. Researchers could study the combinations of approaches to code smell detection as well as propose approaches for currently undetected code smells.

There is an opportunity to assess whether the most cited code smells impact the industry. Recent studies \cite{Taibi:2017, Hozano:2018} also evaluated the developers' perceptions and how they detect smells. Researches should also assess whether practitioners recognize code smells and whether practitioners know that code smells are symptoms of problems.

In the code-smell detection process, researchers should evaluate the developers' participation (\eg when, how) [S2, S11] because they are critical in this process \cite{SantosNeto:2015}. Developers' insights and experiences can be explored in future work to improve the detection process.

Some studies [S24, S35, S37, S38] discussed the impact of code smells on quality attributes (see Subsection \ref{subsec_rq2_impacts}). Other studies [S15, S23, S24] did not establish an explicit connection between smells and quality, which is an opportunity for further work.

Following previous studies [S2, S4, S9, S13, S15, S18, S19, S39, S40], we organized a list of code smell detection tools (see Subsection \ref{subsec_rq3_smells_tools}). However, many of them are obsolete or have severe limitations (scalability,  prioritization, visualization, multi-smells, multi-language , low precision, and recall). Researchers should invest into developing more comprehensive tools.

Research related to developers' knowledge about refactoring is essential [S10, S22] to understand the developers' mental models when refactoring. Although IDEs provide some automated refactoring, some studies showed that developers do not use them \cite{MurphyHill:2007b, Vakilian:2012} (see Subsection \ref{subsec_rq3_refactoring_tools}) because of usability or lack of knowledge of the refactoring process \cite{Murphy-Hill:2007}. Researchers could evaluate whether the decision is rational or subjective, refactoring is a daily activity, factors used in the decision to refactor, learning, among others.

Some studies discussed refactoring and their impact on quality attributes [S4, S5, S7, S22] (see Subsection \ref{subsec_rq1_impact}) but with small numbers of refactoring (Subsection \ref{subsec_rq1_techniques}) and different quality models (Section \ref{sec:qsr}). Researchers could study systematically refactoring and their relationship to quality. Other studies could measure the impact of refactoring on individual attributes (\eg security, understandability, and extensibility) and studies to evaluate most commonly used refactoring by developers that affect the quality (decay or improvement), using quality models such as ISO 25010.

Some studies [S2, S9, S17, S18, S39, S40] presented refactoring tools. Table \ref{tab:main_solutions} and Subsection \ref{subsec_rq3_refactoring_tools} show that there are few refactoring tools and some are obsolete. There is an opportunity to propose and improve refactoring tools, especially tools to predict and evaluate the effects of refactoring.

We showed in Subsection \ref{subsec_rq1_impact} that applying refactoring can cause problems [S21, S23]. Therefore, we identify a research opportunity on refactoring monitoring, to monitor the effects of refactoring on software evolution.

Some studies [S21, S23, S26, S27] discussed architectural refactoring and their benefits as well as whether refactoring should be contextualized (\eg method, class, and package). However, it is an open issues for researchers to explore the frontiers between contexts and their benefits for quality.

Teaching code smells and refactoring is interconnected (see Section \ref{sec:implications}). Researchers could develop studies in the classroom about best practices and tools that enhance learning, training future professionals with this awareness.

Code smells and refactoring are evolving research areas and some studies bring trends, opportunities, and gaps [S5, S15, S17, S20, S21, S23, S24, S26, S27, S34, S37, S40]. We suggest topics such as identifying reliable datasets that can be compared with existing studies, using large-scale academic and industrial projects to generate more reliable conclusions and analyzing industrial and academic studies looking for data inconsistencies.

\begin{table*}[ht]
    \caption{Open issues on smells, refactoring or both}
    \label{tab:open_issues}
    \tiny
    \begin{center}
        \begin{tabular}{p{5cm} p{1.5cm} p{10.5cm} }
            \toprule
            \textbf{Issue} & \textbf{Topic} & \textbf{Observations}  \\
            \hline    
    \textit{I1 - Code smell naming}    & Smell &    There is an apparent problem of code smelling nomenclature [S11, S12, S15, S39, S40] 
    \\
    \myrowcolour
    \textit{I2 - Approaches to code smell detection}    & Smell & We need to explore which approaches are most effective in smell detection [S2, S13, S15, S18, S19, S20, S30, S31, S32, S33]    
    \\
    \textit{I3 - Code smell and industry perception}    & Smell &  We need to assess whether practitioners recognize code smells and whether practitioners know this code smells as symptoms of code problems
    \\
    \myrowcolour
    \textit{I4 - Developers and code smell detection process}    & Smell &  It is essential to evaluate the participation of developers in the code smell detection process [S2, S11], using the developers' insights and experiences to improve the detection process  
    \\
    \textit{I5 - Code smell and impact on quality attributes}    & Smell &  The studies [S15, S23, S24] do not establish an explicit connection between smells and quality, showing there is an opportunity for further studies  
    \\
    \myrowcolour
    \textit{I6 - Code smell tools}  & Smell &  There are many opportunities to explore smell detection tools [S2, S4, S9, S13, S15, S18, S19, S39, S40]
    \\
    \textit{I7 - Developer's refactoring knowledge}    & Refactoring & The use of developers' knowledge about refactoring [S10, S22] can help to improve the refactoring process, refactoring tools, among others
    \\
    \myrowcolour
    \textit{I8 - Refactoring and impact on quality attributes}    & Refactoring &  There are many opportunities to research a low explored refactoring, most commonly used refactoring by developers and their relationships with quality attributes [S4, S5, S7, S22]
    \\
    \textit{I9 - Refactoring tools}    & Refactoring & There is an opportunity to propose/improve refactoring tools [S2, S9, S17, S18, S39, S40]
    \\
    \myrowcolour
    \textit{I10 - Refactoring tracking}    & Refactoring & There is a research opportunity on refactoring tracking and monitoring the effects of refactoring [S21, S23]
    \\
    \textit{I11 - Architectural refactoring versus contextual refactoring}  & Refactoring & It is an open theme for researchers to explore where is this frontier about architectural and contextual refactoring (\eg~package, class, method), as well as its benefits for quality [S21, S23, S26, S27]
    \\
    \myrowcolour
    \textit{I12 - Teaching code smell and refactoring}    & Both & Researchers can develop studies in the classroom about best practices and tools for learning and training
    \\
    \textit{I13 - Research gaps on code smell and refactoring}    & Both &  Production of reliable datasets, use of large-scale academic and industrial projects, industrial and academic analysis looking for data inconsistencies are examples of trends, opportunities, and gaps [S5, S15, S17, S20, S21, S23, S24, S26, S27, S34, S37, S40]
    \\
    \bottomrule
    \end{tabular}
    \end{center}
\end{table*}
\section{Threats to validity}
\label{sec:threats}

This section discusses some threats to the validity and some decisions to mitigate them. The search string of an SLR needs to very well defined to return secondary studies that are relevant to the search topic. In this study, we used several synonyms referring to the main terms of the SLR goal searched. Some pilot searches conducted to find new synonyms for the search string. Therefore, we believe the defined search string has returned as many relevant secondary studies as possible.
Thus, we seek to broaden our research spectrum. However, not all topics covered by secondary studies. It also does not mean that the community does not include this topic.

The choice of electronic databases is another factor that may impact the results of an SLR. In this study, we perform a search for secondary studies in eight different electronic databases. Therefore, other databases not used in the survey may contain work that is relevant to this review. To reduce this threat, we do carry out a snowballing process to find more potentially relevant studies. In this step, the citations of the selected papers verified through a list of references to find pertinent other studies not included initially on our search.

This SLR considered only papers written in English. Some relevant studies may be written in other languages. However, the primary venues of scientific publication in SE accepts papers in English. Therefore, we consider that using English is sufficient to filter the main studies on the subject.

Besides, the classification scheme of studies is another point that considered a threat to validity. The data extraction was performed subjectively and grouped into categories to facilitate the reading and the understanding of the readers. 
To avoid bias, we follow some procedures. However, other reviews can have different classification schemas and ways to group and analyze the papers. Another threat is related to the granularity of the information presented in the reviewed secondary studies. If some information is not described in these studies, it may affect our conclusions. Reliability validity is concerned with issues that affect the ability to draw that the operations of a study can be repeated with the same results. Our research can easily replicated following the steps described and using the search string.
\section{Conclusion and Future Work}
\label{sec:conclusion}

We performed a tertiary study on code smells and refactoring. We systematically analyzed 40 secondary studies, answering five RQs to present the main challenges (\emph{what we do not know}) and observations (\emph{what we know}) related to code smells and refactoring. 
We summarize some of the main findings below. 

We show that the majority of studies discuss smells and refactoring separately (Figure \ref{fig:mindmap_studies}). Only two secondary studies explore both explicitly.
Smells have the majority of secondary studies (62.5\%), with different focuses of study. 
\emph{Duplicated Code/Clones} and \emph{God Class/Large Class} are the most mentioned code smells. Also, \emph{God Class/Large Class} has been the most investigated smells related to technical debt. We observe problems related to smells definition, affecting their detection. 

Another challenge deserving more studies are the co-occurrence of code smells, that is, the appearance of code smell in consequence of another or code smells that are always close. 
There have several detection approaches, being metrics-based, and strategies/rules the most cited among them. Besides, smell detection approaches and the corresponding produced results are highly inconsistent. There is no consensus on the standard threshold values for the detection of smells, which are the cause of the disparity in the results of different approaches. Some approaches, like probabilistic/search-based, have grown and deserve attention.

Extraction refactoring (such as extract classes or methods) have been the most explored in the studies.
Only a small set of refactoring have studied (around 27 of 72), opening up possibilities for studies that evaluate other refactoring, as well as assess the reason for not exploring them.
Although refactoring has been an ally of developers to reduce technical debt, the refactoring do not always improve code quality.

We found 162 distinct smell detection tools, and \emph{CCFinder} is the most cited one. 
Also, we found 24 distinct refactoring tools, and \emph{JDeodorant} is the most cited one. We have also seen that there is a gap in the development of refactoring support tools, such as the extension of techniques not yet exploited (opportunities, execution, developer support, impact on quality).
We cross-referenced the most frequently reported smells with detection approaches, detection tools, suggested refactoring, and refactoring tools (Table \ref{tab:main_solutions}).
We note that even though there are several smell detection tools, many are discontinued or have low accuracy. Similarly, we see that there is room to explore refactoring tools in these smells.

We present some open questions as a basis for future studies on smell detection, refactoring, supporting tools, impact on software quality, as well as research that considers both academia and industry (Section \ref{sec:open_issues}).

Our study shows quality attributes make relationships between refactoring and code smells.
We noticed that the quality attributes affected by code smells were the same affected by refactoring. Code smells and refactoring have a relationship with understandability, maintainability, testability, complexity, functionality, and reusability. Besides, we also observed refactoring affect quality more than code smells.

The relationship between code smell and refactoring have several open questions. For instance, which refactoring could apply in a specific code smell? Which refactoring can be combined to mitigate such code smell? Which refactoring have the most significant impact on quality? Briefly, we suggest more studies to investigate refactoring and code smells jointly as a single phenomenon.

We hope this study could instigate researchers to investigate more deeply both practices and tools to mitigate the code smell and evaluate the impact on quality. In the same way, we suggest studies to explore the goal of refactoring used by practitioners and their effect on quality as well as the development/improvement of refactoring tools to monitor refactoring and its gains.

\bibliographystyle{elsarticle-num}
\bibliography{references}

\end{document}